\documentclass[a4paper,useAMS,usenatbib]{mnras}
\pdfoutput=1

\usepackage{graphicx}
\usepackage[fleqn]{amsmath}
\usepackage{url}
\usepackage[T1]{fontenc}
\usepackage{aecompl}
\usepackage{color}
\usepackage{bm}

\newcommand\bW{{\bf W}}
\newcommand\bF{{\bf F}}
\newcommand\bG{{\bf G}}
\newcommand\bZ{{\bf Z}}
\newcommand\mA{\mathcal{A}}
\newcommand\mB{\mathcal{B}}
\newcommand\mR{\mathcal{R}}
\newcommand\mK{\mathcal{K}}

\title[Multidimensional upwind hydrodynamics on GPUs]{Multidimensional upwind hydrodynamics on unstructured meshes using Graphics Processing Units \\
I. Two-dimensional uniform meshes}
\author[S.-J. Paardekooper]{S.-J. Paardekooper$^{1,2}$\thanks{E-mail:
s.j.paardekooper@qmul.ac.uk}\\
$^1$Astronomy Unit, School of Physics and Astronomy, Queen Mary, University of London, Mile End Road, London E1 4NS,
United Kingdom\\
$^2$DAMTP, University of Cambridge, Wilberforce Road, Cambridge CB3 0WA,
United Kingdom}
\begin{document}

\date{Draft version \today}

\pagerange{\pageref{firstpage}--\pageref{lastpage}} \pubyear{2016}

\maketitle

\label{firstpage}

\begin{abstract}
We present a new method for numerical hydrodynamics  which uses  a multidimensional generalisation of the Roe solver  and operates  on an unstructured triangular mesh. The main advantage over traditional methods  based on Riemann solvers , which commonly use one-dimensional flux estimates as building blocks for a multidimensional integration, is its inherently multidimensional nature, and as a consequence its ability to recognise multidimensional stationary states that are not hydrostatic. A second novelty is the focus on Graphics Processing Units (GPUs). By tailoring the algorithms specifically to GPUs we are able to get speedups of $100-250$ compared to a desktop machine. We compare the multidimensional upwind scheme to a traditional, dimensionally split implementation of the Roe solver on several test problems, and we find that the new method significantly outperforms the Roe solver in almost all cases.  This comes with increased computational costs per time step, which makes the new method approximately a factor of $2$ slower than a dimensionally split scheme acting on a structured grid.
\end{abstract}


\section{Numerical gas dynamics}

The observation that $99\%$ of the visible matter in the  Milky Way  is in gaseous form  \citep{draine07} makes gas dynamics an important part of the study of various systems in astrophysics, from stars and supernovae to accretion discs and gas giant planets. The non-linear nature of the governing equations makes numerical simulations an essential tool to make progress in our understanding of these systems. A wide variety of methods exists for solving the equations of gas dynamics, all of which perform better at some problems than others. Below, we give a very brief overview of the numerical landscape, which serves to put our new method in context.

Numerical methods for gas dynamics solve discretised versions of the governing equations. A first choice when selecting a method to tackle a particular problem in hydrodynamics is whether the resolution elements move with the gas flow (the Lagrangian approach) or are fixed in space (the Eulerian approach). Eulerian methods use a computational mesh to discretise space into small elements (usually squares in two dimensions, cubes in three dimensions). Popular methods include methods based on finite difference approximations such as Pencil \citep{brandenburg02} and {\sc zeus}  \citep{stone92} or  methods based on  Riemann solvers such as {\sc flash} \citep{fryxell00} and {\sc athena} \citep{stone08}. These two classes of Eulerian methods take in some sense opposite viewpoints of the underlying solution. Finite difference methods assume the flow to be smooth, and  in order to prevent unphysical oscillations near discontinuities  add artificial viscosity  in order to smooth these out. Godunov methods based on  Riemann solvers view the underlying solution as a set of discontinuities, for which the time evolution can be computed by solving Riemann problems.  While this means discontinuities in the flow can be handled in a natural way, it also restricts the method to be first order accurate in space and time.  In regions of smooth flow, higher-order methods  can be used safely. In order to avoid unphysical oscillations near discontinuities, the contribution of the high-order method should be limited. If the chosen limiter function is Total Variation Diminishing \citep[TVD,][]{vanLeer74} oscillations near discontinuities can be avoided.

In addition to these two classes of solvers, spectral methods are available \citep[e.g][]{lesur05, burns16, lecoanet14}, that solve the Navier-Stokes equations in spectral (usually Fourier) space. Advantages of Eulerian methods in general are the low intrinsic numerical dissipation, and, in the case of Riemann solvers \citep{falle02},  the automatic addition of the correct amount of dissipation for  shock waves. Main disadvantages include that dissipation is highly non-linear,  making it more difficult to control , and for example velocity-dependent \citep[][hereafter S10]{springel10}, and that it is not trivial to vary the resolution within the computational domain while maintaining low dissipation. For example, adaptive mesh refinement \citep[AMR,][]{berger84} leads to locations in the mesh where dissipation is especially high (where jumps in resolution occur and the update is usually only correct up to first order, see however \cite{schaal15}). Slowly varying spatial resolution can be achieved by choosing an appropriate coordinate transformation, for which second-order methods exist \citep[e.g.][]{eulderink95}. However, in that case it has to be known in advance where high resolution is needed, and, furthermore, if orthogonal coordinates are desired, this limits the ability to achieve high resolution locally.

While  staggered mesh Lagrange plus remap  methods do exist, in which the flow is remapped onto the mesh every time step \citep[e.g.][]{woodward84, pember00}, the most well-known Lagrangian method in astrophysics is the mesh-free method of  Smoothed Particle Hydrodynamics \citep[SPH, ][]{lucy77, gingold77}, where the gas is represented by a set of particles that move with the flow. The Lagrangian nature of SPH gives it two main advantages over traditional grid-based methods: errors associated with large-scale bulk motion of the fluid are virtually non-existent (S10), and resolution automatically follows the concentration of mass. This makes SPH competitive in for example collapse problems in cosmology \citep[e.g.][]{schaye15} and star and planet formation \citep[e.g.][]{bate03,  mayer02}. Disadvantages of SPH include its relatively large numerical dissipation for certain problems, especially when low-density regions are dynamically important \citep[e.g.][]{devalborro06} or in shear flows \citep{agertz07}. A meshless method that can have high resolution in arbitrary locations was presented in \cite{maron12}.

Recent studies have focused on bridging the gap between SPH and grid-based methods, in an effort to get the best of both worlds. Examples include {\sc arepo} (S10), {\sc rich} \citep{yalinewich15} and {\sc gizmo} \citep{hopkins14, hopkins15}. One can arrive at the class of moving mesh codes by starting with SPH, but seeing the particles as mesh generation points, and subsequently solving the Euler equations on this (necessarily unstructured) mesh. If the mesh points are fixed, an Eulerian method on an unstructured grid is obtained. If the points are allowed to move, a moving mesh code results.  The use of these mesh-generating points and their Voronoi tessellation makes the mesh evolve in a continuous manner, and one can avoid mesh-tangling problems of traditional Arbitraly Lagrangian-Eulerian (ALE) methods \citep[e.g.][]{vachal04}.  While in principle one is free to move the mesh points with any velocity, if the mesh velocity is set equal to the local gas velocity, one obtains a Lagrangian method. A cylindrical moving mesh code was recently presented by \cite{duffell16}.  One problem encountered with moving meshes is grid noise \citep{bauer12, hopkins15}, caused by changes in topology as the mesh evolves which lead to volume inconsistency errors \citep{yalinewich15}. Several fixes have been proposed, from smoothing the velocities of mesh points \citep{duffell15}, to regularising the mesh \citep{mocz15}, to directly attacking the volume inconsistency \citep{steinberg16}.

 Given an Eulerian method, for example based on a Riemann solver, acting on a structured grid, adapting it to work on an unstructured grid  is a non-trivial undertaking. Indeed, most multi-dimensional Eulerian methods are built using one-dimensional solvers, combined in such a way, making use of the structured nature of the mesh (often logically rectangular), to yield an accurate approximation to the multidimensional problem \citep[e.g.][]{strang68, colella90, balsara12}. Even more, Fourier spectral methods \emph{require} a regular mesh from the very beginning. On the other hand, unstructured meshes offer desirable properties such as more isotropic numerical diffusion, and the ability to refine the mesh in arbitrary ways without the need for jumps in resolution.

Unstructured meshes usually come in the form of Delaunay triangulations \citep{delaunay34} or Voronoi tesselations \citep{dirichlet50, voronoi07}, for reasons that we will go into in section \ref{secGrid}. Triangular grids, structured or unstructured, have an additional advantage that they  allow for a multidimensional analog of Roe's approximate Riemann solver \citep{struijs94}. That is, it is possible to design a multidimensional solver without having to rely on one-dimensional building blocks. This clearly has advantages over traditional Eulerian methods, especially for flows that are not aligned with any axis of the grid. While these multidimensional upwind, or residual distribution, methods have been around for several decades, initially they were designed to solve for steady flows only. The time-dependent case turned out to be relatively tricky to work out with various different formulations \citep[e.g.][]{ferrante97,depalma05}. Moreover, these were all implicit time integration schemes and therefore relatively expensive.

More recently, an explicit formulation was derived \citep{ricchiuto10}, making multidimensional upwind methods potentially competitive for time dependent astrophysical flows, which are often very compressible but also multidimensional. In this paper, we present and test a two-dimensional version of a multidimensional upwind method in an astrophysical context in the form of {\sc astrix}\footnote{Freely available as an open-source project at \url{https://github.com/SijmeJan/Astrix/}} (AStrophysical fluid dynamics on unstructured TRiangular eXtreme grids). Another advantage of multidimensional upwind methods compared to the more common Riemann solvers is that they employ a very compact stencil: the vast majority of all calculations (in particular calculating the fluxes between cells)  are done using data from one triangle only.  In comparison, a dimensionally split scheme based on a Riemann solver needs four cells in each direction in order to compute a second order accurate interface flux \citep[see e.g.][]{leveque02}. The number of bytes that need to be read in order to compute a flux is therefore much larger in traditional Eulerian codes, which makes multidimensional upwind  methods excellent candidates to port to Graphics Processing Units (GPUs).

The rest of this paper is structured as follows. In section \ref{secGrid} we introduce unstructured grids and how to generate them, while in section \ref{secResDist} we describe the residual distribution methods that are part of {\sc astrix}. In section \ref{secIntroGPU} we  discuss the GPU implementation of both mesh generation and residual distribution . In section \ref{secTest} these methods are tested on one and two dimensional problems and we give a discussion in section \ref{secDisc}. We conclude in section \ref{secCon}.

\section{Unstructured grids}
\label{secGrid}

\subsection{Basic definitions}

\begin{figure}
\centering
\resizebox{\hsize}{!}{\includegraphics[]{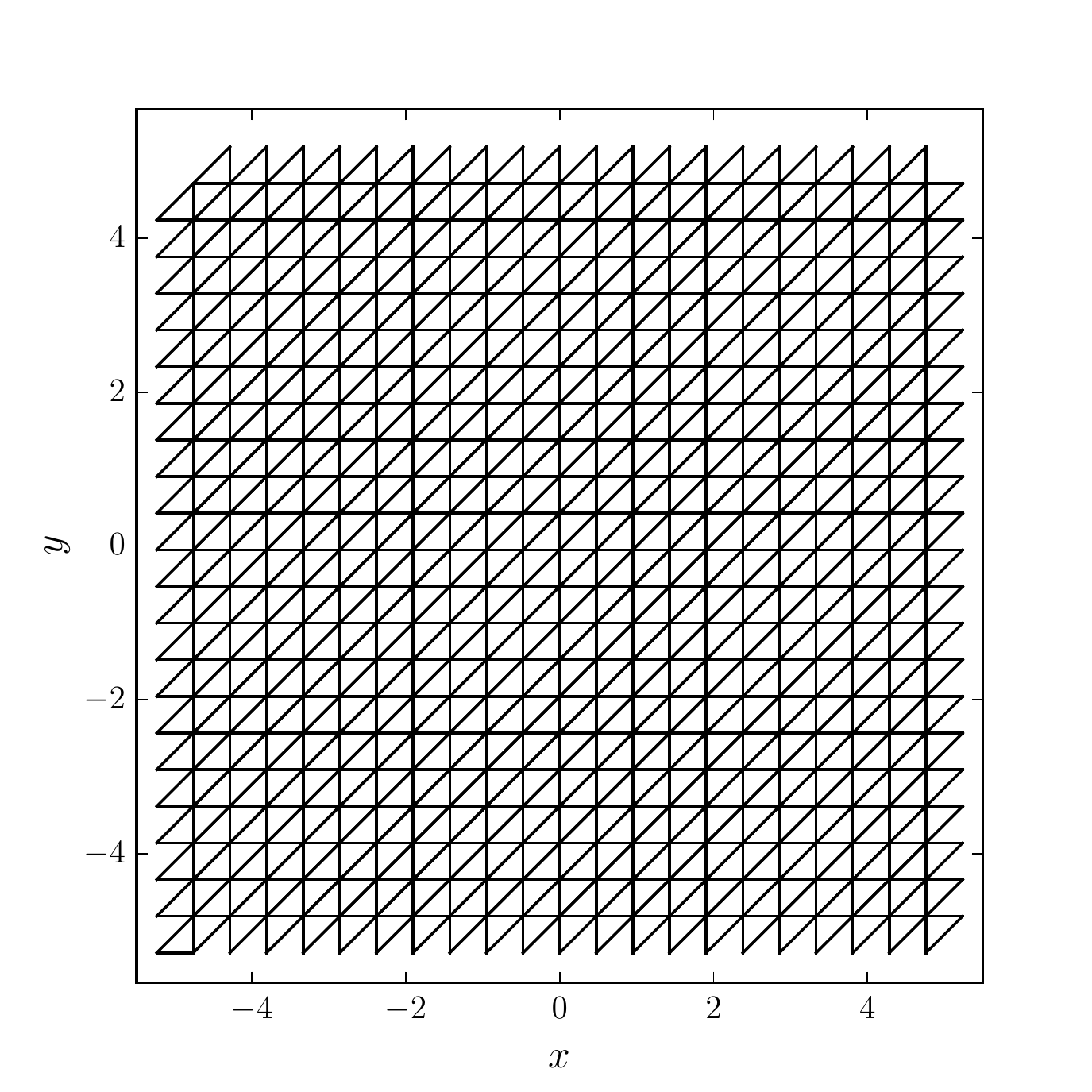}}
\caption{Structured mesh on a periodic domain $-5 \leq (x,y) < 5$}
\label{figMesh2DStruc}
\end{figure}

\begin{figure}
\centering
\resizebox{\hsize}{!}{\includegraphics[]{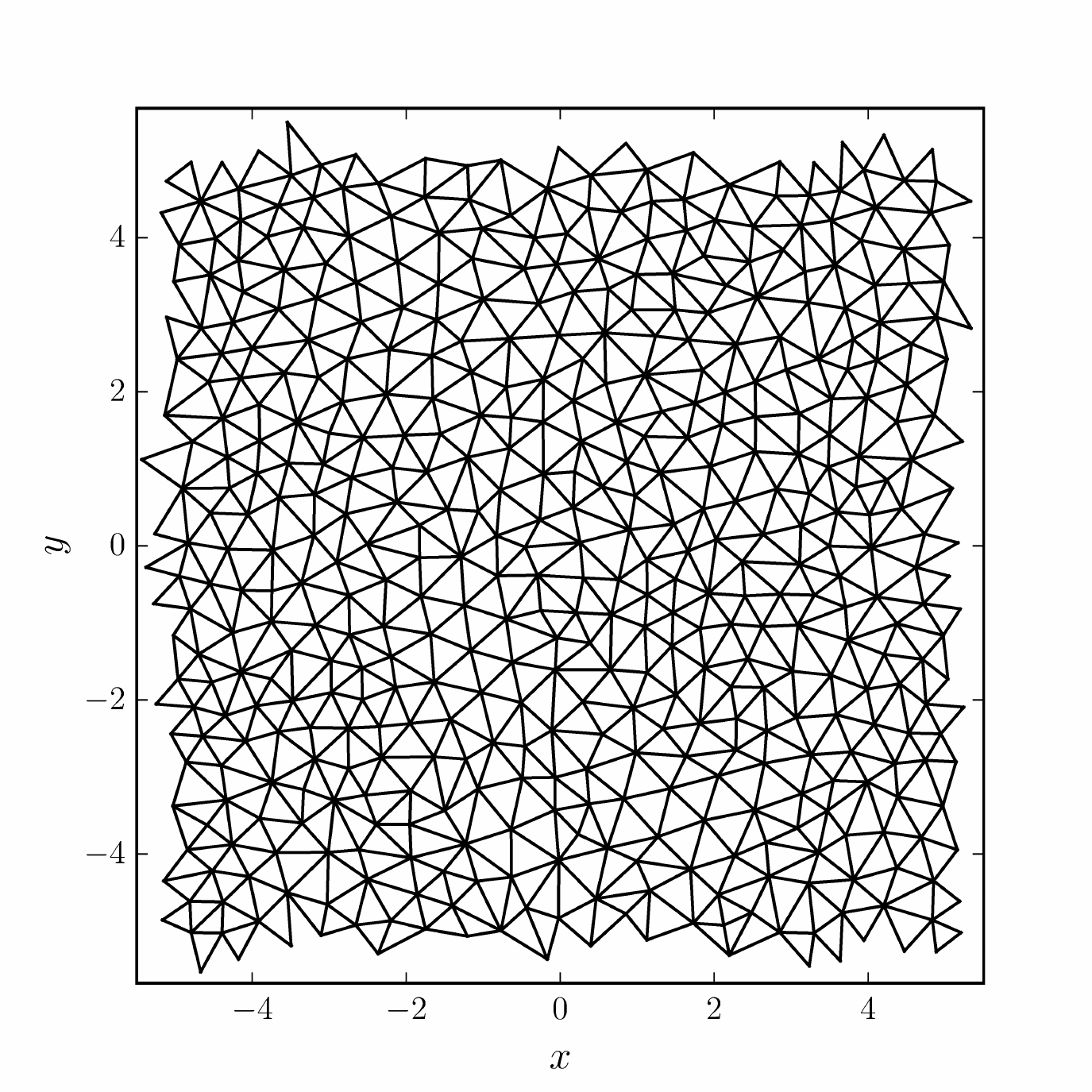}}
\caption{Unstructured mesh consisting of $\sim 400$ vertices on a periodic domain $-5 \leq x < 5$ and $-5 \leq y < 5$.}
\label{figMesh2D}
\end{figure}

A grid, or mesh,  is  defined by a set of nodes, or vertices, together with a recipe for getting from one vertex to its neighbours.  A mesh  can be said to be \emph{structured} if the location of the vertices and their interconnections follow a simple pattern,  which usually leads to a high degree of symmetry.  In two dimensions, an $m\times n$ structured Cartesian mesh on the unit square can be defined as a collection of vertices $(i,j)$ with coordinates
\begin{eqnarray}
x_i &=& i/m, \nonumber\\
y_j &=& j/n,
\label{eqCart}
\end{eqnarray}
where $0 \leq i < m$ and $0 \leq j < n$, together with the connectivity rules that vertex $(i,j)$ is connected to $(i-1,j)$ and $(i+1,j)$ in the $x$-direction, and to $(i,j-1)$ and $(i,j+1)$ in the $y$-direction. Such a mesh consists of square cells. The same collection of vertices with different connectivity rules could for example lead to triangular cells (for an example see Fig. \ref{figMesh2DStruc}).

On the other hand, for an \emph{unstructured} mesh no simple rules exist for the location of the vertices. An example is shown in Fig. \ref{figMesh2D}. Obviously this makes the implementation more complicated and more memory intensive, since the location of all vertices has to be stored explicitly, rather than using simple rules to work out the coordinates. However, unstructured grids provide absolute freedom on where to place the vertices, which has three main advantages:
\begin{itemize}
\item{It is possible to handle complex geometries. When studying the flow across an aircraft wing, for example, it is necessary to get  as close to the real  shape of the wing  as possible . Embedding a shape  that is not rectangular  in a regular Cartesian mesh is hopeless.}
\item{It is possible to construct meshes  with no preferred directions , which leads to more uniform numerical dissipation. For a 2D structured Cartesian mesh, numerical dissipation  will strongly depend on the direction of the flow with respect to the coordinate axes. This leads to the famous carbuncle instability \citep{peery88, quirk94}. }
\item{ One has much more freedom to vary  the resolution of the mesh from place to place in a smooth way.  For structured meshes the options are limited. For a structured cylindrical mesh for example, it is possible to increase the resolution towards small radii by choosing a logarithmic radial coordinate. However, this affects {\emph all} cells in the inner parts, while the region where high resolution is required may be very limited in azimuthal extent.  Of course, for structured meshes there exists the powerful technique of AMR to obtain high resolution locally, but this technique leads to  boundaries between regions of coarse and fine resolution \citep{berger84} where  additional interpolation errors occur, which in some cases may be unacceptable.}
\end{itemize}

While it is possible to generate quadrilateral unstructured meshes, the more  common  cell choice is the triangle. For a given set of  points , there exist many ways of interconnecting them using triangles. This means we can choose a triangulation that is in some sense optimal. A useful  criterion  of the quality of the grid is the minimum opening angle of any triangle. Meshes with small angles often lead to numerical problems for simulations, since numerical diffusion will be very non-isotropic  \citep[e.g.][]{babuska76} . This former problem happens in Cartesian structured meshes when  the cells have a very large aspect ratio ; for example $m \ll n$ in equation (\ref{eqCart}). For Cartesian structured meshes, we would usually like the cells to be as square as possible, i.e. $m=n$ in equation (\ref{eqCart}). For a triangular mesh this translates into having triangles that are as close to equilateral as possible. While it is not possible to achieve this limit in practice, for example because of boundary constraints, we would still like to maximise the minimum angle for any triangle in the grid. For a given set of vertices, this fixes the triangulation, since it is the Delaunay triangulation \citep{delaunay34} that achieves this.

\subsection{Delaunay triangulation}

\subsubsection{Circumcircles}

\begin{figure}
\centering
\resizebox{\hsize}{!}{\includegraphics[]{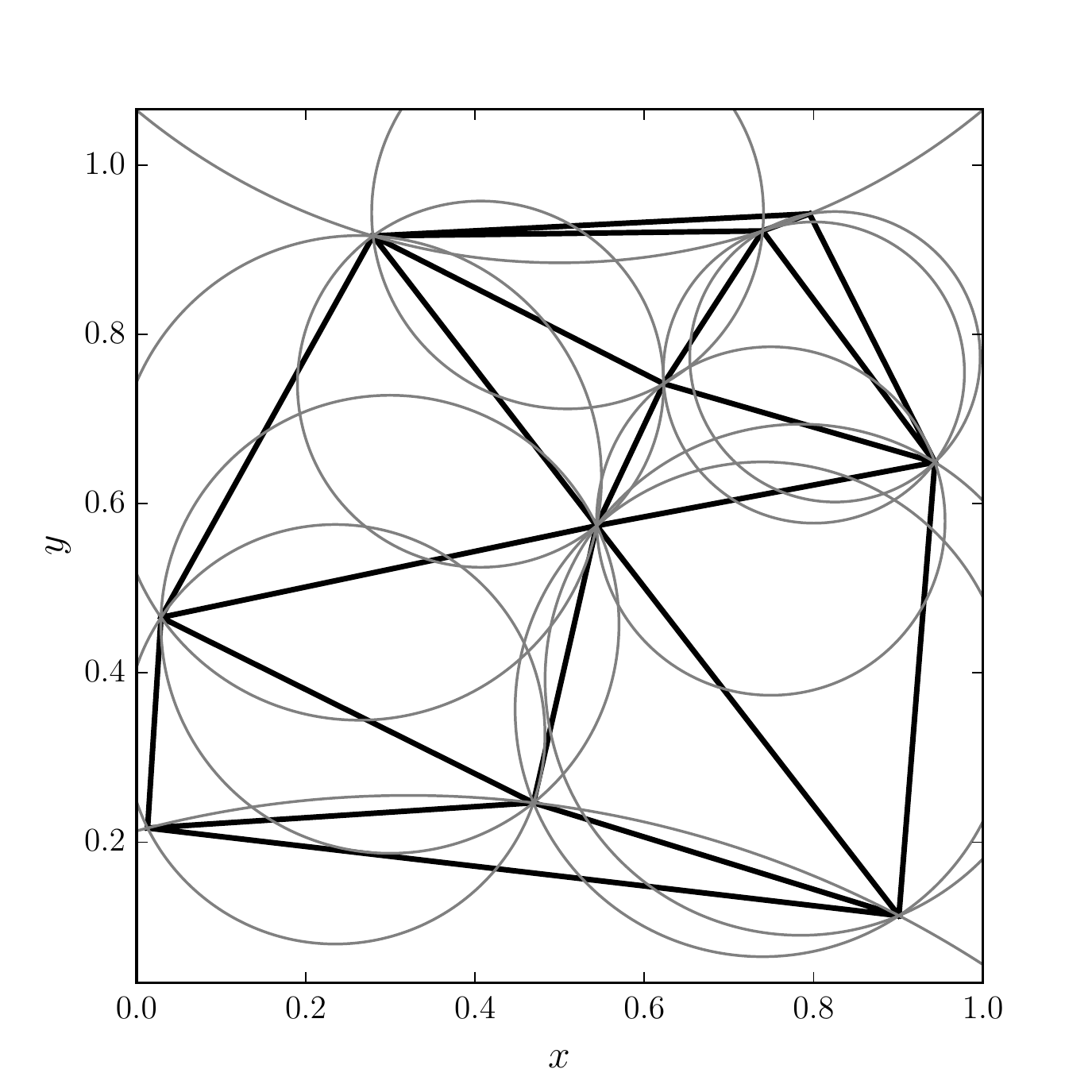}}
\caption{Delaunay triangulation of 10 randomly placed vertices. For every triangle, its circumcircle is shown in grey.}
\label{figCircum}
\end{figure}

The \emph{circumcircle} of a set of two or three vertices is a circle that passes through all vertices in the set. For a set of three vertices, this circle is unique. For a set of two vertices, there are infinitely many circumcircles. A triangle $T$ is called Delaunay if its circumcircle  $C$ is empty, i.e. there are no mesh-generating points inside $C$ . An edge $E$, consisting of vertices $u$ and $v$, is called Delaunay if there exists a circumcircle of $u$ and $v$ that is empty. Note that if a triangle is Delaunay, all of its three edges are automatically Delaunay, since there exists a circumcircle $C$ of any of the edges of $T$ that is empty (take $C$ to be the circumcircle of $T$). The reverse is also true, but less easy to show formally.

A triangulation of a two-dimensional space where all triangles and all edges are Delaunay is guaranteed to exist, and  in addition it is  unique if no four vertices lie on the same circle (and, more trivially, not all vertices are collinear). If this condition is violated, the triangulation is no longer unique: edges that are Delaunay can be crossing, and in order to obtain a valid triangulation a selection of edges has to be made. In practice, this happens automatically during mesh construction (see below), but it is a first indication that numerical roundoff errors will play a prominent role in mesh construction. If we bring four vertices closer and closer to being on the same circle, at some point the triangulation will become degenerate because of roundoff errors. This situation has to be handled with care, since if some parts of the algorithm detect a degeneracy while other parts do not, which can easily happen when working close to roundoff limits, the whole algorithm will break down. There is therefore a need for exact geometric predicates (see section \ref{secPredicates}). In Fig. \ref{figCircum} a Delaunay triangulation is shown together with each triangle's circumcircle.

\subsubsection{Edge flipping}

\begin{figure}
\centering
\resizebox{\hsize}{!}{\includegraphics[]{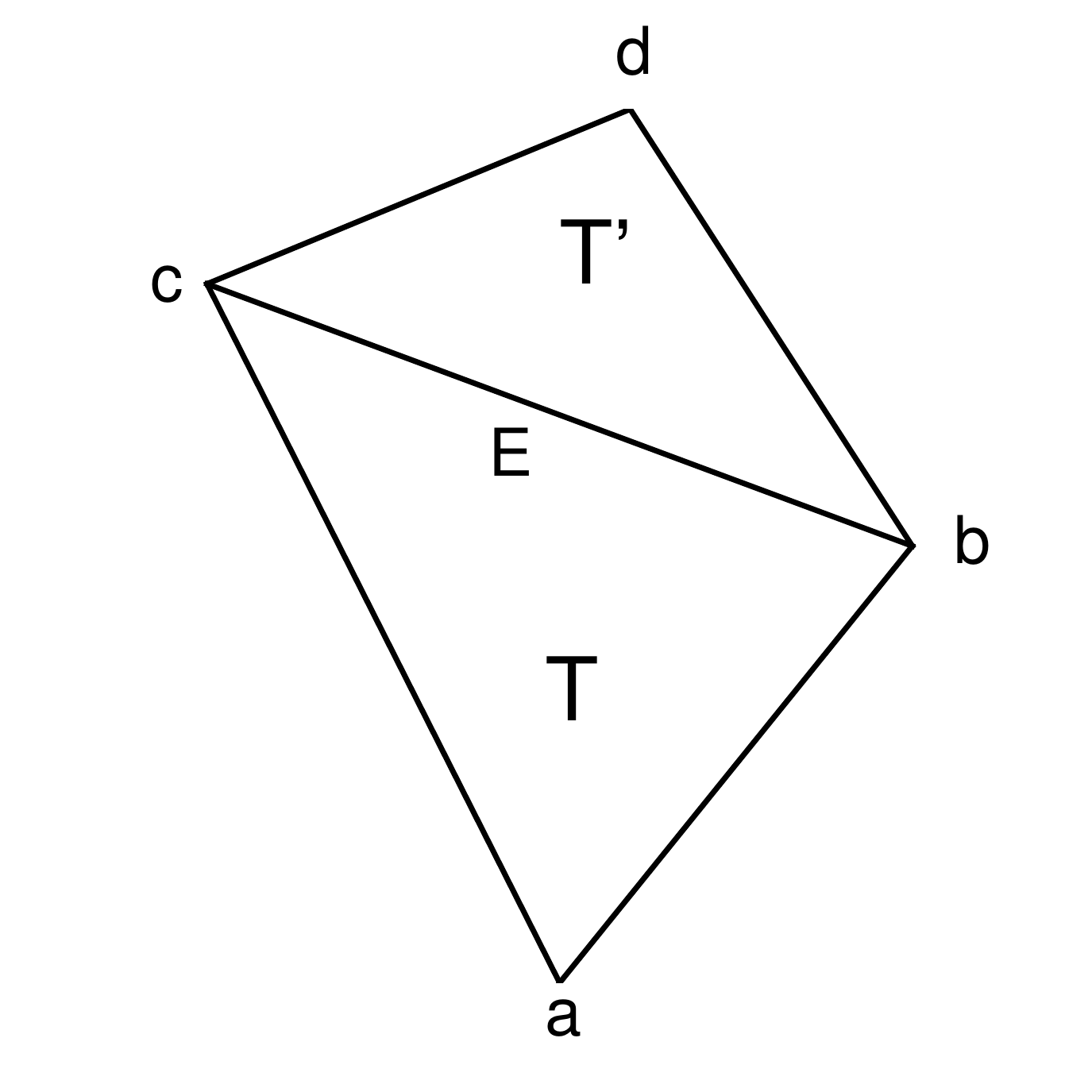}}
\caption{Containing quadrilateral of edge $E$, consisting of triangles $T$ and $T'$ and vertices $a$, $b$, $c$ and $d$.}
\label{figEdgeFlip}
\end{figure}

Consider an edge $E$ in a triangulation (not necessarily Delaunay), together with only its two neighbouring triangles (see Fig. \ref{figEdgeFlip}). The four vertices define a \emph{containing quadrilateral}. Define $E$ to be \emph{locally Delaunay} if  empty  circumcircle of $E$ exists. Obviously, if $E$ is not locally Delaunay it is not Delaunay. The converse is not true: $E$ may be locally Delaunay but not Delaunay. However, if \emph{all} edges in the triangulation are locally Delaunay this means that all edges are Delaunay.

Define $E'$ as the edge that would exist if we connected the two vertices of the containing quadrilateral not part of $E$ ($a$ and $d$ in Fig. \ref{figEdgeFlip}). If $E$ is not locally Delaunay, it follows that $E'$ would be locally Delaunay. This means that we can remove an edge that is not locally Delaunay by \emph{flipping} $E$ (i.e. replacing it with $E'$). Not all edges are flippable, for example if the containing quadrilateral is not convex. However, it can be shown that all edges that are not locally Delaunay can be flipped.

It follows that if a triangulation is not Delaunay, at least one edge is not locally Delaunay and can be flipped. It is intuitively clear, but slightly more difficult to show, that each flip moves the triangulation closer to the Delaunay triangulation\footnote{This is only true in 2D; in 3D, convergence is not guaranteed.}. When there are no more edges to flip, the resulting triangulation is the Delaunay triangulation.

Note that some edges can not be flipped because it would make the triangulation invalid. In Fig. \ref{figCircum}, this is true for the almost horizontal edge near the top. The reason is that the containing quadrilateral is not convex. However, it can be shown that if an edge can not be flipped, it has to be Delaunay.

\subsubsection{Triangle quality}

The main reason for the popularity of Delaunay triangulations for mesh generation is that the resulting triangles are of high quality in the sense that small angles can be avoided  as long as the boundary of the domain does not require them . In fact, among all possible triangulations of a set of vertices, the Delaunay triangulation maximises the minimum angle present in the mesh\footnote{Unfortunately, this result only holds in two dimensions} \citep{lawson77}. This can be shown in a straightforward way by first noting that flipping an edge to make it locally Delaunay always increases the minimum angle present in the containing quadrilateral. Now any valid triangulation can be transformed into a Delaunay triangulation by a sequence of edge flips. Since any edge flip increases the minimum angle, this means that the Delaunay triangulation maximises the minimum angle.

\subsubsection{Incremental insertion algorithm}

The discussion on edge flipping suggests the following algorithm for finding the Delaunay triangulation of a set of vertices: start with \emph{any} valid triangulation, and perform edge flips until all edges are locally Delaunay. The resulting triangulation is the Delaunay triangulation.

Finding an initial triangulation for a large set of input vertices is cumbersome. Therefore, in practice vertices are inserted one by one, starting from an initial triangle large enough to contain all subsequent vertices \citep{lawson77}. Such algorithms are usually called \emph{incremental insertion algorithms}. When implemented in its simplest form, incremental insertion can be slow, because in principle the addition of a single vertex may lead to the flipping of \emph{all} edges in the mesh. However, if the vertices are inserted in random order \citep{guibas92}, incremental insertion becomes a competitive method \citep{su97}. Other algorithms include \emph{divide and conquer} \citep{guibas85, dwyer87}, \emph{sweepline} \citep{fortune86}, \emph{gift-wrapping} \citep{dwyer91} and algorithms based on the convex hull \citep{barber96}. The reason for choosing incremental insertion over any of the other methods is that by its very nature is very well suited to deal with cases where we do not know in advance the position of all vertices in the mesh. This is the situation we are in when generating unstructured meshes.

\subsection{Delaunay refinement}

Even though the Delaunay triangulation is optimal in the sense that it maximises the minimum angle, low-quality triangles can still be seen in Fig. \ref{figCircum}. This is because the location of the vertices were chosen randomly. In practice, we are free to choose the majority of the positions of the vertices. Possible constraints on the location of the vertices include any fixed boundaries (a wall, an aircraft wing, a planet) and any resolution constraints on the local density of vertices. Within these constraints, there is still a lot of freedom in choosing vertex locations. This can be done in an optimal way as to guarantee a mesh of a certain quality, that is, a minimum opening angle that is larger than a certain value. This way of choosing vertices, often called \emph{Delaunay refinement}, is discussed next. For a more detailed discussion, see \cite{shewchuk02}.

\subsubsection{Removing low-quality triangles}
\label{secRemoveLowQuality}

First of all, the idea of a triangle of low quality can be made quantitative. Define $\beta$ to be the ratio of the circumradius to the shortest edge of a triangle. A straightforward calculation shows that $2\beta = 1/\sin\alpha$, where $\alpha$ is the smallest angle of the triangle. Therefore, in order to avoid small angles, we need to avoid triangles with large values of $\beta$, say we require $\beta \leq B$ for some appropriate bound $B$.

A low-quality triangle $T$ can be removed by adding a new vertex $v$ to the mesh, exactly at the circumcentre of $T$. It is clear that $T$ can not be part of the new mesh, since its circumcircle is no longer empty. However, it is also clear that any new triangle created by inserting $v$, will have a shortest edge that is at least the radius of the circumcircle of $T$. Since $\beta > B$ for $T$, we have that the shortest edge of any new triangle is at least $B$ times the shortest edge of $T$. For $B\geq 1$, this means that new edges will have at least the length of the minimum edge length in the initial triangulation $|E|_{\mathrm{min}}$. Therefore, an algorithm that removes triangles with $\beta > B$ must terminate eventually if $B \geq 1$, since at some point the vertex density becomes so high that no new triangles can be created with minimum edge length larger than $|E|_{\mathrm{min}}$. The three most well-known Delaunay refinement algorithms use $B=\sqrt{2}$ \citep{ruppert95} and $B=1$ \citep{chew89, chew93}.

In addition to the quality constraint $\beta \leq B$, we can impose a size constraint, by removing triangles that are too large in the same way as removing low-quality triangles. The maximum size $|T|_\mathrm{max}$ can be a function of space, allowing for non-uniform meshes. As long as $|T|_\mathrm{max}({\bf x}) > M$ for some positive constant $M$, the algorithm is still guaranteed to terminate.

\subsubsection{Splitting segments}

The main distinction between the algorithms of \cite{chew89}, \cite{chew93} and \cite{ruppert95} lies in the way mesh boundaries are treated. These boundaries are part of the input of the Delaunay refinement algorithm: it takes as input the vertices that make up the boundary together with their interconnection. The simplest case would be if the computational domain is a rectangle, which can be specified with four vertices (the corners) and four edges (the sides). More complicated cases would include an outer boundary that is a circle, or a hole in the computational domain (an inner boundary) in the form of an aircraft wing. These edges are special in the sense that they will \emph{have to} be part of the final mesh. Such edges will be referred to as \emph{segments}\footnote{Segments do not have to be part of the boundary: in principle the term refers to any input edge that needs to be part of the final triangulation}.

\cite{ruppert95} provides an elegant way of dealing with segments. Define the \emph{diametric circle} of a segment to be the smallest circle that encloses the segment. A segment is defined to be \emph{encroached} if any vertex lies inside the diametric circle. During Delaunay refinement, if a new vertex would lead to encroached segments, the vertex is not inserted: instead the segments in question are split by inserting vertices at their midpoints. Note that a non-encroached segment is an edge that is Delaunay since there exists a circumcircle, namely the diametric circle, that is empty. This way, the algorithm ensures that boundary segments will be part of the final triangulation. An additional advantage is that it is also guaranteed that a new vertex will be \emph{inside} the existing triangulation: if the circumcentre of a triangle happens to be outside the triangulation, this must mean that somewhere a segment is encroached.

\subsection{Grid generation}

We now go over the different steps in our Delaunay refinement algorithm. Every iteration consists of six steps, which are described below and are repeated until no more vertices need to be inserted. It is important to realise that the resulting mesh, defined by the input boundary vertices and the desired quality, is not unique. In particular, the locations of the vertices depend on the order in which they were inserted.

\subsubsection{Finding low-quality triangles}

First of all, for every triangle $T$ with vertices $(a,b,c)$ in the mesh we compute its circumradius $r$:
\begin{equation}
r = \frac{l_al_bl_c}{2\left|\begin{array}{cc} a_x - c_x & a_y - c_y \\ b_x - c_x & b_y - c_y \end{array}\right|},
\label{eqCircumRadius}
\end{equation}
where $l_a$, $l_b$ and $l_c$ are the length of the edges opposite vertices $a$, $b$ and $c$, respectively, and its circumradius-to-shortest-edge ratio $\beta$. If either $r > R$ (triangle too big) or $\beta > B$ (triangle too low quality), a vertex will be inserted at its circumcentre. Here, $R$ and $B$ are user input quantities specifying the desired triangle size ($R$) and the desired triangle quality ($B$). This step leads to a list of vertices to add into the mesh.

\subsubsection{Finding triangles containing the new vertices}
\label{secFindTriangle}

Next, for every new vertex we find the triangle containing this vertex (often this is not the original triangle, especially if this triangle has a large value of $\beta$). A triangle $T$ with vertices $a$, $b$ and $c$ contains vertex $v$ if $v$ lies on the `correct' side of all three edges of $T$. If $(a, b, c)$ are in counterclockwise order, then $v$ lies in $T$ if both $(a, b, v)$, $(b, c, v)$ and $(c, a, v)$ are in counterclockwise order. Following \cite{shewchuk97}, we will refer to the counterclockwise test as {\sc Orient2D}$(a, b, c)$, a function that returns a positive value if $(a, b, c)$ are in counterclockwise order, negative if they are in clockwise order, and zero if they are collinear. This function can be implemented as a matrix determinant\footnote{Note that the same determinant appears in the denominator in equation (\ref{eqCircumRadius})}:
\begin{equation}
\textsc{Orient2D}(a, b,c)=\left|\begin{array}{cc} a_x - c_x & a_y - c_y \\ b_x - c_x & b_y - c_y \end{array}\right|.
\label{eqOrient}
\end{equation}
It should be clear that in order to maintain a valid triangulation, it is extremely important that {\sc Orient2D}$(a, b, c)$ gives the correct result, even in the presence of round-off errors. This is difficult when $(a, b, c)$ are close to collinear. If we try and place a vertex $v$ in triangle $T$ but extremely close to the edge that is shared by $T$ and $T'$, due to round-off errors a naive implementation of {\sc Orient2D} may decide that $v$ lies in both $T$ and $T'$, or not in either of them. Such mutually contradictory results inevitably lead to nonsensical results and invalid triangulations. One might hope that for straightforward computational domains (no complicated boundaries) such cases never show up, but in practice they always do. Therefore, we evaluate {\sc Orient2D} using exact geometric predicates (see section \ref{secPredicates}).

While the triangle containing $v$ may not be the original triangle $T_0$ leading to the insertion of $v$, often it is close to it. Therefore, we start searching in $T_0$, and move to a neighbouring triangle in the direction of $v$ until we have found a triangle containing $v$. In some cases, $v$ will be located exactly on an edge and those have to be dealt with separately.

\subsubsection{Testing if any new vertex encroaches upon a segment}

Vertices can not be inserted if they lead to an encroached segment. A vertex $v$ with coordinates ${\bf r}_v$ encroaches upon a segment with vertices $a$ and $b$ with coordinates ${\bf r}_a$ and ${\bf r}_b$ if
\begin{equation}
\left({\bf r}_a - {\bf r}_v\right)\cdot \left({\bf r}_b - {\bf r}_v\right) < 0
\end{equation}
There is usually no need for exact predicates in this computation.

If vertex $v$ is to be inserted in triangle $T$ with vertices $(a, b, c)$ we check all triangles that have either $a$, $b$ or $c$ as a vertex. Should $v$ be inserted onto an existing edge $E$, we check all triangles that share a vertex with the two triangles sharing $E$ as an edge. If any of these triangles has a segment for an edge upon which $v$ encroaches, $v$ is not inserted at its original location but on the middle of this particular segment, thereby splitting the segment. Of course, inserting $v$ might have lead to multiple encroached segments, but this approach is computationally advantageous as the number of vertices to be inserted does not change. Any difficulties arising from this approach are dealt with in section \ref{secSplitSegment}.

Note that by checking all triangles that have either $a$, $b$ or $c$ as a vertex, we make sure that $v$ does not encroach upon \emph{any} segment. For if $v$ encroaches upon any other segment, it automatically follows that either $a$, $b$ or $c$ encroaches upon this segment as well, which can not be the case if the original mesh had no encroached segments.

\subsubsection{Insert new vertices into mesh}

Next, we actually insert the vertices into the mesh. If vertex $v$ is to be inserted in triangle $T$, $T$ is split into three triangles that all have $v$ as a vertex. This operation therefore adds two triangles to the mesh. If $v$ is inserted on an edge $E$, the two triangles sharing $E$ are each split into two, thereby again adding two triangles to the mesh. When splitting a segment, the single triangle of which the segment is part is split into two, thereby adding one triangle to the mesh. Note that at this point, we do not maintain a Delaunay triangulation.

\subsubsection{Split any encroached segments}
\label{secSplitSegment}

Occasionally, splitting a segment leads to another encroached segment. In order to deal with this, we check any vertex that was inserted on a segment if it encroaches any other segment. If this is the case, the segment in question is split by adding an extra vertex to the mesh. In practice this only very rarely happens.

\subsubsection{Maintain Delaunay triangulation}
\label{secMaintainDelaunay}

With the insertion of new vertices and therefore the addition of new triangles the current triangulation will usually no longer be Delaunay. The final step in the iteration is to transform the current triangulation into a Delaunay triangulation, which is achieved by edge-flipping.

First of all, all relevant edges in the mesh are checked for Delaunay-hood. Consider edge $E$ and its containing quadrilateral, which consists of triangles $T$ and $T'$ (see Fig. \ref{figEdgeFlip}). Let the vertices of $T$ be $(a, b,c)$, and let $d$ be the one vertex of $T'$ that is not part of $T$. Then $E$ is Delaunay if $d$ does not lie inside the circumcircle of $T$. Following \cite{shewchuk97}, define the {\sc InCircle2D}$(a,b,c,d)$ test to return a positive value when $d$ lies inside the circle defined by $(a,b,c)$ (assuming $(a,b,c)$ are in counterclockwise order), a negative value when $d$ lies outside this circle, and zero if $d$ lies exactly on this circle. As with {\sc Orient2D}, {\sc InCircle2D} can be expressed as a determinant:
\begin{eqnarray}
&\textsc{InCircle2D}(a,b,c,d)=\nonumber\\
&\left|\begin{array}{ccc}
a_x-d_x & a_y-d_y & (a_x-d_x)^2+(a_y-d_y)^2\\
b_x-d_x & b_y-d_y & (b_x-d_x)^2+(b_y-d_y)^2\\
c_x-d_x & c_y-d_y & (c_x-d_x)^2+(c_y-d_y)^2\end{array}\right|.
\label{eqInCircle}
\end{eqnarray}
Again, we need the {\sc InCircle2D} test to give the exact result, even in the case of finite precision arithmetic. Otherwise, not only could the flip algorithm get stuck, continuously flipping the same edge, but it could also flip an edge that is not flippable because the containing quadrilateral is concave, yielding an invalid triangulation and therefore breaking the Delaunay refinement algorithm.

Flipping an edge may create new edges that are not Delaunay. Therefore, the process of checking and flipping is iterated until all edges are Delaunay. When the triangulation is Delaunay again, the next iteration can be started by testing all triangles if they match the quality constraints. When there are no more vertices to add, the algorithm exits.

\section{Residual distribution schemes}
\label{secResDist}

We now turn to the hydrodynamic solver acting on our unstructured grid. The equations to be solved are conservation of mass, momentum and energy in two spatial dimensions:
\begin{eqnarray}
\frac{\partial\rho}{\partial t} + \frac{\partial}{\partial x}(\rho u)+  \frac{\partial}{\partial y}(\rho v)&=&0,\\
\frac{\partial}{\partial t}(\rho u) + \frac{\partial}{\partial x}(\rho u^2 + p) +  \frac{\partial}{\partial y}(\rho uv)&=&0,\\
\frac{\partial}{\partial t}(\rho v) + \frac{\partial}{\partial x}(\rho uv) +  \frac{\partial}{\partial y}(\rho v^2 + p)&=&0,\\
\frac{\partial e}{\partial t} + \frac{\partial}{\partial x}(\rho hu) +  \frac{\partial}{\partial y}(\rho hv)&=&0,
\end{eqnarray}
where $\rho$ is the density, ${\bf v} = (u,v)^T$ is the velocity vector, $p$ is the pressure and $e$ is the total energy  for an ideal gas equation of state :
\begin{equation}
e = \frac{1}{2}\rho\left(u^2+v^2\right) + \frac{p}{\gamma-1},
\end{equation}
in which the last term denotes the internal energy, which is specified by the pressure under the assumption of a perfect gas, with ratio of specific heats $\gamma$. Finally, $h=(e+p)/\rho$ is the fluid enthalpy.

Above conservation laws can be written concisely as:
\begin{equation}
\frac{\partial \bW}{\partial t} + \nabla\cdot \mathcal{F}= 0,
\label{eqConsTot}
\end{equation}
where $\bW$ is the state vector and $\mathcal{F}=(\bF, \bG)$ is the flux term, containing the flux in the $x$ direction $\bF$ and the flux in the $y$ direction $\bG$. Integrating over a volume $V$ and using Gauss' theorem shows that the time evolution of $\bW$ is governed by the \emph{residual} $\phi$:
\begin{equation}
\phi = \oint_{\partial V} \mathcal{F}\cdot {\bf n} dS,
\end{equation}
where ${\bf n}$ is the outward-pointing unit normal vector of surface element $dS$. If $\phi=0$, the state is stationary. A residual distribution scheme, as the name suggests, is a numerical scheme that solves a discrete version of (\ref{eqConsTot}) by taking $\phi$ and distributing it over neighbouring cells.

\subsection{Roe solver}

The Roe solver is a well-known approximate Riemann solver that is part of many astrophysical fluid dynamics packages \citep[e.g.][]{mignone07,stone08}. However, it can also be viewed as a residual distribution scheme in one spatial dimension, and thereby making a connection between more traditional methods and the framework presented in the next sections.

\subsubsection{Brief derivation}

Consider a system of $q$ hyperbolic conservation laws in one spatial dimension:
\begin{equation}
\frac{\partial \bW}{\partial t} + \frac{\partial \bF}{\partial x} = 0,
\end{equation}
where $\bW$ is the state vector and $\bF$ the flux vector. Write the conservation laws in quasi-linear form:
\begin{equation}
\frac{\partial \bW}{\partial t} + \mA(\bW) \frac{\partial \bW}{\partial x}= 0,
\end{equation}
where $\mA=\partial\bF/\partial\bW$ is the Jacobian.

Now consider a uniform grid with cell centres $x_i$ and spacing $\Delta x$.  Consider  two neighbouring grid cells $i-1$ and $i$, and corresponding states $\bW_{i-1}$ and $\bW_i$. The interaction of these two grid cells can be seen to arise from a state jump at the cell interface.  The setup of two constant states separated by a discontinuity is known as a Riemann problem and has an analytic solution \citep[e.g.][]{lax57}.  Every cell interface has its own Riemann problem defined by the neighbouring states, and for small enough time steps these Riemann problems will be independent. Solving the Riemann problems then yield for example interface fluxes that can then be used to update the state in the cells \citep[see e.g.][]{leveque02}.

\begin{figure}
\centering
\resizebox{\hsize}{!}{\includegraphics[]{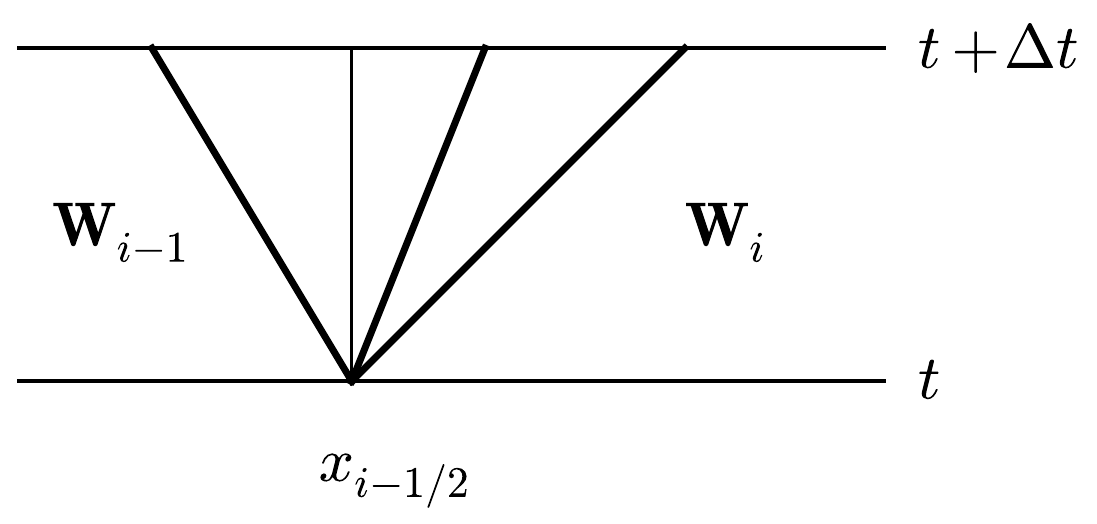}}
\caption{Space-time diagram of the solution to a linear Riemann problem between states $\bW_{i-1}$ and $\bW_i$ at cell interface $x_{i-1/2}$. The solution is assumed to consist of three waves, their propagation denoted by the bold lines. There are two intermediate states between $\bW_{i-1}$ and $\bW_i$, and the jumps between them are found by projecting the initial state difference onto the eigenvectors of $\mA$.}
\label{figLinearRiemann}
\end{figure}

For a \emph{linear} system, i.e. $\mA$ does not depend on $\bW$, the solution to the Riemann problem consists of a set of discontinuities travelling at speeds given by the eigenvalues of $\mA$ (see Fig. \ref{figLinearRiemann}). The strength of each discontinuity can be found by projecting the initial state jump onto the right eigenvectors of $\mA$:
\begin{equation}
\bW_i-\bW_{i-1}=\sum_{p=1}^q \alpha_{i-1/2,p} {\bf r}_{i-1/2,p},
\end{equation}
where ${\bf r}_{i-1/2,p}$ is the $p$th eigenvector of $\mA$ and $\alpha_{i-1/2,p}$ is the corresponding projection coefficient,  which are found by
\begin{equation}
\bm{\alpha}_{i-1/2}=\mathcal{R}^{-1}_{i-1/2}\left(\bW_i-\bW_{i-1}\right),
\end{equation}
where $\mathcal{R}$ is the matrix containing the right eigenvectors of $\mathcal{A}$ as columns.

This completes the solution $\bW^*(t,x)$, which we can use to do a time step from $t$ to $t+\Delta t$ by averaging the solution over a grid cell:
\begin{equation}
\bW_i(t + \Delta t) = \frac{1}{\Delta x}\int_{x_{i-1/2}}^{x_i+1/2} \bW^*(x,t+\Delta t)dx.
\end{equation}
Since $\bW^*$ is piecewise constant, the integration is easily done, yielding
\begin{equation}
\bW_i(t + \Delta t) =\bW_i(t)-\frac{\Delta t}{\Delta x} \sum_{p=1}^q \left(\lambda^p\right)^+\alpha^p {\bf r}^p,
\label{eqUpdateRoe}
\end{equation}
where $\left(\lambda^p\right)^+$ stands for $\mathrm{max}(0,\lambda^p)$, with $\lambda^p$ the $p$th eigenvalue of $\mA$. Note that we have only considered the cell interface between $i$ and $i-1$: there will be a similar contribution to the update of $\bW_i$ from the interface with cell $i+1$.

Of course, the governing equations of gas dynamics are non-linear. The Roe solver \citep{roe81} is defined by a suitable linearisation of $\mA(\bW)$. For the one-dimensional Euler equations, with \begin{equation}
\bW=(\rho, \rho u, e)^T,
\end{equation}
where $\rho$ is the density, $u$ the velocity and $e$ the total energy, \cite{roe81} found that by using a parameter vector
\begin{equation}
\bZ = \sqrt{\rho}(1,u,h)^T,
\end{equation}
where $h=(e+p)/\rho$ is the fluid enthalpy, the matrix $\mA$ evaluated at $\bar \bZ = (\bZ_i + \bZ_{i-1})/2$ provides a linearisation with desirable properties. In particular
\begin{equation}
\mA(\bar{\bZ}) (\bW_i - \bW_{i-1})=\bF_i-\bF_{i-1};
\end{equation}
a property necessary for a conservative scheme.

\subsubsection{Second-order accuracy}
\label{secSecond1D}

The scheme presented above is only first-order accurate, i.e. the largest error term is proportional to $\Delta x$. It is possible to increase the order of accuracy by considering higher-order terms in the Taylor expansion of the solution:
\begin{eqnarray}
\bW(t + \Delta t, x)&=& \bW(t,x) + \Delta t \frac{\partial \bW}{\partial t} + \nonumber\\
& &\frac{\Delta t^2}{2}\frac{\partial^2\bW}{\partial t^2}+O(\Delta t^3)\nonumber\\
&=&\bW(t,x) -\Delta t \mA\frac{\partial \bW}{\partial x} + \nonumber\\
& &\frac{\Delta t^2}{2}\mA^2\frac{\partial^2\bW}{\partial x^2}+O(\Delta t^3).
\label{eqTaylor1D}
\end{eqnarray}
The term proportional to $\Delta t$ is dealt with by the first order scheme above, while the term proportional to $\Delta t^2$ must come from considering a linear reconstruction of the solution \citep[e.g.][]{leveque02}. The contribution of this last term involves information from cells further away from the interface under consideration  (i.e. $i-2$ or $i+1$) and has to be limited in order to avoid spurious oscillations near shocks.  TVD limiter functions come in many flavours, from the least compressive minmod limiter to the very compressive superbee limiter \citep[e.g.][]{sweby84}.  In regions of smooth flow, the update is done using the Lax-Wendroff scheme  \citep{lax60} , and is second-order in both space and time.

A different approach to improve the order of accuracy is not based on (\ref{eqTaylor1D}), but in stead separates discretization in space and time. Dealing with space first leads to an ordinary differential equation (ODE)
\begin{equation}
\frac{d\bW}{dt}=\mathcal{L}(\bW),
\end{equation}
where $\mathcal{L}$ is the operator governing the spatial discretization. Above equation can now be solved by a second-order ODE solver, leading to a second-order accurate method if the spatial discretization is second order as well. Again, a limiter has to be applied to avoid oscillations near discontinuities, and care must be taken that the ODE integrator does not introduce oscillations. This \emph{method of lines} has the advantage that, unlike the Taylor series approach, it is straightforward to extend the method to higher than second order. The main disadvantage is that it needs to solve more than one Riemann problem per cell interface per time step, again unlike the Taylor series approach.

\subsubsection{Approaches for more than one spatial dimension}

The one-dimensional Roe solver, or any other Riemann solver, can be used to build a multidimensional numerical method. Here we highlight some of the problems that arise, since they are pertinent to our discussion later.

Consider the two-dimensional hyperbolic system
\begin{equation}
\frac{\partial \bW}{\partial t} + \frac{\partial \bF}{\partial x}+ \frac{\partial \bG}{\partial y} = 0,
\end{equation}
with quasi-linear form
\begin{equation}
\frac{\partial \bW}{\partial t} + \mA \frac{\partial \bW}{\partial x}+ \mB \frac{\partial \bW}{\partial y}= 0,
\end{equation}
with $\mB=\partial \bG/\partial \bW$. In this section, we can afford to deal only with the linear problem and therefore take $\mA$ and $\mB$ to be constant. A Taylor expansion of the solution is given by
\begin{eqnarray}
\bW(t + \Delta t,x)&=& \bW(t,x) + \Delta t \frac{\partial \bW}{\partial t} + \nonumber\\
& &\frac{\Delta t^2}{2}\frac{\partial^2\bW}{\partial t^2}+O(\Delta t^3)\nonumber\\
&=&\bW(t,x) -\Delta t \left(\mA\frac{\partial \bW}{\partial x} + \mB\frac{\partial \bW}{\partial y}\right)  +\nonumber\\
& & \frac{\Delta t^2}{2}\left(\mA^2\frac{\partial^2\bW}{\partial x^2} + \mA\mB\frac{\partial^2\bW}{\partial y\partial x}+\right.\nonumber\\
& &\left.\mB\mA\frac{\partial^2\bW}{\partial x\partial y}+\mB^2\frac{\partial^2\bW}{\partial y^2}\right)+O(\Delta t^3).
\label{eqTaylor}
\end{eqnarray}
The first-order Roe solver takes care of the terms proportional to $\Delta t$. If only a first order method is required, all interactions between grid cells can be taken into account simultaneously through applying (\ref{eqUpdateRoe}) four times for every cell (one for every neighbour).

Terms proportional to $\Delta t^2$ come in two flavours: those containing $\mA^2$ and $\mB^2$, which in the one-dimensional case are taken care of by correction fluxes or slope limiters, and cross-derivative terms proportional to $\mA\mB$ and $\mB\mA$. The latter come about because flow at an angle to the grid may take material from cell $(i,j)$ directly to for example cell $(i+1, j+1)$: there can be transport across the corners of the grid.

The simplest way of dealing with these cross-derivatives is by using dimensional splitting: treat each dimension separately, varying the order in such a way as to minimize the splitting error \citep[e.g.][]{strang68}. Alternatively, one can adopt an unsplit method such as the Corner Transport Upwind method \citep[CTU,][]{colella90}, which then has to take  care  of the cross-derivatives directly, an approach that is taken for example in ATHENA \citep{stone08}, while PLUTO \citep{mignone07} offers both options. Note that neither of these options are available for unstructured meshes, which makes a code such as {\sc arepo} \citep{springel10} formally only first-order accurate, although in practice it shows second-order convergence \citep[e.g.][]{pakmor16}.

While both dimensional splitting and CTU offer formal second-order accuracy for smooth flows, in some cases it can be advantageous to adopt a fully multidimensional approach, in particular in regions where the flow is not extremely well-resolved \citep{balsara10}. One concern is that in many implementations, higher-order corrections are basically one-dimensional: when deciding if the interaction between cell $(i,j)$ and $(i-1,j)$ can be second order, information from cells $(i-2,j)$ and $(i+1,j)$ is used, all at the same value of $j$ \citep[e.g.][]{leveque02}. It is therefore interesting to look at alternatives, such as fully multidimensional Riemann solvers \citep{balsara10}. Here, we explore \emph{multidimensional upwind} methods in the framework of \emph{residual distribution}.

\subsubsection{Residual distribution formulation}

Before diving into the residual distribution framework, we first show how the familiar one-dimensional Roe solver can be formulated as a residual distribution scheme. This sets the scene for the next few sections.

A different way of looking at the one-dimensional linear problem ($\mA$ does not depend on $\bW$) involves defining the \emph{residual} associated with cell interface $i-1/2$:
\begin{equation}
\phi=\bF_i-\bF_{i-1}=\bar{\mA}\left(\bW_i - \bW_{i-1}\right).
\label{eq1Dresidual}
\end{equation}
If the residual is zero, the flux is constant and there should be no evolution of the state. Note that, in Fig. \ref{figLinearRiemann}, the left state $\bW_{i-1}$ is only modified by the jump associated with a negative propagation speed, and therefore a negative eigenvalue of $\bar{\mA}$, while the right state is modified by the two jumps associated with positive eigenvalues of $\bar{\mA}$. The residual is therefore split between the two neighbouring cells according to the sign of the eigenvalues of $\bar{\mA}$\footnote{\cite{leveque02} uses the term \emph{fluctuation splitting} rather than residual distribution.}. This can be appreciated even more when writing the update (\ref{eqUpdateRoe}) in matrix form, introducing $\Lambda^+$ as the diagonal matrix with $\lambda^+$ as entries on the diagonal:
\begin{eqnarray}
\sum_{p=1}^q \left(\lambda^p\right)^+\alpha^p {\bf r}^p&=&\mR\Lambda^+\alpha\nonumber\\
&=&\mR\Lambda^+\mR^{-1}\left(\bW_i-\bW_{i-1}\right)\nonumber\\
&\equiv& \bar{\mA}^+\left(\bW_i-\bW_{i-1}\right),
\label{eqContRight}
\end{eqnarray}
where $\mR$ denotes the matrix of right eigenvectors of $\bar{\mA}$. The contribution of the Riemann problem at interface $i-1/2$ to cell $i-1$ is, using the same notation:
\begin{equation}
\sum_{p=1}^q \left(\lambda^p\right)^-\alpha^p {\bf r}^p= \bar{\mA}^-\left(\bW_i-\bW_{i-1}\right).
\label{eqContLeft}
\end{equation}
Note that the contributions to the two neighbouring cells (\ref{eqContRight}) and (\ref{eqContLeft}) sum up to $\phi$. The residual $\phi$ is split, or redistributed, amongst the neighbouring cells, according to the recipe:
\begin{eqnarray}
\phi_i&=&\bar{\mA}^+\bar{\mA}^{-1}\phi, \nonumber\\
\phi_{i-1}&=&\bar{\mA}^-\bar{\mA}^{-1}\phi.
\label{eqResDist1D}
\end{eqnarray}
This formulation of the Roe solver is interesting because it has higher dimension counterparts, which we introduce next.

\subsection{Residual distribution basics}

Consider a system of $q$ hyperbolic conservation laws in $d$ spatial dimensions:
\begin{equation}
\frac{\partial \bW}{\partial t} + \sum_{j=1}^d \frac{\partial \bF_j}{\partial x_j} = 0,
\label{eqW}
\end{equation}
Write in quasi-linear form, introducing a parameter vector $\bZ$, to be specified later:
\begin{equation}
\frac{\partial \bW}{\partial \bZ} \frac{\partial \bZ}{\partial t} + \sum_{j=1}^d\frac{\partial\bF_j}{\partial \bZ} \frac{\partial \bZ}{\partial x_j} = 0.
\end{equation}
Now we introduce our triangular mesh. Assume we know $\bZ$ at the vertices (nodes), and since the nodes are connected by triangles there exists a piecewise linear interpolation of the nodal values:
\begin{equation}
\bZ^h({\bf x}, t) = \sum_{i=1}^N \bZ_i ({\bf x}_i,t)\omega_i^h({\bf x}),
\label{eqZh}
\end{equation}
where $N$ is the total number of nodes in the mesh and $\omega_i^h$ is the piecewise linear shape function equal to unity at node $i$ and vanishing outside of the triangles sharing $i$ as a vertex.

The triangle residual is given by (cf. the one-dimensional case (\ref{eq1Dresidual})):
\begin{equation}
\phi^T = \int_T \sum_{j=1}^d \mA_j\frac{\partial \bZ}{\partial x_j}dV,
\label{eqRes}
\end{equation}
where $\mA_j = \partial \bF_j/\partial \bZ$. For the piecewise linear interpolation (\ref{eqZh}) we have that
\begin{equation}
\frac{\partial \bZ}{\partial x_j}=\frac{1}{d}\frac{1}{|T|}\left(\sum_{i=1}^{d+1}\bZ_i {\bf n}_i\right)\cdot {\bf \hat x}_j,
\end{equation}
where $|T|$ denotes the area of triangle $T$, ${\bf n}_i$ is the inward pointing normal to the edge opposite node $i$ of triangle $T$ and ${\bf \hat x}_j$ is the unit vector in direction $j$. Plugging this into (\ref{eqRes}) yields
\begin{equation}
\phi^T = \sum_{i=1}^{d+1}\mK_i \bZ_i,
\end{equation}
with
\begin{equation}
\mK_i =  \frac{1}{d}\left[\sum_{j=1}^d \mathcal{\bar A}_j {\bf \hat x}_j \right] \cdot {\bf n}_i,
\end{equation}
with
\begin{equation}
\bar \mA_j=\frac{1}{|T|}\int_T \mA_j dV.
\end{equation}
In order for the resulting scheme to be conservative, the average matrix $\bar \mA_j$ must obey above relation given $\bZ^h$. While this is difficult in general, if the fluxes $\bF$ are at most quadratic functions of $\bZ$, this means that the entries of $\mA_j$ are at most linear in $\bZ$, making the integrals trivial to evaluate so that the average matrix $\bar \mA_j$ is just $\mA_j$ evaluated at the nodal average of $\bZ$:
\begin{equation}
\bar\mA_j=\mA_j\left(\frac{1}{d+1}\sum_{i=1}^{d+1} \bZ_i\right).
\label{eqLinearA}
\end{equation}
For the Euler equations, a parameter vector can be found that leads to a quadratic flux function, which is basically a multidimensional analogue of Roe's original parameter vector \citep{deconinck93}.

The hyperbolic nature of equation (\ref{eqW}) guarantees that $\mK_i$ has $q$ real eigenvalues and a complete set of linearly independent eigenvectors. Diagonalization yields
\begin{equation}
\mK_i = \mR_i\Lambda_i \mathcal{L}_i,
\end{equation}
where $\mR_i$ is a matrix whose columns are the right eigenvectors of $\mK_i$, $\mathcal{L}_i = \mR_i^{-1}$, and $\Lambda_i$ is a diagonal matrix containing the eigenvalues of $\mK_i$. We can now define the multidimensional upwind parameter as
\begin{equation}
\mK_i^{\pm} = \mR_i\Lambda_i^{\pm}\mathcal{L}_i,
\end{equation}
with
\begin{equation}
\Lambda_i^{\pm}=\frac{\Lambda_i\pm \left|\Lambda_i\right|}{2}.
\end{equation}
Explicit expressions for the matrix elements are provided in appendix \ref{appMat}.

\begin{figure}
\centering
\resizebox{\hsize}{!}{\includegraphics[]{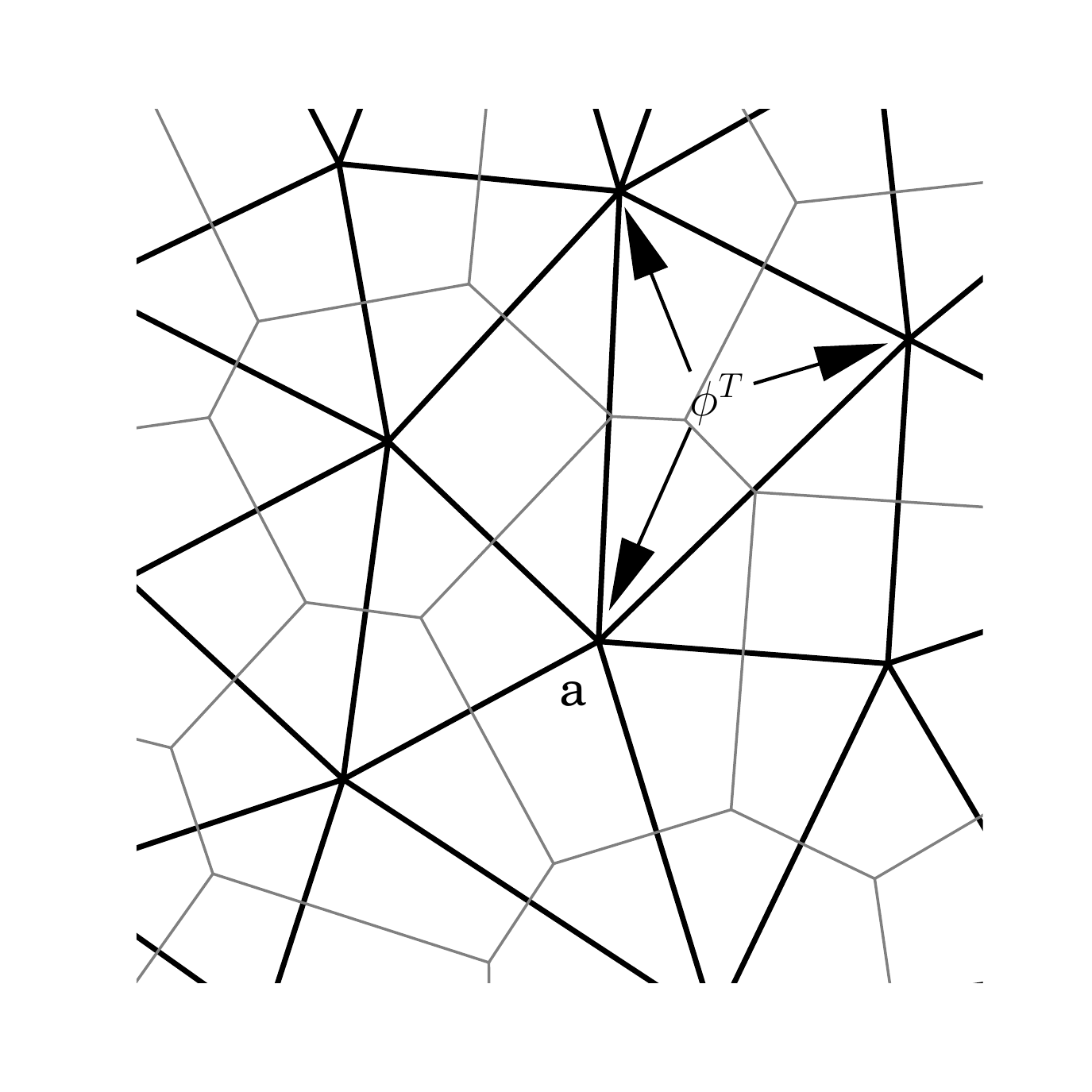}}
\caption{Illustration of how the state at vertex $a$ is updated. Each triangle $T$ has a residual $\phi^T$, which is redistributed among its vertices. The total update to vertex $a$ is the sum of the contributions of all seven triangles that have $a$ as a vertex. Each vertex is associated with a Voronoi cell, shown in grey lines, defined by the Delaunay triangulation.}
\label{figUpdate}
\end{figure}

Residual distribution schemes take the cell residual, and, as the name suggests, redistributes it over neighbouring nodes (cf. the one-dimensional case (\ref{eqResDist1D})). They take into account the hyperbolic nature of the equations by looking at upwind directions, and do so in a multidimensional way by looking at the eigenvalues of $\mK$ rather than $\mA$ and $\mB$ separately. All nodes gather parts of the residuals of all triangles that have that specific node as a vertex. This process is illustrated in Fig. \ref{figUpdate}. Triangle $T$ has residual $\phi^T$, which is redistributed over the three vertices of $T$. The total update at vertex $a$ is the sum of the contributions of the seven triangles that have $a$ as a vertex. Defining $\phi^T_i$ as the part of the residual of triangle $T$ to be sent to node $i$, an update would look like
\begin{equation}
\bW^{n+1}_i = \bW^n_i - \frac{\Delta t}{V_i}\sum_{T, i\in T} \phi^T_i,
\label{eqUpdate}
\end{equation}
where $n$ denotes the number of time steps taken so far, and $V_i$ is the volume associated with node $i$. For a Delaunay triangulation, $V_i$ is the volume of the Voronoi cell centred on $i$ (see Fig. \ref{figUpdate}). For stability, the time step is limited by a CFL condition:
\begin{equation}
\Delta t \leq \min_i \frac{2V_i}{\sum_{T:i\in T} l^T_\mathrm{max} \lambda^T_{\mathrm{max}}},
\end{equation}
where the minimum is taken over all vertices in the triangulation, $l^T_\mathrm{max}$ is the length of the longest edge of $T$ and $\lambda^T_\mathrm{max}$ is the maximum possible signal speed. For the Euler equations,
\begin{equation}
\lambda^T_\mathrm{max}=\max_{j\in T}\left(|{\bf v}_j| + c_j\right),
\end{equation}
where ${\bf v}_j$ and $c_j$ are the velocity and sound speed at vertex $j$.

The update (\ref{eqUpdate}) is only first order accurate in time. Higher order temporal accuracy is possible but will depend on the exact scheme used and will be discussed in section \ref{secRK2}. A residual distribution scheme is defined by how it defines $\phi_i$, or, in other words, how the residual is distributed over the neighbouring nodes. Below, we discuss two possible choices for the distribution function.

\subsection{Distribution schemes}

Traditionally, residual distribution schemes have been used to find \emph{steady} solutions to the Euler equations, with (\ref{eqUpdate}) used only to reach the required steady state. In this case, temporal accuracy is not an issue. The final steady state will depend on the distribution coefficients $\phi_i$. Several design criteria have been identified \citep[e.g.][]{vanderweide98}, of which we have encountered two already: conservation and multi-dimensional upwinding. The former sets the linearization (\ref{eqLinearA}), while the latter was introduced through the use of  $\mK$ rather than using $\mA$ and $\mB$ separately. All schemes discussed below are both conservative and multidimensional upwind.

Two other important considerations are \emph{positivity}, or monotinicity, and \emph{linearity preservation}. A scheme is said to be positive when no new extrema are introduced in the solution when going from one time step to the next. It is therefore related to the concept of total variation diminishing \citep{leveque02}, and is especially important in compressible flows since a positive scheme does not introduce oscillations near discontinuities. If a scheme is linearity preserving it means that exact linear solutions are recovered by the scheme. In a steady state, this means such a scheme is second-order accurate in space \citep{abgrall01}.

A scheme is said to be linear if, when applied to a linear partial differential equation such as
\begin{equation}
\frac{\partial u}{\partial t} + {\bf a}\cdot \nabla u=0,
\end{equation}
the solution update can be expressed as
\begin{equation}
u_i^{n+1} = \sum_{j=1}^N c_j u_j^n,
\end{equation}
where $N$ is the total number of grid points and the coefficients $c_j$ are independent of $u$. For example, the one-dimensional first-order Roe scheme is linear. Unfortunately, as a consequence of Godunov's theorem, a linear scheme can not be both positive (monotone) and linearity preserving (second-order) \citep{struijs94}. Just as in the one-dimensional case, non-linear schemes have to be designed in order to get the best of both worlds.

\subsubsection{Linear $N$ scheme}

The $N$ scheme \citep[$N$ for narrow,][]{struijs94,vanderweide98} is a monotonic scheme that is at most first-order accurate in space. The distribution function is given by
\begin {equation}
\phi^N_i = \mK_i^+\left(\bZ_i -\mathcal{\hat N}\sum_{j=1,j\in E}^{d+1}\mK_j^- \bZ_j\right),
\end{equation}
where
\begin{equation}
\mathcal{\hat N}=\left(\sum_{i=1,i\in E}^{d+1}\mK_i^-\right)^{-1}
\end{equation}
There are certain cases for which the inverse matrix $\mathcal{\hat N}$ does not exist, for example at stagnation points. However, the product $\mK^+_i \mathcal{\hat N}$ always has meaning, making the $N$ scheme always well-defined.

\subsubsection{Linear $LDA$ scheme}

While the $N$ scheme can deal with shocks in a stable and satisfactory way, its first-order nature makes it too diffusive in smooth flows to be of practical use. A popular second-order scheme is $LDA$ \citep[Low Diffusion A,][]{struijs94, vanderweide98}, for which the distribution function is given by:
\begin{equation}
\phi^{LDA}_i=\beta_i\phi^T.
\end{equation}
with distribution coefficients
\begin{equation}
\beta_i=-\mK_i^+\mathcal{\hat N}.
\label{eqBetaLDA}
\end{equation}
Using the $LDA$ scheme in the presence of discontinuities leads to unphysical oscillations, as expected.

\subsubsection{Non-linear blended schemes}
\label{secBlend}

In order to get the best of both worlds, second-order accuracy in regions of smooth flow while remaining monotone in the presence of discontinuities, schemes that blend the $N$ and $LDA$ residue have been designed:
\begin{equation}
\phi^{B}_i=\Theta^E\phi_i^N + \left(I-\Theta^E\right)\phi_i^{LDA},
\end{equation}
where $\Theta^E$ is a diagonal non-linear blending matrix
\begin{equation}
\Theta^E_{k,k}=\frac{\left|\phi^E_k\right|}{\sum_{j=1,j\in E}^{d+1}\left|\phi^N_{j,k}\right|},
\end{equation}
where subscript $k$ indicates the $k$th equation of the system \citep[e.g.][]{csik02}. This scheme will be referred to as the $B$ (for blended) scheme. Two useful variants can be obtained by setting all diagonal values of $\Theta^E$ to $\theta$, there $\theta$ can be taken to be either the maximum ($B$max) or the minimum ($B$min) over all $\Theta_{k,k}$. $B$max favours the $N$ scheme when in doubt, and is therefore a good choice when strong shocks are present in the solution, while $B$min favours the $LDA$ scheme, and is therefore a better choice for solutions that are relatively smooth.

A different blending scheme was proposed in \cite{dobes05}, where the blending coefficient is taken to be to be a scalar $\theta$:
\begin{equation}
\phi^{Bx}_i=\theta\phi_i^N + \left(1-\theta\right)\phi_i^{LDA},
\end{equation}
which is based on a shock sensor $s$:
\begin{equation}
\theta=s^2 h,
\end{equation}
where $h$ is a measure of the size of the element and
\begin{equation}
s=\left(\frac{-L\nabla\cdot {\bf v}}{\left|{\bf v}\right|_\mathrm{max}-\left|{\bf v}\right|_\mathrm{min}}\right)^+,
\end{equation}
where $L$ is the domain size, is a sensor that is non-zero only in regions of compression and is $O(1)$ in regions of smooth flow, which makes $\theta$ $O(h)$ and the scheme second-order accurate \citep{dobes05}. This scheme will be referred to as the $Bx$ scheme.

\subsection{Second order temporal accuracy}
\label{secRK2}

When time-dependent problems are considered, temporal accuracy becomes an issue. While there exists a residual distribution scheme that is second-order in space and time, the Lax-Wendroff scheme (cf. section \ref{secSecond1D}), this has to be mixed with a first-order scheme whenever discontinuities are present, and such schemes have been only moderately successful \citep{maerz96, ferrante97, hubbard00}.

The second method for improving the order of accuracy of section \ref{secSecond1D}, the method of lines, has seen much more progress in recent years. Unfortunately, a straightforward implementation of the method of lines is impossible due to the fact that residual distribution schemes in their basic formulation suffer from an inconsistent spatial discretization \citep{maerz96}. For linearity preserving schemes such as LDA,  a consistent update can be obtained  through a Petrov-Galerkin formulation  well known from finite element analysis , but for positive schemes another approach is needed \citep{abgrall03}.

In the time-dependent case, the \emph{total} residual $\Phi^T$has both a space and a time component:
\begin{equation}
\Phi^T=\int_T \frac{\partial \bW^h}{\partial t}dV + \phi^T=\sum_{i=1}^{d+1} \frac{|T|}{3}\frac{d\bW_i}{dt} + \phi^T.
\end{equation}
As for the steady case, the question is how to redistribute this residual over the nodes. In order for the scheme not to suffer from an inconsistent discretization we introduce a mass matrix $m_{ij}$, well known from finite element analysis:
\begin{equation}
\Phi_i^T=\sum_{j=1}^{d+1}m_{ij}\frac{d\bW_j}{dt} + \phi^T_i.
\end{equation}
The particular form of $m_{ij}$ is discussed below. A time step can now be taken by requiring that for every node
\begin{equation}
\sum_{T, i \in T} \Phi_i^T=0.
\end{equation}
For a general choice of $m_{ij}$, this leads to an implicit solver. Explicit schemes were derived only very recently \citep{rossiello09,ricchiuto10}. In the following, we present the second-order scheme of \cite{ricchiuto10}, which is a two-stage Runge-Kutta method:
\begin{eqnarray}
\bW_i^* &=& \bW_i^n - \frac{\Delta t}{V_i}\sum_{T|i\in T}\phi_i(\bW_h^n) \label{eqRK1}\\
\bW_i^{n+1}&=& \bW_i^* - \frac{\Delta t}{V_i}\sum_{T|i\in T} \Phi_i,\label{eqRK2}
\end{eqnarray}
where we have indicated specifically that the spatial residue in the first step is based on the interpolation $\bW^h$ at time level $n$. The total residue $\Phi_i$ appearing in the second step is for the $N$ scheme given by
\begin{equation}
\Phi_i^N=\frac{|T|}{3}\frac{\bW_i^*-\bW_i^n}{\Delta t}+\frac{1}{2}\left(\phi^N_i(\bW^*_h)+\phi^N_i(\bW^n_h)\right),
\end{equation}
while for the $LDA$ scheme it reads:
\begin{equation}
\Phi^{LDA}_i=\sum_{j \in T} m_{ij}^{LDA} \frac{\bW_j^*-\bW_j^n}{\Delta t}+\frac{\beta_i}{2}\left(\phi(\bW_h^n)+\phi(\bW_h^*)\right).
\end{equation}
Note that this integration scheme is closely related to total-variation-diminishing time integration schemes \citep{shu88}. For the $N$ residual, the mass matrix is simply
\begin{equation}
m_{ij}^N=\frac{|T|}{3}\delta_{ij},
\end{equation}
which is not consistent but since the $N$ scheme is only first order anyway this is not a problem. Several choices can be made for the mass matrix in the $LDA$ residue \citep{ricchiuto10}, from which we choose\footnote{While different versions of $m_{ij}$ can formally be ranked according to their dissipative nature, much less is known about their stability. The choice of (\ref{eqMassMatrixLDA}) is based on simplicity and apparent stability.}
\begin{equation}
m_{ij}^{LDA} =\frac{|T|}{3}\beta_i,
\label{eqMassMatrixLDA}
\end{equation}
where $\beta_i$ is given by (\ref{eqBetaLDA}). A slightly more complicated update, called \emph{selective lumping} \citep{ricchiuto10} adds an anti-diffusive term to the residual:
\begin{equation}
\Phi^{SL}_i = \Phi_i + \sum_{j\in T} \left(\frac{|T|\delta_{ij}}{3}-m^G_{ij}\right)\frac{\bW^*_j-\bW^n_j}{\Delta t},
\end{equation}
where $\Phi_i$ can be either $\Phi^{LDA}_i$ or $\Phi^N_i$ and $m^G_{ij}$ is the Galerkin mass matrix
\begin{equation}
m_{ij}^G=\frac{|T|}{12}\left(\delta_{ij}+1\right).
\end{equation}
For blended schemes, the $N$ and $LDA$ total residue are mixed in the usual way (see section \ref{secBlend}). The resulting scheme is second order accurate in space and time wherever the blending procedure favours the $LDA$ scheme \citep{ricchiuto10}.

To summarise, a complete integration scheme is defined by the residual distribution scheme ($N$, $LDA$, $B$, etc), a mass matrix, and a choice of lumping. In all test problems discussed below, we stick to the mass matrix of equation (\ref{eqMassMatrixLDA}) together with global lumping. The only choice left is the residual distribution scheme, with the annotation that when using the first order $N$ scheme we also use a first order time integration.

\subsection{Boundary conditions}

On structured grids, boundary conditions are often imposed by adding a layer of ghost cells to the computational domain, whose states are set in such a way to achieve the desired boundary condition. Periodic boundary conditions, for example, are simply achieved by copying the relevant states from the other side of the computational domain into the ghost cells. A reflecting boundary can be achieved by copying the states next to the boundary from the computational domain into the ghost cells but reversing the velocity normal to the boundary. One reason why this approach is very effective in the case of structured grids is that the boundaries always  align with one  of the coordinate axis. A second reason is that all cells have the same shape and volume so that copying the state is trivial\footnote{Exceptions include for example structured grids in curvilinear coordinates, where cells at different spherical radii $r$ will have different volumes $\propto r^2$.}.

For an unstructured grid, we do not necessarily have the boundary aligned with one of the coordinate axis, and, since all computational cells are slightly different, having a layer of ghost cells would mean copying part of the grid structure, which is expensive both in terms of computational effort and memory requirement, and should therefore be avoided. An exception to this rule are \emph{non-reflecting} boundaries, where the boundaries are taken to be so far away from the region of interest that their exact shape does not matter. In this case, we promote the boundary vertices to ghost vertices, whose states never changes from the initial conditions, which are taken to be a stationary state. All vertices connected to the ghost vertices, because of multidimensional upwinding, `see' waves through the usual characteristic decomposition. A wave trying to leave the computational domain can do so, but if information needs to be drawn from outside the computational domain because one of the characteristics points inward, this information is drawn from the ghost cells containing no wave. Therefore, no waves enter the computational domain and therefore we call these boundary conditions \emph{non-reflecting}, and they are relatively trivial to implement: at the start of a time step, all boundary cells have to be set to the initial condition. This procedure can also be used to specify an \emph{inflow} boundary.

Periodic boundary conditions are completely handled by the mesh. If the mesh is periodic in both $x$ and $y$, all vertices and triangles have neighbours in all directions (see Fig. \ref{figMesh2D}). Therefore, for periodic boundaries, no extra work is required.

The final boundary discussed here is a reflecting wall. Such a boundary is most easily enforced in a weak sense, which means specifying the flux across the boundary rather than the state at the boundary. For the implementation we follow \cite{vanderweide98}. Consider a triangle of which one edge $e$ with vertices $a$ and $b$, belongs to a reflecting wall, and let the normalized normal of this edge be ${\bf n}=(n_x,n_y)^T$. The desired flux at a reflecting wall is such that the velocity normal to the wall vanishes: $v_n={\bf v} \cdot {\bf n}=0$. Usually the flux as computed as if there was no wall does not obey this condition, and therefore a correction flux has to be applied:
\begin{eqnarray}
{\bf F}^\mathrm{c} n_x + {\bf G}^\mathrm{c} n_y= -\left(\begin{array}{c} \rho v_n\\ \rho u v_n \\ \rho v v_n \\ \rho h v_n\end{array}\right).
\end{eqnarray}
The correction residual is then given by
\begin{eqnarray}
\Phi^{c} &=& \int_e \left({\bf F}^\mathrm{c} n_x + {\bf G}^\mathrm{c} n_y\right)de \nonumber\\
& \approx& \frac{|e|}{2}\left({\bf F}_a^\mathrm{c}n_x + {\bf G}_a^\mathrm{c}n_y + {\bf F}_b^\mathrm{c}n_x + {\bf G}_b^\mathrm{c}n_y\right),
\end{eqnarray}
where the trapezium rule was used to approximate the integral. This residual is distributed over the nodes $a$ and $b$ using a parameter $\alpha \in [0,1]$:
\begin{eqnarray}
\Phi_a^{c} = \frac{|e|}{2}\left(\alpha\left({\bf F}_a^\mathrm{c}n_x + {\bf G}_a^\mathrm{c}n_y\right) + (1-\alpha)\left({\bf F}_b^\mathrm{c}n_x + {\bf G}_b^\mathrm{c}n_y\right)\right),\\
\Phi_b^{c} = \frac{|e|}{2}\left((1-\alpha)\left({\bf F}_a^\mathrm{c}n_x + {\bf G}_a^\mathrm{c}n_y\right) + \alpha\left({\bf F}_b^\mathrm{c}n_x + {\bf G}_b^\mathrm{c}n_y\right)\right),
\end{eqnarray}
Following \cite{vanderweide98}, we choose $\alpha=0.75$.

\section{GPU implementation}
\label{secIntroGPU}

GPUs have emerged relatively recently as viable alternatives to large distributed-memory machines. Modern single GPU cards are capable of Teraflops performance, comparable in speed to a CPU cluster of a few $100$ cores but at a fraction of the cost. However, getting close to peak performance of a GPU is not straightforward, even though in recent years is has become much easier.

The computational intensity of numerical fluid dynamics made it a prime candidate to be ported to GPUs. Both SPH \citep{herault10} and structured grid-based methods \citep{hagen06} were successfully run on GPUs, with unstructured grid methods for magneto-hydrodynamics following later \citep{lani14}. The latter was based on an implementation of the Rusanov flux \citep{rusanov61}. A method specific for turbulence calculations on hybrid grids was presented in \cite{asouti11}. To the best of  our  knowledge, {\sc astrix} is the first implementation of an explicit residual distribution method on GPUs.

{\sc astrix} is written using NVIDIA's CUDA (Compute Unified Device Architecture) programming model. CUDA-capable cards come in different generations or \emph{compute capabilities}. The higher the compute capability, the newer the card and this means more features may be available. The first generation of CUDA-capable cards were built to comply with single precision IEEE requirements\footnote{The first generation CUDA cards with compute capability $1.$x does not support denormal numbers, and the precision of division and square root operations are slightly below IEEE 754 standards \citep{whitehead11}. Cards with compute capability $2.$x and higher do not suffer from these issues.}, which, as we saw  turns out to be important for generating unstructured meshes as at several stages we need exact geometric predicates, while newer cards (compute capability $2.$x and higher) fully support double precision arithmetic.

Designing algorithms for the GPU is fundamentally different from designing traditional parallel CPU algorithms, simply because the GPU is a very different beast. A useful analogy is that the CPU is a single genius, while the GPU represents millions of unskilled labourers. The GPU gets its performance not by doing single computations fast, but by taking on a lot of single computations at the same time and switching between them if necessary. Of course, this can only be done if the computations are independent. Therefore, we need to expose enough parallellism: we need to load the GPU with millions of independent relatively small tasks in order to get close to peak performance. This is done by launching many threads (typically millions) that each work independently. These threads are organised into \emph{warps} that can be moved between computing units very efficiently, thereby hiding instruction and memory latency. Exposing enough parallellism is straightforward in the residual distribution schemes discussed in section \ref{secResDist}: the bulk of the computations can be done independently for all triangles, and one expects a two-dimensional mesh to contain typically millions of triangles. The situation is more tricky for mesh generation as will be explained in section \ref{secMeshGPU} below.

A second condition for good GPU performance is adequate use of GPU memory. Data transfer from CPU to GPU is slow since it has to go through the PCI bus (typical speed $6$ GB/s). This has to be kept in mind when porting only part of an application to GPU. While the GPU may speed up a particular computation 100 times, if this computation is preceded and followed by a few seconds of data transfer the overall speedup may be negligible or even negative. The converse is also true: if a particular part of an algorithm is not very well suited to run on a GPU, the benefit of running this part on the CPU may be outweighed by the increased memory traffic. {\sc astrix} is designed with minimal CPU-GPU memory transfers: all data reside on the GPU and stay on the GPU, unless explicit output is required.

Finally, this brings us to performance metrics. From a user's point of view, it is important to know how much the code speeds up when running on the GPU compared to the CPU. Unfortunately, this is extremely dependent on the GPU/CPU combination. Moreover, one should compare an algorithm optimised for the GPU to an algorithm optimised for the CPU, and usually these are very different algorithms because the GPU works differently from a CPU. A fair comparison therefore requires designing and optimising two different algorithms performing the same task. While for completeness we do mention GPU/CPU speedups when measuring performance of {\sc astrix}, above considerations should be kept in mind.

\subsection{Mesh generation}
\label{secMeshGPU}

As discussed above, efficient use of a GPU is non-trivial. In order to expose as much parallelism as possible, most steps in the Delaunay refinement algorithm are done launching one CUDA thread per element. For example, when finding low-quality triangles, each thread will check a single triangle if the quality and size constraints are met. This leads to a list of vertices to add to the mesh, for which we can find their containing triangles independently, and also check independently whether they encroach upon any segment.

\subsubsection{Parallel insertion set}
\label{secParallel}

Inserting new vertices requires more attention, since not all vertices can be inserted independently. For example, two new vertices may find themselves in the same triangle, in which case only one of them can be inserted at a time. If a new vertex is to be inserted on an edge, no vertices can be inserted in the neighbouring triangles. A more stringent constraint arises from the demand that it must be possible to generate the mesh by inserting the vertices one at a time. This is important because all proofs of quality and termination of the algorithm rely on inserting new vertices sequentially. Therefore, we have to select a subset of circumcentres that can be inserted independent from each other.

\begin{figure}
\centering
\resizebox{\hsize}{!}{\includegraphics[]{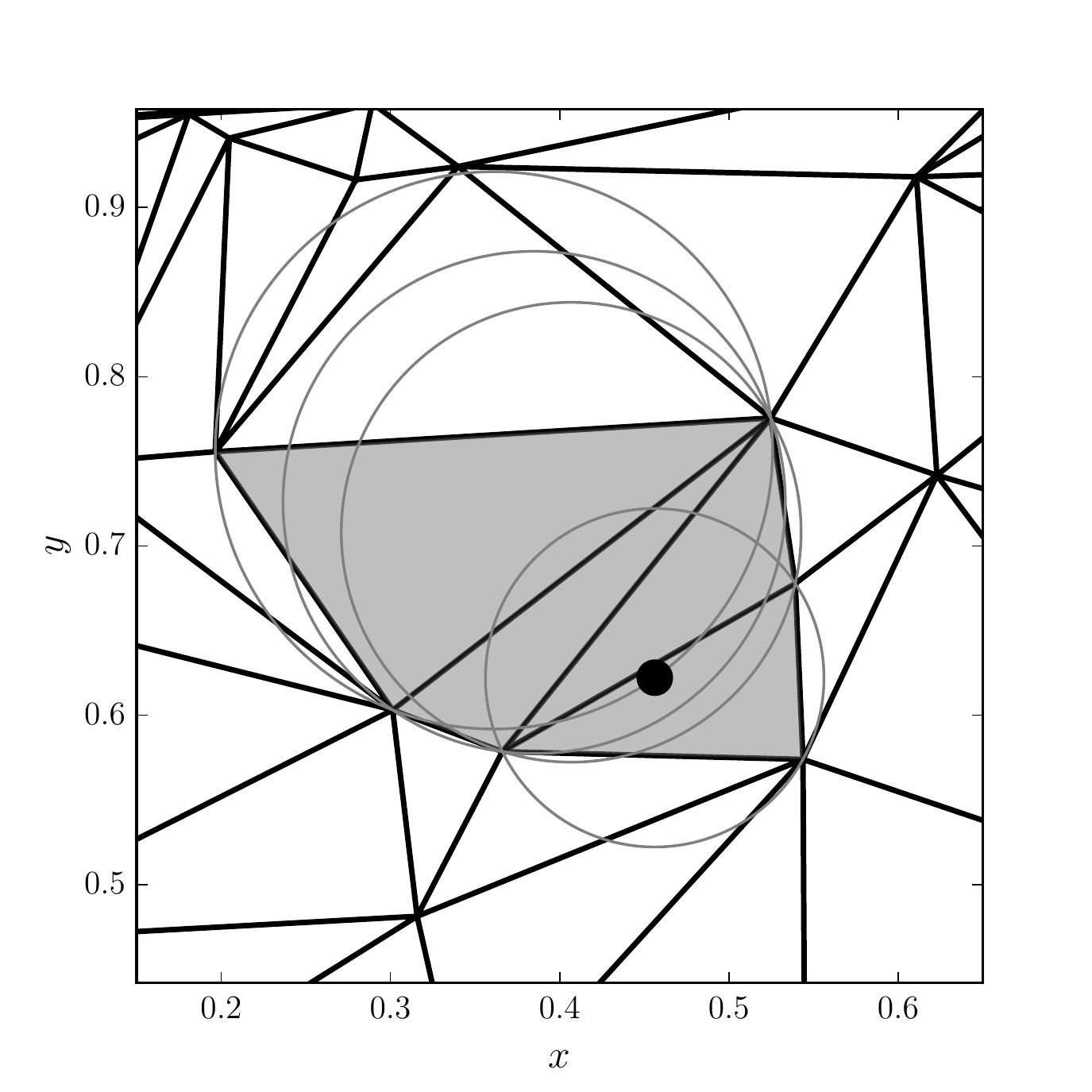}}
\caption{Insertion cavity of a new vertex. The vertex to be inserted is denoted by the black dot, and all triangles belonging to the cavity are coloured grey. Also shown are the circumcircles of the cavity triangles.}
\label{figCavity}
\end{figure}

It is obvious that a single new vertex $v$ will only affect those triangles in the mesh whose circumcircles contain $v$. All other triangles have empty circumcircles and will therefore be part of the new triangulation. This leads to the definition of the \emph{cavity} of $v$: the set of triangles whose circumcircles contain $v$. Then two vertices can be inserted independently from each other if their cavities do not overlap. An example of a cavity is shown in Fig. \ref{figCavity}.

Finding all triangles belonging to the cavity of $v$ is relatively straightforward. We already have found the insertion triangle (see section \ref{secFindTriangle}). Starting from this triangle, which is of course part of the cavity, move into one of the neighbouring triangles if it also belongs to the cavity. By always checking the neighbours in anti-clockwise order, it is possible to find all triangles belonging to the cavity in relatively few steps.

A simple algorithm for selecting non-overlapping cavities is the following. For every new vertex $v$, flag all triangles belonging to its cavity with a unique integer $i(v)$. If a triangle has already been flagged with $i'(v')$, take $\mathrm{max}(i,i')$ as the flag so that a triangle will always be associated with at most one new vertex, with higher values of $i(v)$ being given priority. Taking the maximum as mentioned above involves three steps: 1) reading the current value of the flag, 2) comparing it with the new value, 3) write back the new value if it is higher than the old value. In order to prevent a race condition, these three operations have to be done without interference from other GPU threads, which can be done within CUDA through so-called atomic functions, in this case \texttt{atomicMax}. Once all new vertices are processed in this way, we walk through the cavities a second time, and check for every new vertex $v$ if \emph{all} triangles belonging to its cavity are flagged with $i(v)$. If so, the vertex can be inserted.

Of course, we want to insert as many vertices in one parallel step as possible. There exists what is called a maximal independent set \citep[e.g.][]{luby86}; a maximum number of vertices that can be inserted in a single parallel step. While \cite{spielman07} present an algorithm for calculating the maximal independent set for this specific problem, the simple algorithm presented above performs remarkably well if the integers $i(v)$ are chosen to be random. That is, if there are $N$ bad triangles in the mesh, and therefore potentially $N$ vertices to insert $(v_1...v_N)$, give each vertex a unique random integer as its $i(v)$. This leads to independent cavities covering the whole domain relatively uniformly.

\begin{figure}
\centering
\resizebox{\hsize}{!}{\includegraphics[]{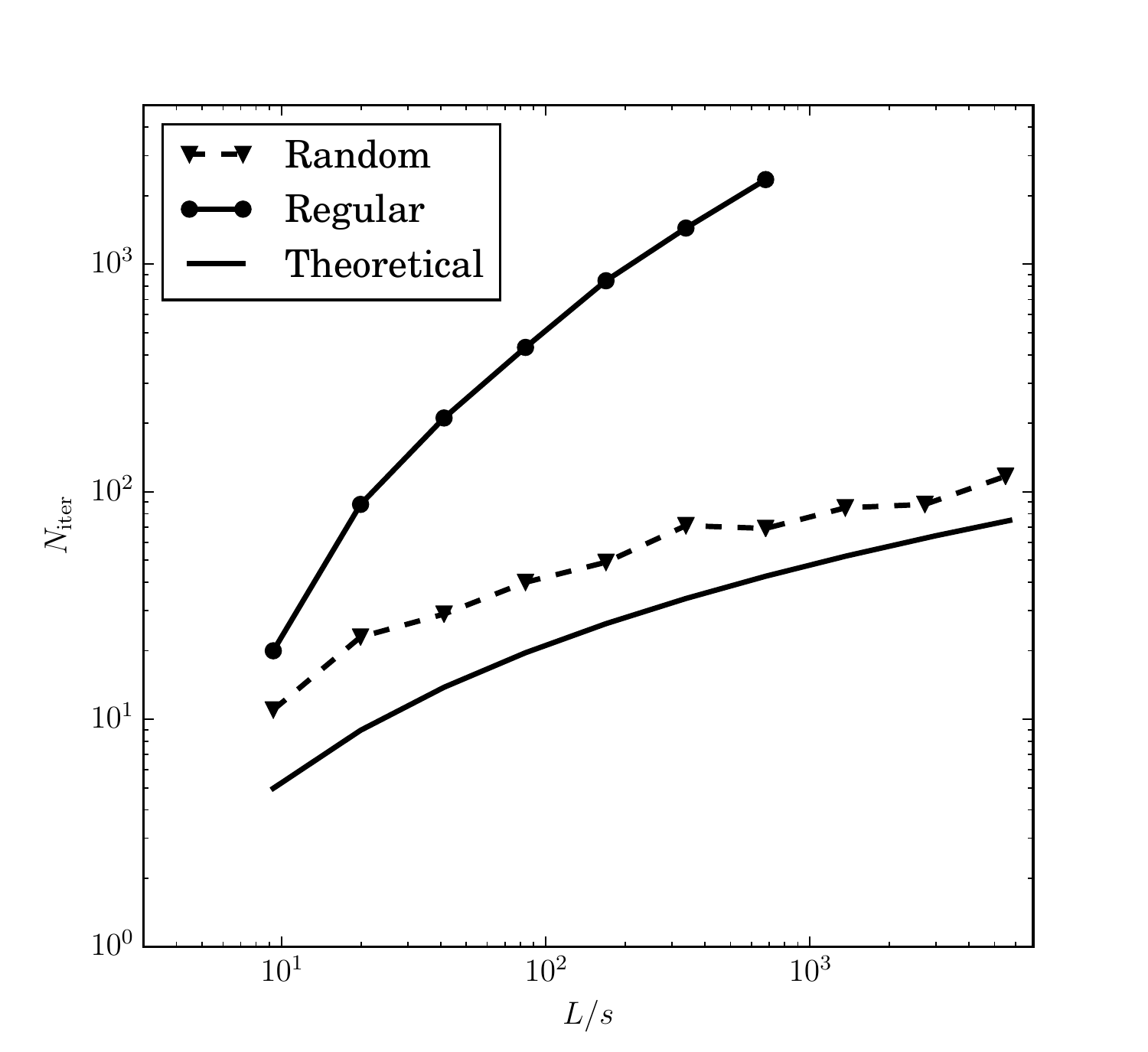}}
\caption{Number of insertion iterations needed for an unstructured mesh of size $L$ with quality constraint $B=\sqrt{2}$ and minimum circumradius $s$. The solid curve shows the regular case, while the dashed curve shows the case where we assign random integers to the new vertices. Also shown is the theoretical bound of $O(\log^2(L/s))$ \citep{spielman07}.}
\label{figParallel}
\end{figure}

The algorithm presented in \cite{spielman07} takes $O(\log^2(L/s))$ iterations, where $L$ is the domain size and $s$ the smallest circumradius in the final mesh. In Fig. \ref{figParallel}, we compare our simple algorithm to this theoretical limit, both for regular integer assignment (i.e. $i(v_n)=n$) and random integer assignment. Note that because of Morton ordering (see section \ref{secMorton} below), vertices with large values of $i$ are located very close to each other, making the parallel selection method very inefficient. However, for random integer assignment, the number of iterations necessary follows the theoretical limit rather nicely. In addition, finding the cavities associated with the new vertices has the additional advantage that we know which edges may need flipping: only edges part of a cavity of a new vertex plus any newly created edges need to be checked for the Delaunay property. This saves a lot of redundant checking of edges.

\subsubsection{Data structure}
The mesh contains \texttt{nVertex} vertices, \texttt{nTriangle} triangles and \texttt{nEdge} edges. The basic structure of the grid is stored in four arrays, using CUDA intrinsics:
\begin{itemize}
\item{\texttt{vertexCoordinates}; a \texttt{float2} array of size \texttt{nVertex} containing the $x$ and $y$ coordinates of all vertices in the mesh.}
\item{\texttt{triangleVertices}; an \texttt{int3} array of size \texttt{nTriangle} containing for every triangle the three vertices that make up the triangle.}
\item{\texttt{triangleEdges}; an \texttt{int3} array of size \texttt{nTriangle} containing for every triangle its three edges.}
\item{\texttt{edgeTriangles}; an \texttt{int2} array of size \texttt{nEdge} containing for every edge the two neighbouring triangles (or just one if the edge is part of the boundary and therefore a segment).}
\end{itemize}

\subsubsection{Exact geometric predicates}
\label{secPredicates}

As indicated in the previous sections, at several stages (finding insertion triangles and testing edges for the Delaunay property) we need exact geometric predicates, i.e. the exact sign of the determinants $\textsc{Orient2D}$ (\ref{eqOrient}) and $\textsc{InCircle2D}$ (\ref{eqInCircle}). While this can be done in principle using exact arithmetic, the price is quite high: up to two orders of magnitude reduction in speed. Fortunately, an adaptive method was designed by \cite{shewchuk97}, based on earlier work by \cite{priest91}. The key insight is that the exact determinant is not needed: all we are interested in is the sign. If we can be sure that a calculation at finite precision gives the correct sign, there is no need to make it more precise. These algorithms work on most processors, in particular those complying to the IEEE 754 standard, and can therefore be ported in a straightforward way to modern GPUs.

\subsubsection{Morton ordering}
\label{secMorton}

Data locality has always been critical for efficient use of GPUs. On older cards (compute capability 1.x), when reading an array from global device memory, it was critical for neighbouring threads to read neighbouring data: if thread 0 reads $\texttt{array}[0]$, it was necessary for thread 1 to read $\texttt{array}[1]$ and so on; any other order would incur a speed penalty of up to 2 orders of magnitude. More recent GPUs have relaxed these requirements by introducing on-chip cache, but this still means that data locality is highly desirable: if neighbouring threads read data that is close together in memory, chances are it can be found in the cache, which means a read from global device memory is unnecessary.

Unstructured meshes pose a challenge for maintaining data locality due to the non-trivial interconnections between vertices. Moreover, in the process of creating the mesh new vertices, edges and triangles are added, quickly destroying data locality even if it was present at some stage. In order to mitigate this, after every parallel insertion step, we reorder vertices, edges and triangles as to maintain as much data locality as possible. This is done by assigning a Morton value \citep{morton66} to for example each vertex, and then sorting the vertices according to their Morton value. The same for edges and triangles. While sorting itself is non-local and not trivial to implement of a GPU, CUDA has fast built-in sorting algorithms so that the overall effect on execution speed of Morton ordering is positive.

\subsubsection{Delaunay triangulator}

For efficient use of the GPU in maintaining a Delaunay triangulation, we want to flip as many edges as possible in parallel. First, we use the robust $\textsc{InCircle2D}$ (\ref{eqInCircle}) test to generate a list of edges that do not satisfy the Delaunay requirement. From this list, edges can be flipped independently if they are not part of the same triangle. We select an independent set in much the same way as done in section \ref{secParallel}, where the `cavity' of an edge is now defined by the two neighbouring triangles. Randomization was not found to be necessary in this case, since because the `cavities' are so small a large independent set can always be found.

Unfortunately, edge flipping can corrupt the data structure, in particular \texttt{edgeTriangles} (\texttt{triangleVertices} and \texttt{triangleEdges} are updated during the flip). While the flipped edge still has the same neighbouring triangles $t_1$ and $t_2$, other edges belonging to for example the original $t_1$ may suddenly have $t_2$ rather than $t_1$ as a neighbour. Fortunately, this is straightforward to correct. For every edge $e$, we look at its neighbouring triangles $t_1$ and $t_2$ through \texttt{edgeTriangles}. If none of the edges of $t_1$, as per \texttt{triangleEdges}, is equal to $e$ then this means that $e$ now neighbours $t_2$ rather than $t_1$. Therefore, before flipping edges, we create a triangle substitution array \texttt{triangleSub} so that for every edge to be flipped $\texttt{triangleSub}[t_1]=t_2$ and vice versa. Note that no conflicts can arise since any triangle can only be associated with one edge that will be flipped (otherwise these edges can not be flipped in parallel). After a parallel step of edge flipping, we can then replace $t_1$ or $t_2$ with $\texttt{triangleSub}[t_1]$ or $\texttt{triangleSub}[t_2]$, respectively, where necessary. See \cite{navarro11} for more details.

\subsubsection{GPU performance}
\label{secGPUperformMesh}

\begin{figure}
\centering
\resizebox{\hsize}{!}{\includegraphics[]{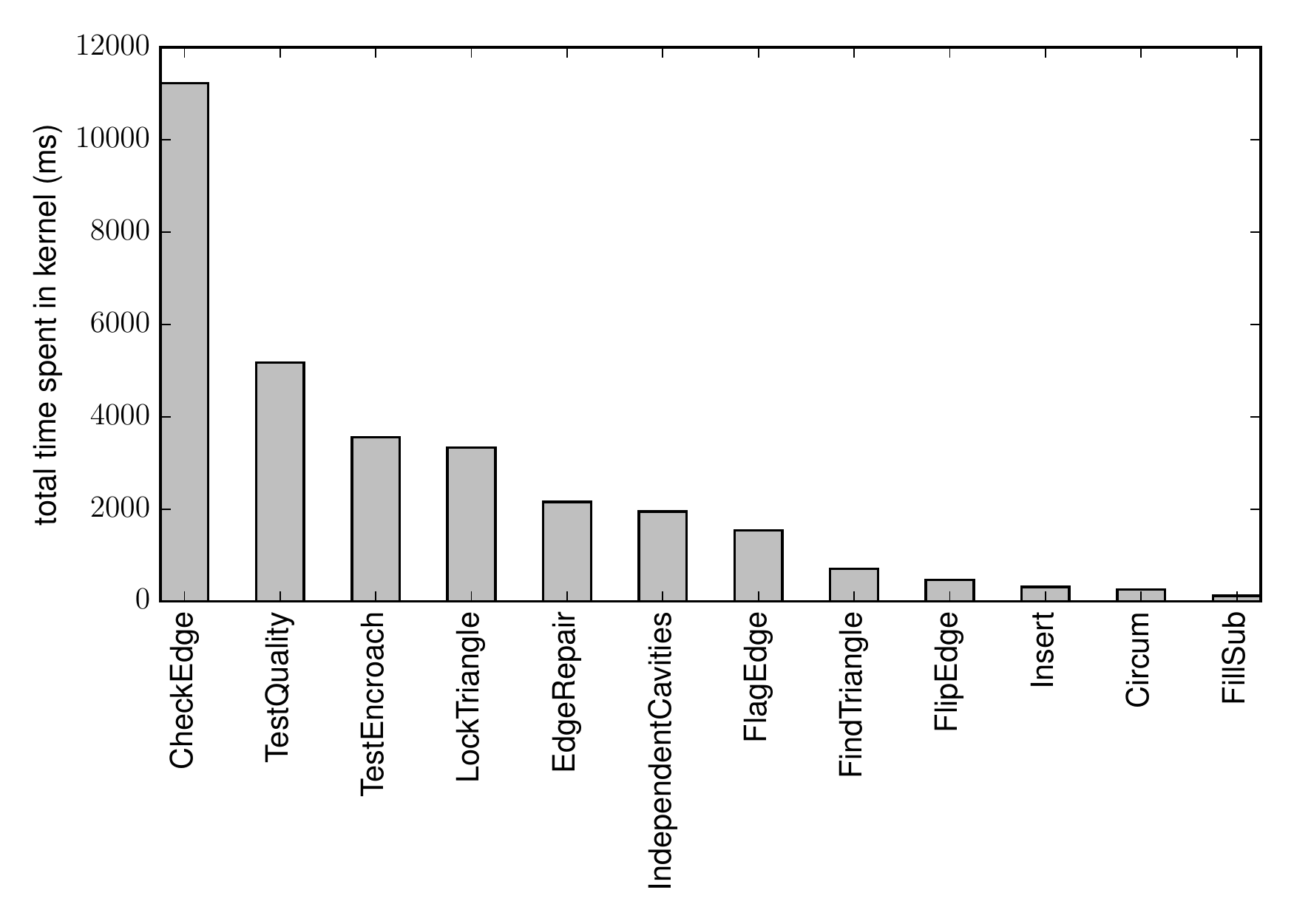}}
\caption{Total CPU time in milliseconds spent in kernels while generating a uniform unstructured periodic mesh with $1.3$ million vertices. From left to right, they represent checking edges for Delaunay-hood, testing which triangles are of low quality, testing if new vertices lead to encroached segments, lock all triangles in insertion cavities, repair edges after flipping, find independent insertion cavities, flag edges for checking Delaunay-hood, find insertion triangles, flip edges, insert new vertices, finding circumcentres of triangles and fill the triangle substitution array.}
\label{figCPUtimeMesh}
\end{figure}

\begin{figure}
\centering
\resizebox{\hsize}{!}{\includegraphics[]{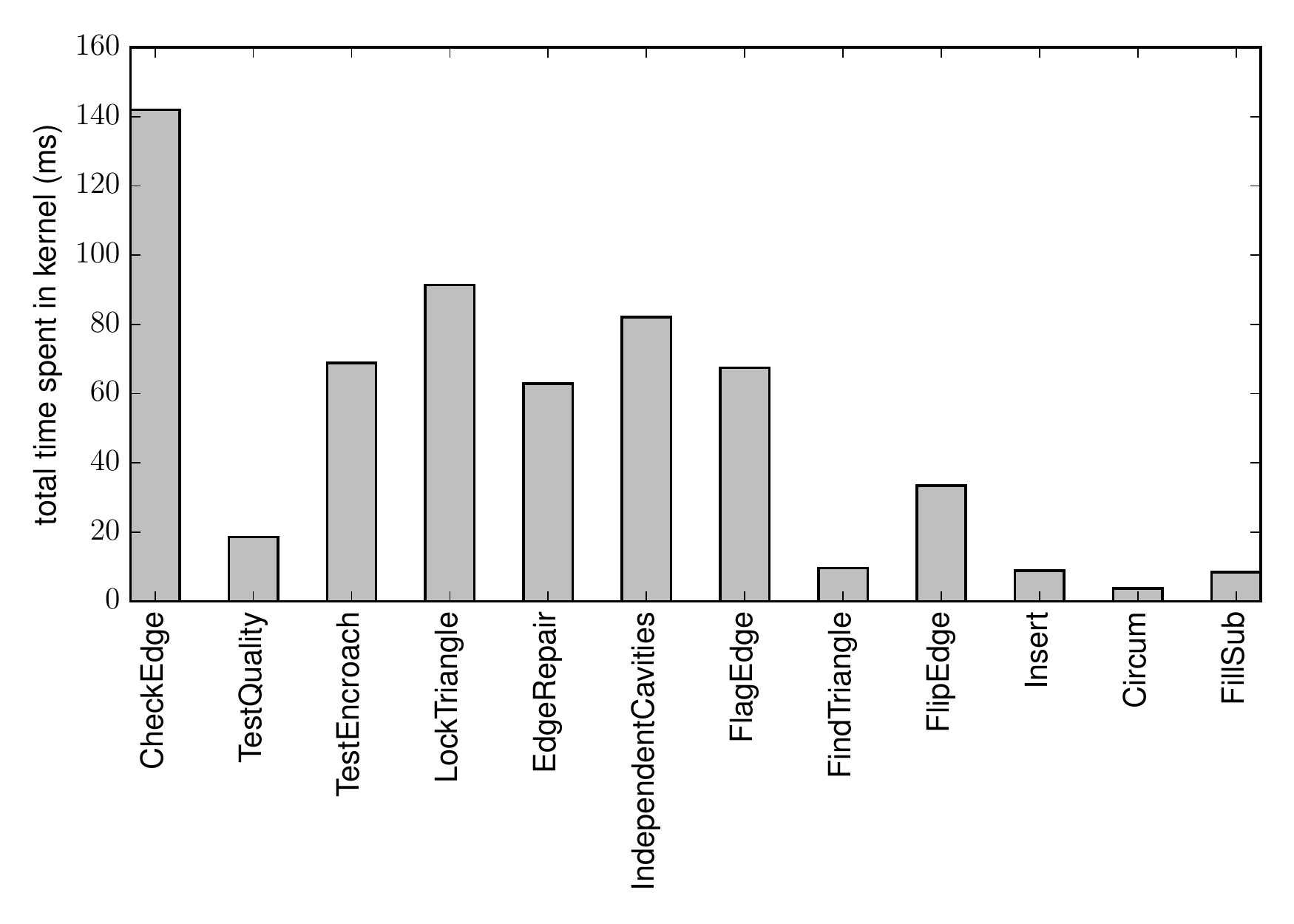}}
\caption{Total GPU time in milliseconds spent in kernels while generating a uniform unstructured periodic mesh with $1.3$ million vertices. Kernels are displayed in the same order as in Fig. \ref{figCPUtimeMesh}.}
\label{figGPUtimeMesh}
\end{figure}

As a test case, we consider the generation of a uniform unstructured mesh, periodic in both $x$ and $y$, with $1.3$ million vertices. We compare a GPU version to a CPU version, using exactly the same algorithms using single precision floating points. The test was run on a system consisting of an Intel Xeon $1.8$ GHz CPU and a NVIDIA Tesla K20m GPU, which has CUDA compute capability $3.5$.

In Fig. \ref{figCPUtimeMesh} the total time spent on the CPU in each `kernel'\footnote{When running on the CPU a kernel is replaced by a $\texttt{for}$-loop performing exactly the same task} is shown. Most time is spent checking if edges are Delaunay, followed by the quality check of triangles. It should be noted that the kernel $\texttt{TestQuality}$ is called once every refine cycle, while the kernel $\texttt{CheckEdge}$ is called multiple times in the same cycle until the mesh is Delaunay. The time spent in individual instances of $\texttt{CheckEdge}$ is actually smaller than that for $\texttt{CheckTriangle}$, but the number of kernel calls make $\texttt{CheckEdge}$ the most time-consuming kernel.

The corresponding timings for the GPU are shown in Fig. \ref{figGPUtimeMesh}, in the same order from left to right as Fig. \ref{figCPUtimeMesh}. Checking edges for Delaunay-hood is still the most expensive operation, but the costs have been reduced by a factor of $\sim 80$ compared to the CPU version. While this may seem as a healthy speedup, it is nowhere near the maximum capability of the Tesla K20m GPU, as we will se below. This is partly due to the fact that many calls to $\texttt{CheckEdge}$ involve only very few edges to be processed, hence limiting the parallelization. The maximum number of edges to be checked in a single kernel call is $\sim 650000$ for this particular mesh, giving a speedup of $\sim 100$ compared to the CPU. The brute force approach of checking \emph{all} edges in every Delaunay iteration therefore gives a bigger speedup, but the overall computational costs would still increase. A second reason for the relatively poor performance is that the kernel requires a lot of memory traffic. For every edge checked, we need to know all coordinates and all edges of the two neighbouring triangles, and because of the unstructured nature of the grid the memory access involved is not ideal for the GPU. The third reason is that the kernel has to make use of exact geometric predicates, which first of all makes the algorithm more complicated, which increases the number of registers used and therefore limits the amount of blocks that can be run simultaneously, and at the same time leads to warp divergence: the amount of computation performed can differ significantly for different edges.

On the other hand, the kernel $\texttt{TestQuality}$ show a much better speedup, from $\sim 220$ on average to $\sim 270$ maximum. The achieved bandwidth of $120$ Gb/s comes reasonably close to the theoretical maximum of the Tesla K20m GPU of $200$ Gb/s, given the unfavourable memory access pattern due to the unstructured nature of the mesh. Nevertheless, even here there is room for improvement, although the focus should of course be on the $\texttt{CheckEdge}$ kernel.

The other kernels worth mentioning are $\texttt{TestEncroach}$, $\texttt{LockTriangle}$, $\texttt{IndependentCavities}$ and $\texttt{FlagEdges}$. These have in common that they walk through the grid around a certain point, visiting an unknown number of triangles, for example the insertion cavity in the case of $\texttt{LockTriangle}$. These kernels show the worst speedup on the GPU ($\sim 50$), first of all for similar reasons as $\texttt{CheckEdge}$ mentioned above. In addition, there is the extra complication of insertion cavities having different sizes, which leads to different work loads for different GPU threads. Moreover, the size of the cavity is unknown beforehand, making optimisations more difficult for the compiler.

Overall, the creation of the $1.3$ million vertex mesh has sped up by roughly a factor of $100$ compared to the CPU, on a graphics card that costs only a fraction of a CPU compute cluster, making the effort of specialising to the GPU worthwhile.

\subsection{Hydrodynamics}

\subsubsection{Residual distribution}

The two-stage Runge Kutta update (\ref{eqRK1})-(\ref{eqRK2}) consists of four steps:
\begin{itemize}
\item{Calculate $\phi(\bW_h^n)$, $\phi_i^N(\bW_h^n)$ and $\phi_i^{LDA}(\bW_h^n)$}
\item{Blend into $\phi_i(\bW_h^n)$ and calculate $\bW^*$}
\item{Calculate $\phi(\bW_h^*)$, $\Phi_i^N$ and $\Phi_i^{LDA}$}
\item{Blend into $\Phi_i$ and calculate $\bW^{n+1}$}
\end{itemize}
These steps are distributed over the following GPU kernels:
\begin{itemize}
\item{$\texttt{CalcResidual}$: calculate $\phi(\bW_h^n)$, $\phi_i^N(\bW_h^n)$, $\phi_i^{LDA}(\bW_h^n)$}
\item{$\texttt{AddResidual}$: blend into $\phi_i(\bW_h^n)$ and calculate $\bW^*$}
\item{$\texttt{CalcTotalResNtot}$: calculate $\phi(\bW_h^*)$, $\Phi_i^N$}
\item{$\texttt{CalcTotalResLDA}$: calculate $\Phi_i^{LDA}$}
\item{$\texttt{AddResidual}$: blend into $\Phi_i$ and calculate $\bW^{n+1}$}
\end{itemize}
In addition, there are kernels for calculating the allowed time step, the parameter vector and to set the boundary conditions, but the vast majority of the computational time is spent in the kernels mentioned above. Note that the kernel $\texttt{AddResidual}$ performs exactly the same task twice but with different residuals.

Calculations of the residuals are independent for each triangle and can therefore be parallelized very efficiently. The node updates involve the contributions from all triangles sharing a particular node. While this could be parallelized over the nodes, we do not have direct information on which triangles share for example node $i$ from the mesh data structure. This would be difficult to achieve, since the number of triangles per node can vary quite a lot. It would be possible to assign one triangle to every node, and walk around the node collecting the contribution from all triangles sharing the node, but since \emph{all} nodes have to be updated we found it more efficient to again paralellize over triangles and update the nodal values using atomic operations.

\subsubsection{GPU performance}
\label{secGPUperformSim}

\begin{figure}
\centering
\resizebox{\hsize}{!}{\includegraphics[]{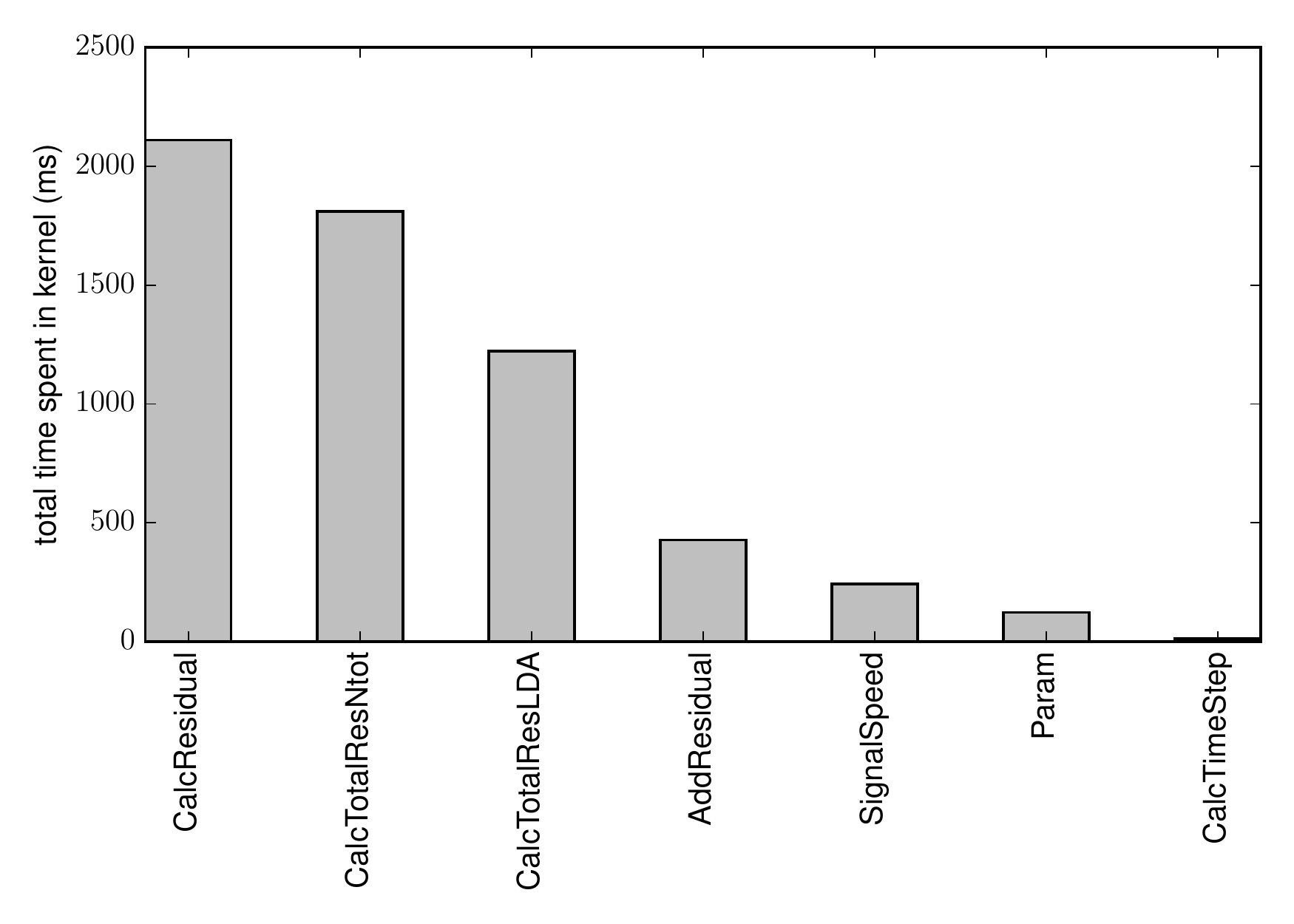}}
\caption{Total CPU time in milliseconds spent in kernels during a single time step on a uniform unstructured periodic mesh with $1.3$ million vertices. From left to right, they represent calculating the spatial residuals $\phi(\bW_h^n)$, $\phi_i^N(\bW_h^n)$ and $\phi_i^{LDA}(\bW_h^n)$, calculating the total residuals $\Phi_i^N$ and $\Phi^T$, calculating the total residual $\Phi_i^{LDA}$, adding the residuals to the vertices, calculating the maximum signal speeds within a triangle, calculating the blend parameter, calculating the parameter vector and calculating the time step.}
\label{figCPUtimeSim}
\end{figure}

\begin{figure}
\centering
\resizebox{\hsize}{!}{\includegraphics[]{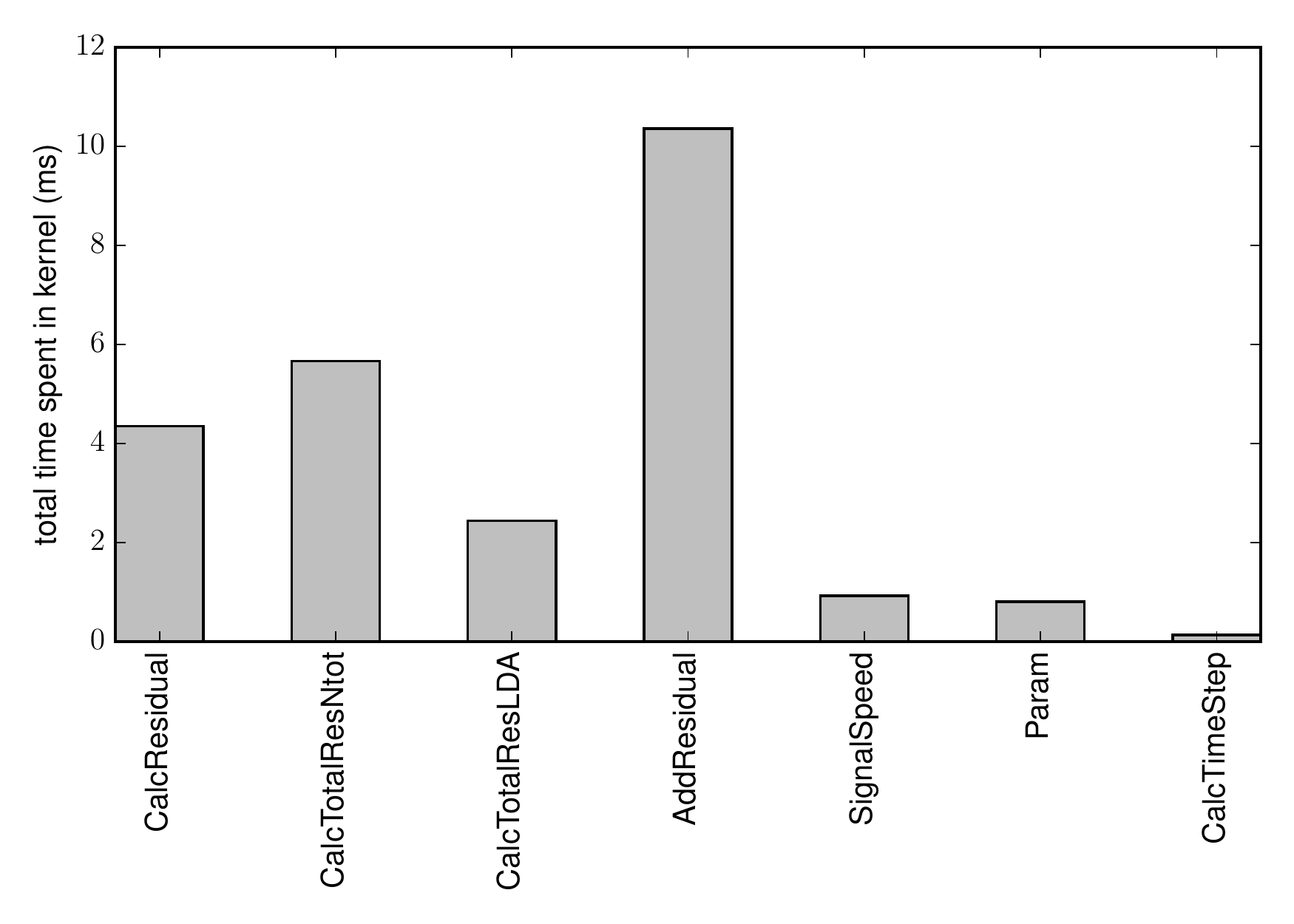}}
\caption{Total GPU time in milliseconds spent in kernels during a single time step on a uniform unstructured periodic mesh with $1.3$ million vertices. Kernels are displayed in the same order as in Fig. \ref{figCPUtimeSim}.}
\label{figGPUtimeSim}
\end{figure}

As a test case, we consider the mesh generated in section \ref{secGPUperformMesh} and consider the cost of a single time step, using the same CPU/GPU combination as in section \ref{secGPUperformMesh}. The results for the CPU are shown in Fig. \ref{figCPUtimeSim}. It is clear that the bulk of the computational time is spent calculating the residuals. Comparing with Fig. \ref{figCPUtimeMesh}, we see that the cost of setting up the mesh is roughly $100$ time steps. While in this paper we are considering static meshes only, this high cost of generating the mesh should be kept in mind when contemplating dynamic meshes. We will see below that this issue is even more important on the GPU.

In Fig. \ref{figGPUtimeSim} we show the corresponding timings for the GPU. Now most time is spent in adding the residuals to the vertices. This kernel shows a modest speedup with respect to the CPU of roughly $40$.This is not because the kernel makes inefficient use of the GPU: the achieved bandwidth is $\sim 150$ Gb/s, which is good compared to the theoretical maximum of $200$ Gb/s considering that all additions have to be atomic. Rather, it is the relatively low amount of computations compared to the memory traffic in this kernel that limits the speedup.

The situation is much better for the computationally intensive kernels calculating the residuals. The kernels $\texttt{CalcResidual}$ and $\texttt{CalcTotalResLDA}$ show a speedup of $500$ at a bandwidth of $110$ Gb/s and $130$ Gb/s, respectively, while $\texttt{CalcTotalResNtot}$ shows a speedup of $320$ at a bandwidth of $135$ Gb/s. This shows the true power of the GPU as these kernels do a lot of computations per memory element. Overall, this makes a GPU time step $250$ times faster compared to the CPU. Since this skews the performance on the GPU towards the hydrodynamics compared to the generation of the mesh, a dynamic mesh will hurt performance more on the GPU than on the CPU.

\section{Test problems}
\label{secTest}

In this section we discuss the performance of {\sc astrix} in various standard test problems. Since the schemes implemented in {\sc astrix} can be seen as multidimensional variants of the Roe solver, we will use a dimensionally split version of the Roe solver, as implemented in {\sc rodeo} \citep[see e.g.][]{paardekooper06, paardekooper12}, as a benchmark. In addition, we also show results obtained with a different approximate Riemann solver \citep[HLLC,][]{toro92}. In all cases, except where explicitly mentioned, the minmod limiter was used as a flux limiter.

\subsection{One-dimensional tests}

\begin{figure}
\centering
\resizebox{\hsize}{!}{\includegraphics[]{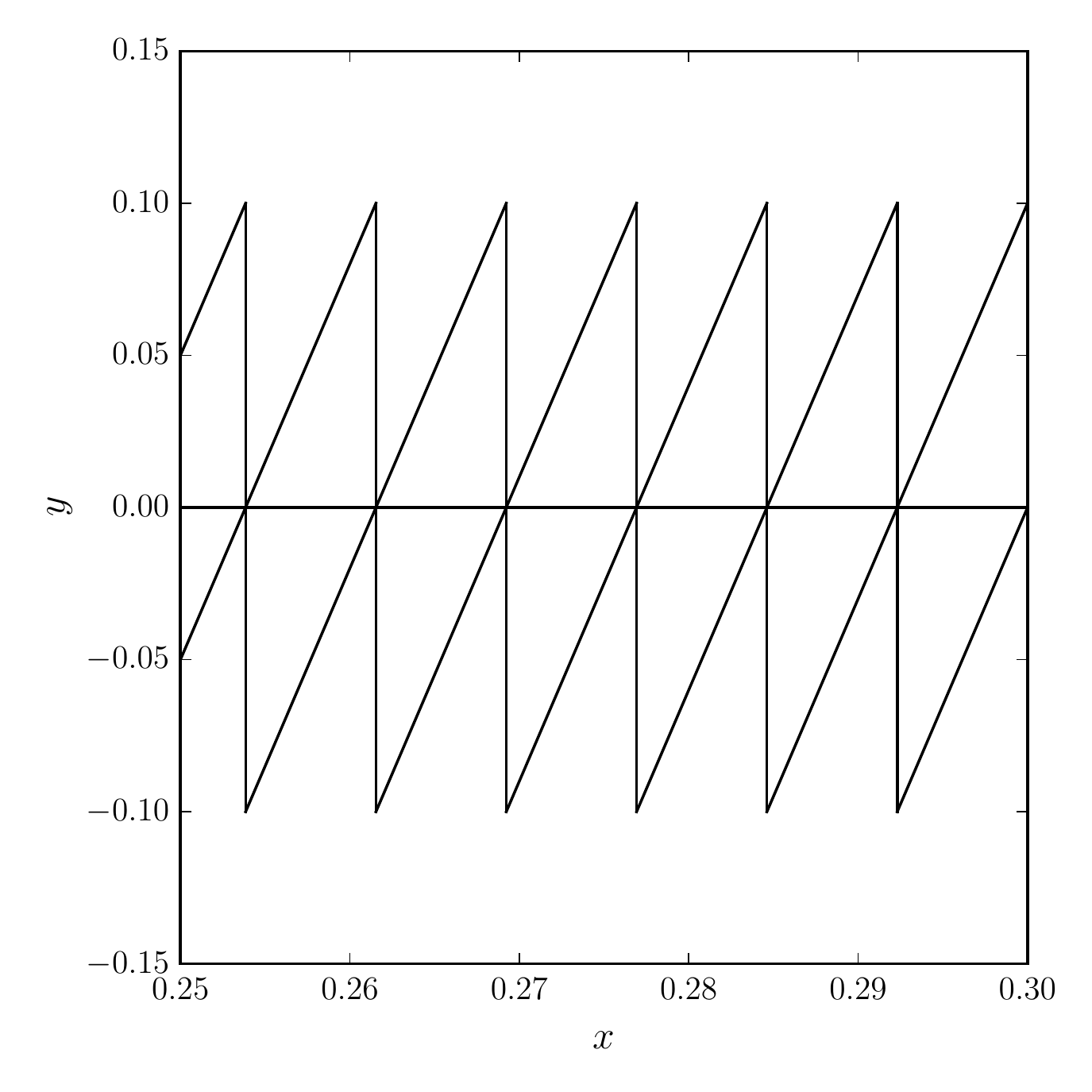}}
\caption{Detail of the mesh used for one-dimensional calculations. The mesh is periodic in the $y$-direction, so that all vertices map onto $y=0$.}
\label{figMesh1D}
\end{figure}

We begin by discussing some standard one-dimensional test problems. Since {\sc astrix} is a purely multidimensional method, we have to choose a two-dimensional mesh that does not allow for variations in the state in one of the coordinate directions. This can be achieved by making a structured  mesh periodic in $y$ with one cell covering the whole domain, see Fig. \ref{figMesh1D}. The `one-dimensional' domain is located at $y=0$, and the domain is periodic in $y$ with period $0.1$. All vertices are located either at $y=0$ or at $y=\pm 0.1$, which means all vertices map onto $y=0$, making the calculation effectively one-dimensional.

\subsubsection{Linear sound wave}

The first problem consists of a linear sound wave of amplitude $10^{-4}$, in a uniform medium of $\rho=1$, ${\bf v} = {\bf 0}$ and adiabatic sound speed $c=1$. The ratio of specific heats is set to $\gamma=1.4$. The domain size in $x$ as well as the wavelength of the perturbation is set to unity. Boundary conditions are periodic, and the wave is evolved to $t=1$. We compute the $L_1$ error norm in the density as
\begin{equation}
L_1=\frac{1}{A}\sum_i A_i|\rho_i - \rho_{0,i}|,
\end{equation}
where $\rho_i$ is the density in the $i$th cell after the final time step, $\rho_{0,i}$ is the exact solution at that particular location, $A_i$ is the volume of the $i$th cell, and $A=\sum_i A_i$ is the total volume of the computational domain. Note that the scaling with volume effectively makes this an estimate for the mass error. In the case of a 1D mesh, all cells have the same volume and the scaling with volume has no effect.

\begin{figure}
\centering
\resizebox{\hsize}{!}{\includegraphics[]{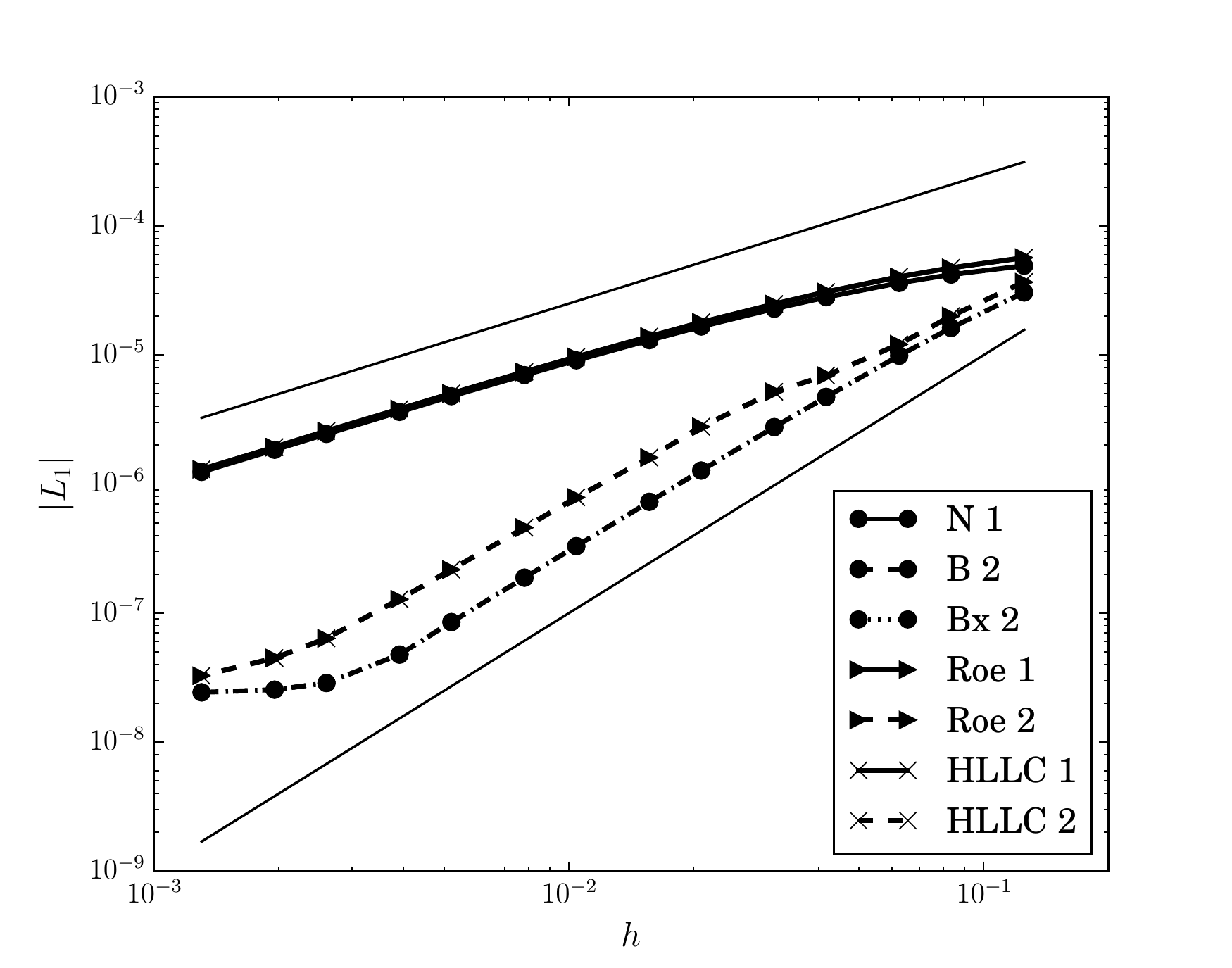}}
\caption{$L_1$ error norm for the 1D linear sound wave as a function of cell size $h$ for different algorithms.  In the case of {\sc astrix}, the curves are labeled by distribution scheme ($N$, $B$, $Bx$), while for the structured grid methods the curves are labeled by Riemann solver (Roe or HLLC). In all cases, the order of the scheme is shown (1 or 2).  The thin solid lines are included to guide the eye and indicate errors $\propto h$ and $\propto h^2$.}
\label{figL1Linear}
\end{figure}

 The results for three first-order schemes, the $N$ scheme, the first order Roe scheme and the first order HLLC scheme are shown by the solid curves in Fig. \ref{figL1Linear}. The thin solid lines indicate errors $\propto h$ and $\propto h^2$. The differences between the first order schemes are so small they are hardly visible, and in particular the Roe scheme and the HLLC scheme give the same $L_1$ error up to three significant digits. For high enough resolution, all schemes show linear convergence as expected. Second order Roe and HLLC, shown in Fig. \ref{figL1Linear} by dashed curves are again indistinguishable. The same is true for the two second order {\sc astrix} schemes $B$ and $Bx$, although they do show smaller errors than both Roe and HLLC. Second order convergence is obtained for all second order schemes around $h\sim 0.01$. Of course, the linear sound wave itself is only an approximation to the true solution, and towards higher resolution, the errors become dominated by the departure from linearity, which removes the second order convergence for all schemes. We note that results obtained with the $LDA$ scheme are indistinguishable from those obtained with both $B$ and $Bx$.

\subsubsection{Sod shock tube}

This one-dimensional Riemann problem is a well-known test case for gas dynamics codes \citep{sod78}. The initial conditions consist of two constant states, separated by a membrane at $x=0.5$. The left state has density $\rho_L=1$, pressure $p_L=1$, while the right state has density $\rho_R=0.125$ and pressure $p_R=0.1$. The velocity is zero everywhere initially, and the ratio of specific heats is set to $\gamma=1.4$. The domain $0 < x < 1$ is covered by $100$ grid cells, and the solution is evolved until $t=0.2$.

\begin{figure}
\centering
\resizebox{\hsize}{!}{\includegraphics[]{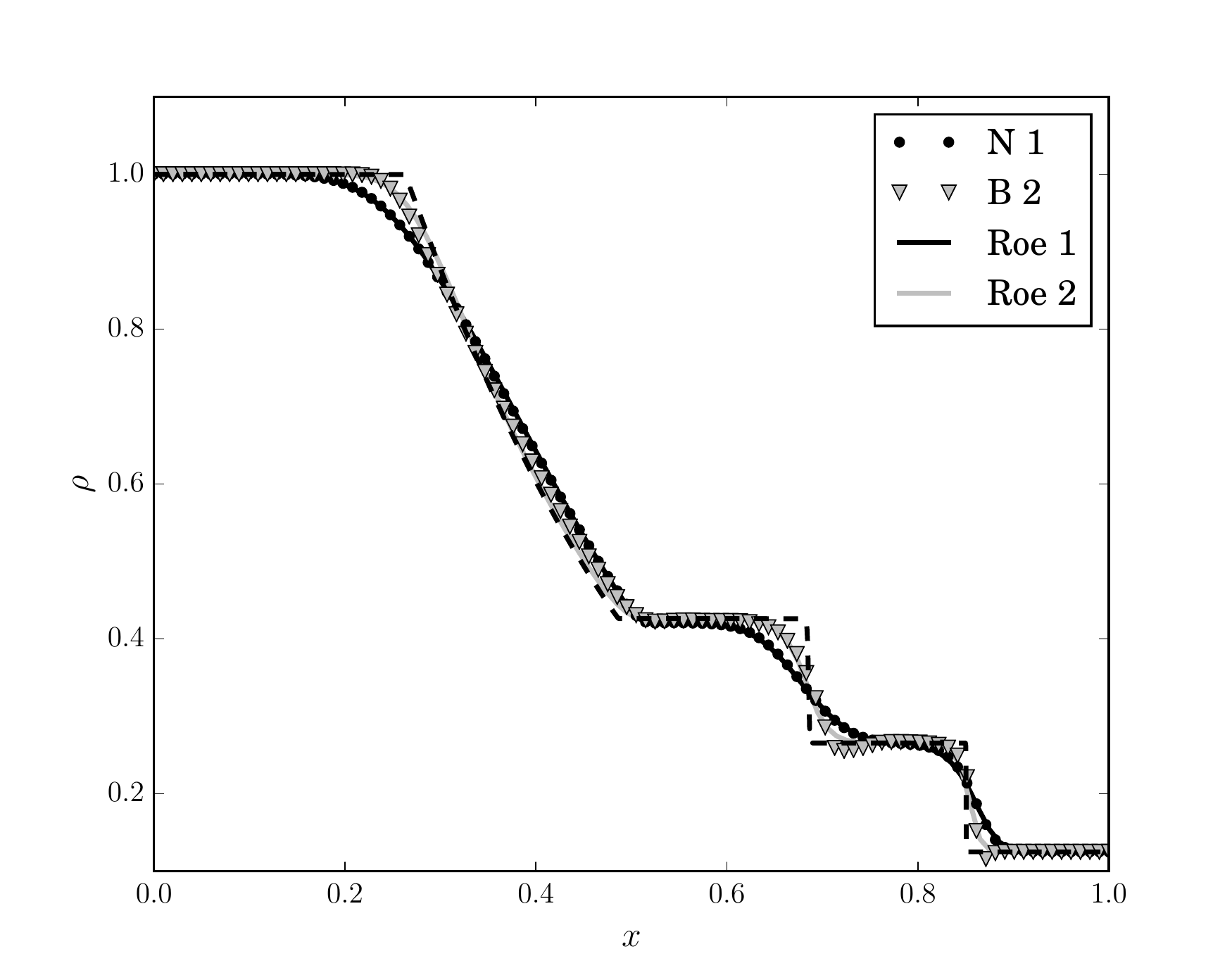}}
\caption{Density at $t=0.2$ for the Sod shock tube problem. Results obtained with {\sc astrix} are shown with symbols  (for both the first order $N$ scheme and the second order $B$ scheme) , results obtained with the Roe solver are shown with solid curves  (first as well as second order) . The dashed curve indicates the exact solution.}
\label{figSod}
\end{figure}

\begin{figure}
\centering
\resizebox{\hsize}{!}{\includegraphics[]{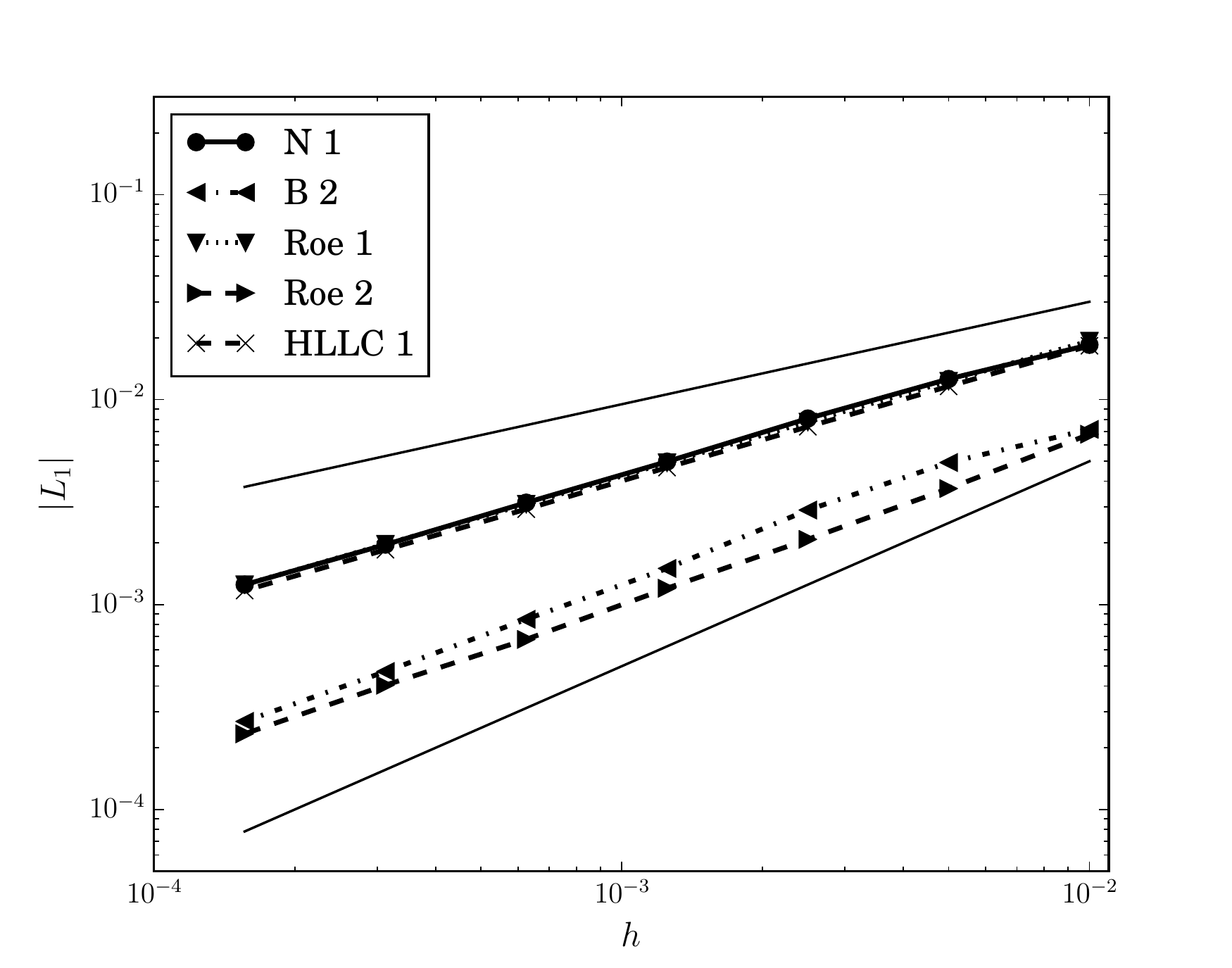}}
\caption{$L_1$ density error norm for Sod's shock tube problem as a function of cell size $h$ for different algorithms.  In the case of {\sc astrix}, the curves are labeled by distribution scheme ($N$, $B$), while for the structured grid methods the curves are labeled by Riemann solver (Roe or HLLC). In all cases, the order of the scheme is shown (1 or 2).  The thin solid lines are included to guide the eye and indicate errors $\propto \sqrt{h}$ and $\propto h$.}
\label{figSodError}
\end{figure}

The results are shown in Fig. \ref{figSod}. The analytic solution, shown with the dashed curve, consists of a shockwave and a contact discontinuity traveling to the right, and a left-traveling rarefaction wave. First thing to note is that the results obtained with {\sc astrix} are almost indistinguishable from those obtained with the Roe solver. The second-order {\sc astrix} scheme produces very minor overshoots near discontinuities, but the first-order $N$ scheme is monotone as expected. The first order schemes show more numerical diffusion, but in all cases the correct signal speeds are recovered. In the case of the Roe scheme, the sharpness of the shocks mildly depends on the choice of the flux limiter, while for {\sc astrix}, they depend on the exact blending procedure. The $Bx$ scheme gives slightly sharper shocks compared to the $B$ scheme, at the expense of slightly stronger overshoots.

The $L_1$ density error is shown in Fig. \ref{figSodError} for different resolutions. The convergence obtained is less than first order for all algorithms. While this may be surprising at first, this is due to the presence of a contact discontinuity for which it is well-known that shock-capturing schemes show sublinear convergence rates\footnote{In fact, the theoretical convergence rate is $1/2$ \citep{hedstrom68}. Figure \ref{figSodError} shows a slightly higher convergence rate because of the additional presence of both a shock and a rarefaction wave.} \citep[e.g.][]{hedstrom68, orszag74}. It should be noted in particular that this behaviour is not limited to the Roe solver: the HLLC scheme shows similar sublinear convergence and gives results almost indistinguishable from the results obtained with the Roe solver. We note that this behaviour for second order methods is sensitive to the choice of flux limiter, and, in the case of {\sc astrix}, one of the few cases where both the chosen mass matrix and the lumping strategy matter. Both a more compressive flux limiter than the minmod limiter used in Fig. \ref{figSodError} for the Roe solver \citep{orszag74} and a less diffusive mass matrix combined with selective lumping in the case of {\sc astrix} give slightly better results than given in Fig. \ref{figSodError}.

\subsubsection{Blast wave}
\label{secBlastPlanar}

\begin{figure}
\centering
\resizebox{\hsize}{!}{\includegraphics[]{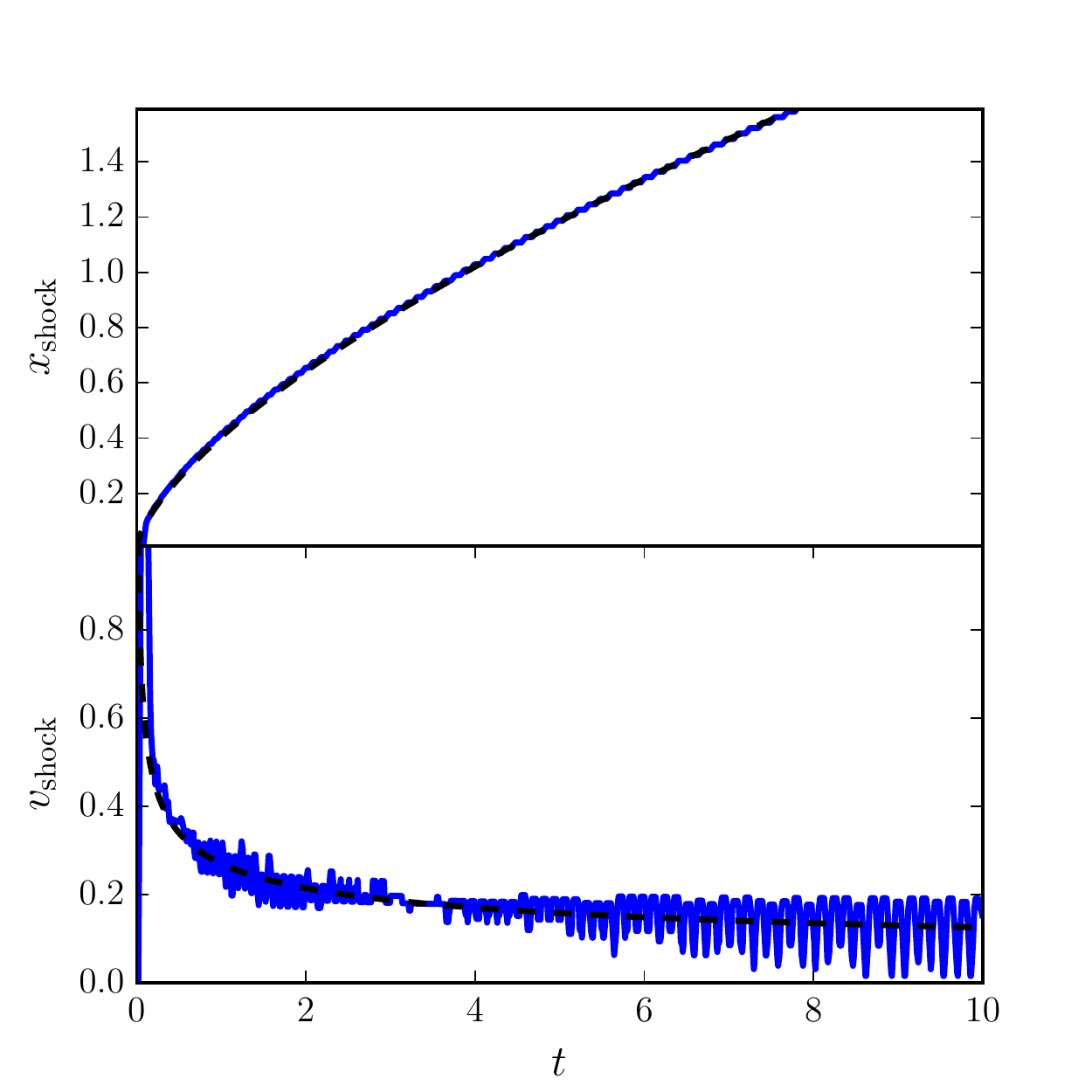}}
\caption{Shock position (top panel) and shock velocity (bottom panel) for the Cartesian blast wave problem. Shown are the results obtained with the $N$-scheme with 200 grid cells within $|x|<2$ (blue solid curve), together with the analytical expectations (black dashed curve).}
\label{figSedovShock}
\end{figure}

\begin{figure}
\centering
\resizebox{\hsize}{!}{\includegraphics[]{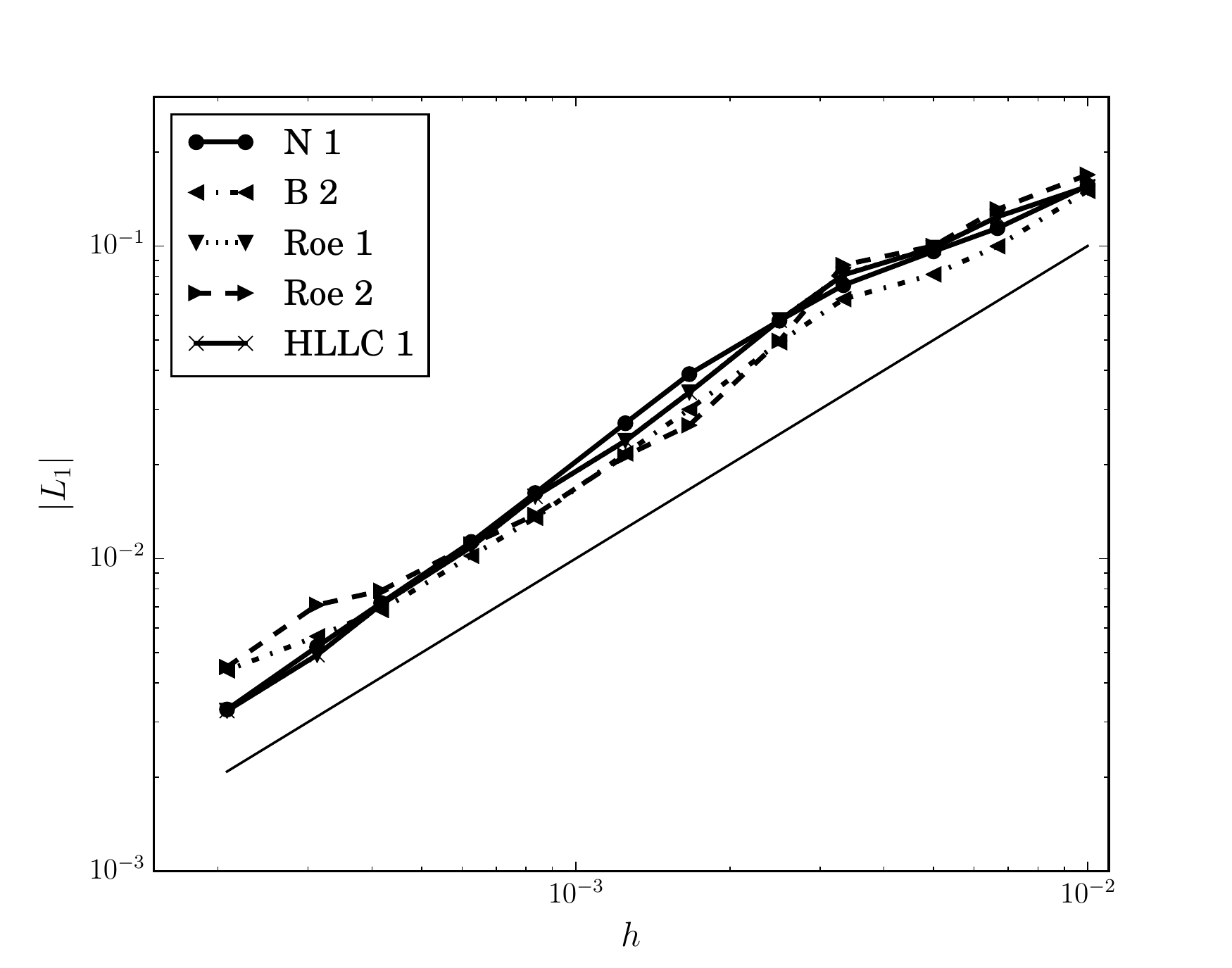}}
\caption{$L_1$ error norm for the planar blast wave as a function of cell size $h$ for different algorithms.  In the case of {\sc astrix}, the curves are labeled by distribution scheme ($N$, $B$), while for the structured grid methods the curves are labeled by Riemann solver (Roe or HLLC). In all cases, the order of the scheme is shown (1 or 2).  The thin solid line is included to guide the eye and indicate errors $\propto h$.}
\label{figSedovError1D}
\end{figure}

Similarity solutions to strong blast waves have been known for a long time \citep[e.g.][]{bethe47,sedov46,taylor50}. While in the astrophysical community they are already well-known because they are useful models for supernova explosions, they can be used as test problems for numerical hydrodynamical codes as well. Here, we discuss a one dimensional variant in planar geometry, for which the method of obtaining a reference solution is discussed in appendix \ref{secSedovSolution}.

The initial conditions consist of uniform density $\rho_0=1$, zero velocity and negligible pressure $p_0=10^{-6}$. To this we add a Gaussian pressure perturbation of the form
\begin{equation}
p_1 = \frac{\gamma - 1}{500 h  \sqrt{\pi}}\exp\left(-\frac{x^2}{100 h^2}\right)
\end{equation}
with ratio of specific heats $\gamma=5/3$ and cell size $h$. The computational domain is $|x| < 2$ and we evolve the system until $t=10$. Note that the total energy injected is constant for different resolutions and equals $E = 1/50$, which means that for the lowest resolution we consider (200 grid cells) this amounts to an energy perturbation of unity within one grid cell.

A strong shock emerges that propagates away from $x=0$. The shock position and velocity as obtained with the $N$ scheme with 200 grid cells are displayed in Fig. \ref{figSedovShock}. The position of the shock was taken to be the cell where the density passes through $\rho=2.5$ (halfway between the expected post- and pre-shock values) and decreases with $|x|$. Since we can not determine the position of the shock within one grid cell, the shock position shows a stair step pattern. This leads to a lot of noise when we differentiate with respect to time in order to get the shock velocity, hence these were smoothed with a window of $\Delta t = 0.05$ in order to make the results readable. Results for the $B$ and $Bx$ schemes as well as the Roe solver are almost indistinguishable from the results with the $N$ scheme and are therefore not shown.

The $L_1$ density errors are shown in Fig. \ref{figSedovError1D}. All tested methods show linear convergence, and the errors are comparable in magnitude. As in the case of Sod's shock tube, the results with the HLLC solver are indistinguishable form those obtained with the Roe solver. Results for the $Bx$ scheme (not shown) are very similar to those obtained with the $B$ scheme.

\subsection{Two-dimensional tests}

In two dimensions, we can make full use of an unstructured grid. An example mesh is shown in Fig. \ref{figMesh2D}, for a computational domain that is periodic in both directions; as a consequence, the jagged outer right edge slots in the left edge, and similarly the bottom edge slots in the top edge. The number of vertices is $\sim 400$, so the resolution is equivalent so a structured mesh of $20\times 20$. Such a structured triangular mesh is shown in Fig. \ref{figMesh2DStruc}. Note that the structured mesh has clear symmetries, while the unstructured mesh is locally isotropic.

\begin{figure}
\centering
\resizebox{\hsize}{!}{\includegraphics[]{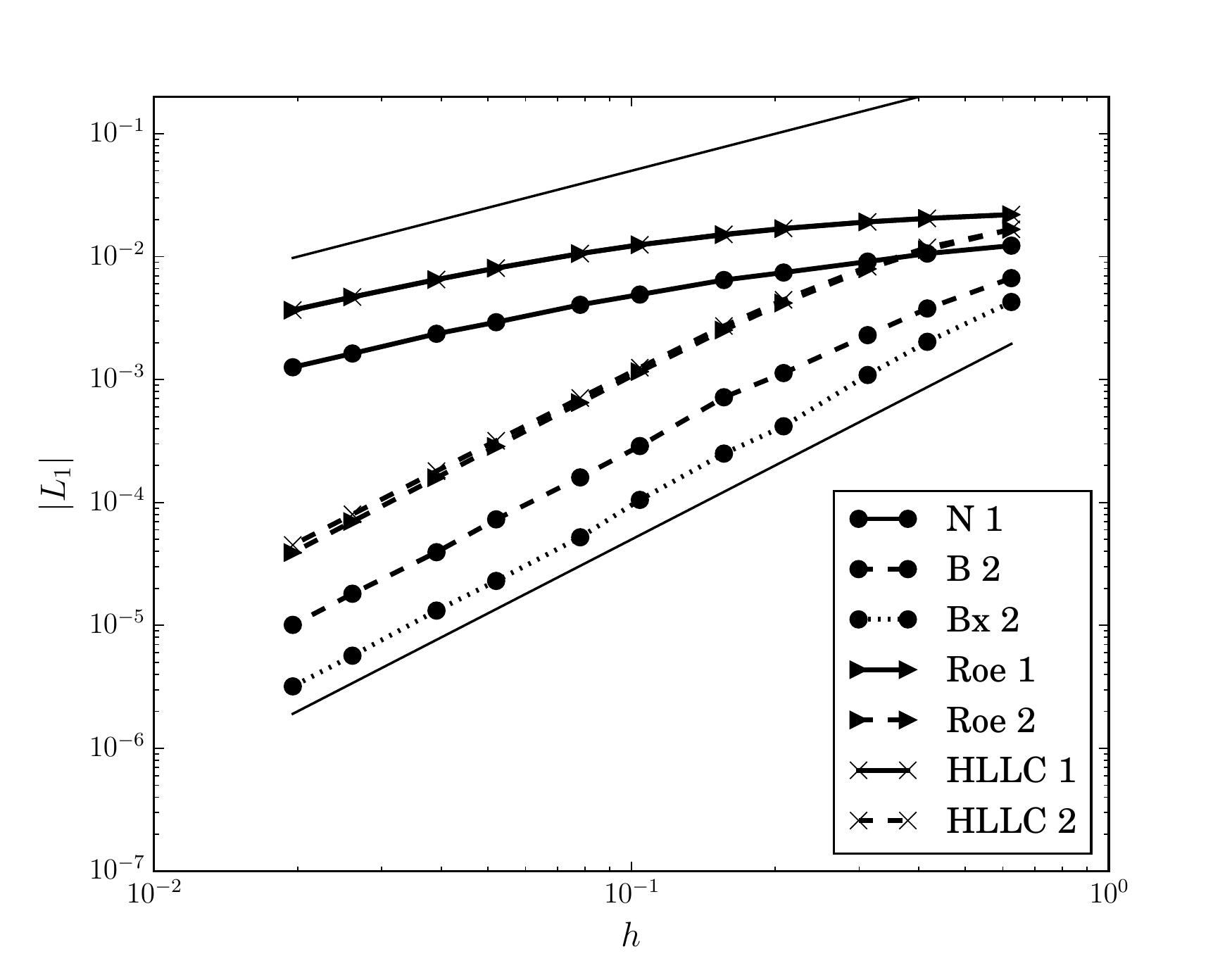}}
\caption{$L_1$ error norm for the stationary vortex problem as a function of cell size $h$ for different algorithms.  In the case of {\sc astrix}, the curves are labeled by distribution scheme ($N$, $B$, $Bx$), while for the structured grid methods the curves are labeled by Riemann solver (Roe or HLLC). In all cases, the order of the scheme is shown (1 or 2).  The thin solid lines are included to guide the eye and indicate errors $\propto h$ and $\propto h^2$.}
\label{figL1Yee}
\end{figure}

\begin{figure}
\centering
\resizebox{\hsize}{!}{\includegraphics[]{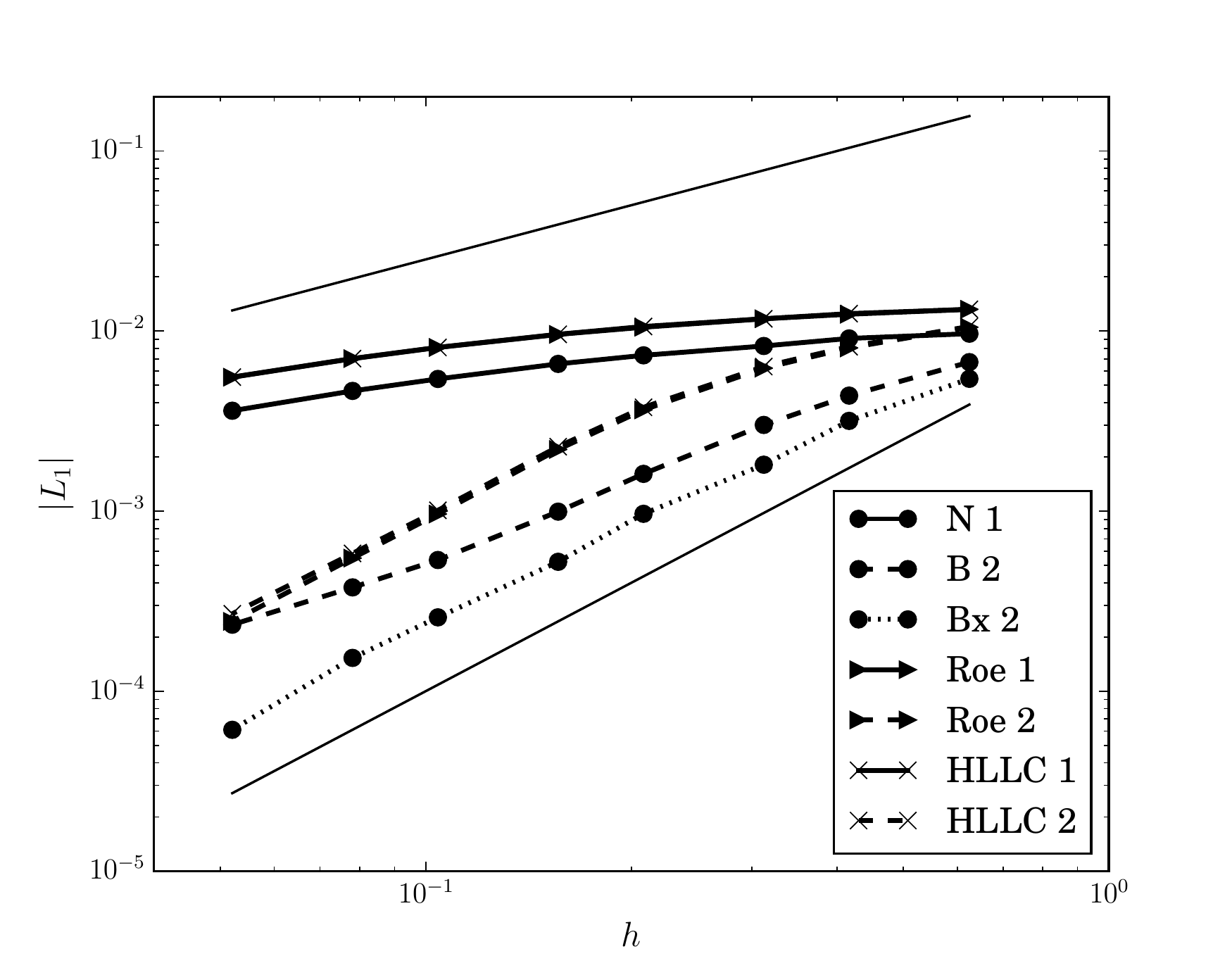}}
\caption{$L_1$ error norm for the moving vortex problem as a function of cell size $h$ for different algorithms.  In the case of {\sc astrix}, the curves are labeled by distribution scheme ($N$, $B$, $Bx$), while for the structured grid methods the curves are labeled by Riemann solver (Roe or HLLC). In all cases, the order of the scheme is shown (1 or 2).  The thin solid lines are included to guide the eye and indicate errors $\propto h$ and $\propto h^2$.}
\label{figL1YeeAdvect}
\end{figure}

\begin{figure*}
\centering
\resizebox{\hsize}{!}{\includegraphics[]{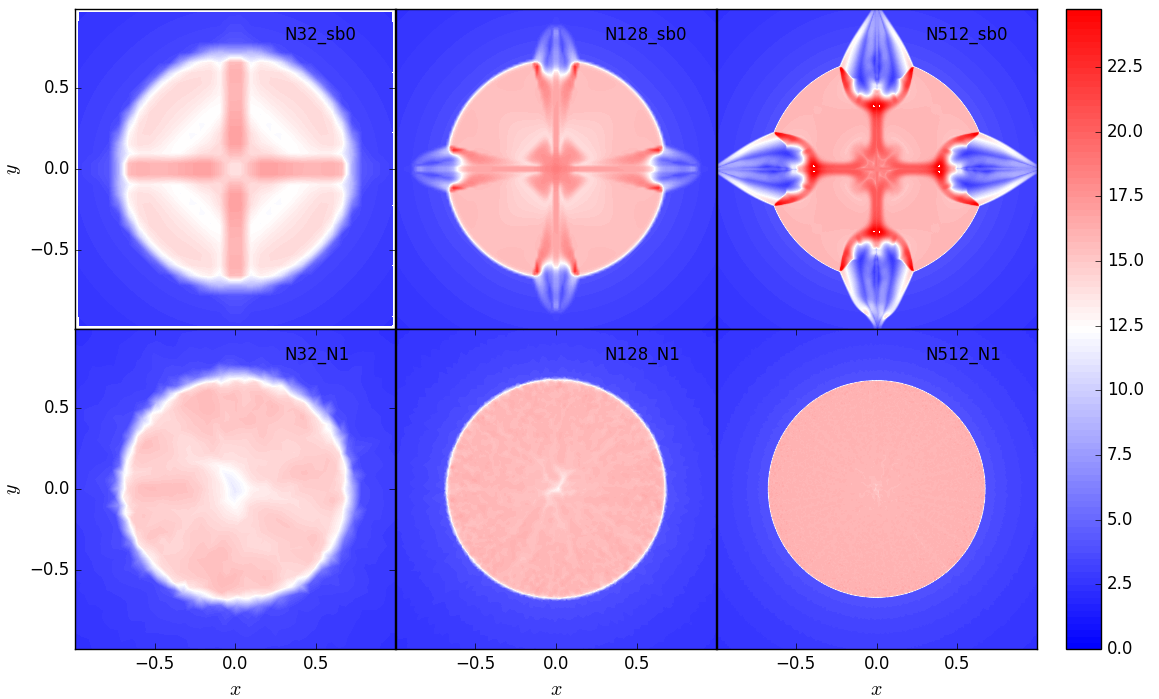}}
\caption{Density at $t=3$ for the Noh problem. Top panels: results obtained with the first-order Roe solver; bottom panels: results obtained with the first-order $N$ scheme. Resolution increases from left to right: $32\times 32$, $128\times 128$, $512\times 512$.}
\label{figNohcont}
\end{figure*}

\subsubsection{Isentropic vortex}

As a first test case, we consider an isentropic stationary vortex located at $x=x_c$, $y=y_c$. The velocity profile is given by
\begin{eqnarray}
v_x=-\Omega(r)(y-y_c) \nonumber \\
v_y = \Omega(r)(x-x_c),
\end{eqnarray}
where $r = \sqrt{(x-x_c)^2 + (y-y_c)^2}$ and the angular velocity is given by
\begin{equation}
\Omega(r) = \frac{\beta\exp\left(\frac{1-r^2}{2}\right)}{2\pi}.
\end{equation}
Equilibrium requires
\begin{equation}
\frac{1}{\rho}\frac{dp}{dr}=r\Omega^2,
\end{equation}
which, together with the isentropic assumption $p=K\rho^\gamma$, leads to
\begin{equation}
\frac{\gamma}{\gamma-1}\frac{d}{dr}\left(\frac{p}{\rho}\right)=r\Omega^2.
\end{equation}
Solving for $T=p/\rho$ we find
\begin{equation}
T(r)=\frac{p_\infty}{\rho_\infty} - \frac{\gamma-1}{\gamma}\frac{\beta^2}{8\pi^2}\exp(1-r^2),
\label{eqTvortex}
\end{equation}
where $p_\infty$ and $\rho_\infty$ are the pressure and density far away from the vortex. Pressure and density distributions follow from (\ref{eqTvortex}):
\begin{eqnarray}
\rho(r)&=&\left(\frac{T(r)}{K}\right)^{\frac{1}{\gamma-1}}\\
p(r)&=&K\rho(r)^\gamma
\end{eqnarray}
We take $p_\infty=\rho_\infty=K=1$ and a vortex with $\beta=5$ placed at $x_c=5$, $y_c=5$ in a computational domain $0 < (x,y) < 10$. We let the vortex evolve until $t=10$, and measure the $L_1$ error in the density.

The results are shown in Fig. \ref{figL1Yee}. It is clear that for this problem, the multidimensional upwind methods really come into their own, outperforming the Roe solver by more than an order of magnitude in the second order case. Note that since this is a stationary problem, the second-order time integration adds nothing and similar results can be obtained using the $Bx$ scheme with only first-order time integration. Comparing first and second order results, we see that part of this big difference in error between Roe and $Bx$ is due to the multidimensional characteristic decomposition, which is the only major difference between first order Roe and the $N$ scheme.  The results obtained with a first order HLLC scheme are indistinguishable from those obtained with the first order Roe scheme.

All second order schemes show second order convergence, as expected. The second order Roe scheme does slightly better ($10-20\%$) than the second order HLLC scheme, which is probably linked to the way the flux limiter is applied when the characteristic waves are not aligned with with eigenvectors of the linearised Jacobian \citep{leveque97}. The blended $B$ scheme significantly outperforms both HLLC and Roe, and the $Bx$ scheme does in fact more than an order of magnitude better than the second order Roe scheme. In this case, since there are no shocks present, the $Bx$ scheme yields exactly the same results as the $LDA$ scheme.

We now make the problem time-dependent by giving the whole domain an $x$ velocity boost of $v_\mathrm{advect}=1$. We enlarge the computational domain in the $x$ direction to $0 < x < 20$, and again run the simulation until $t=10$. This should place the vortex at $x=15$, with velocity and density structure unchanged from $t=0$. The resulting $L_1$ errors in the density are displayed in Fig. \ref{figL1YeeAdvect}. Again, we see that {\sc astrix} outperforms the Roe solver for both first and second order updates, but not by as much as in the previous case of a stationary vortex.  The difference in error has gone down from roughly a factor of $3$ to a factor of $2$ for the first order schemes. Again, results obtained by HLLC and Roe are exactly the same at first order, while the Roe solver shows slightly better results at second order. The $Bx$ scheme, which again gives the same results as the $LDA$ scheme, outperforms Roe and HLLC by roughly a factor of $3$. The $B$ scheme shows a departure from second-order convergence towards high resolution. The cause of this remains to be investigated.

The reason that {\sc astrix} performs so well on a stationary vortex lies in the linearity preserving nature of the $LDA$ scheme: if a stationary solution is linear over an element, it is recovered exactly by the integration scheme. Of course, the isentropic vortex is not a linear solution, but it can up to some precision be represented by a linear solution, which is the solution the $LDA$ scheme will evolve towards. Note that the solution only needs to be linear over every single triangle; it does not have to be linear globally. The difference between the $LDA$ solution and the true solution scales with $h^2$ \citep{abgrall01}. In other words: the scheme tries to find a solution where the residual $\phi^T=0$ everywhere, and since the distribution coefficients satisfy $\phi_i^{LDA}=\beta_i\phi^T=0$ no evolution takes place. It is not obvious, especially with a dimensionally split scheme, whether such a multidimensional numerically stationary state exists, unless it is a hydrostatic solution. Even in one spatial dimension in the presence of source terms one needs to be careful not to evolve away from stationary states \citep[e.g.][]{eulderink95}. What is happening in the case of the Roe solver, is that the solution continues to evolve to the only state that is numerically stationary for this scheme, which is the hydrostatic solution, in which case there is of course no vortex. Note that the $N$ scheme is not linearity-preserving and will therefore behave more like the Roe solver. Nevertheless, there is still an advantage coming from the multidimensional upwinding.

\begin{figure}
\centering
\resizebox{\hsize}{!}{\includegraphics[]{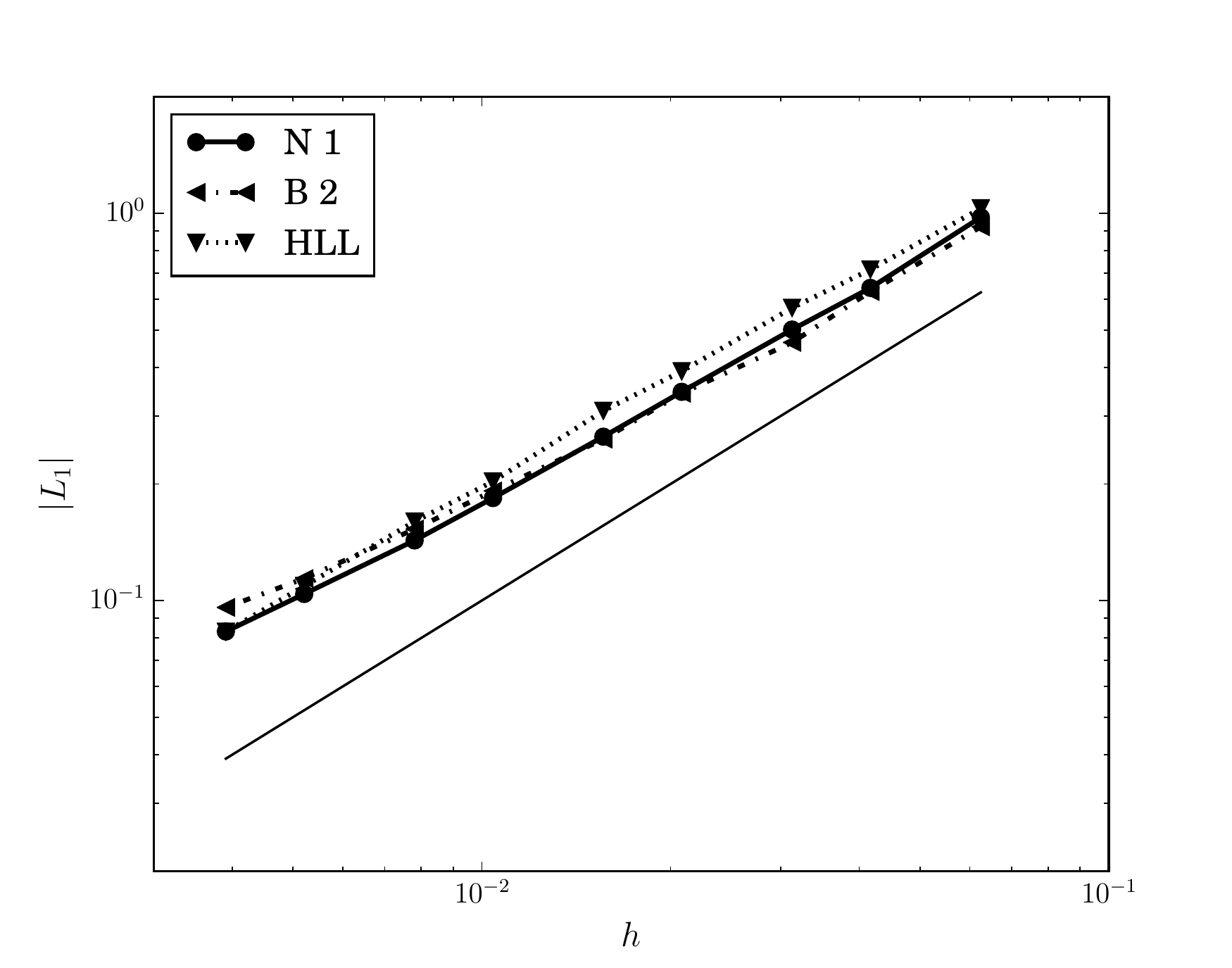}}
\caption{$L_1$ error norm for the Noh problem as a function of cell size $h$ for different algorithms.
 Results obtained with {\sc astrix} are shown for the first-order $N$ scheme as well as the second-order $B$ scheme.
Results obtained with the dimensionally split structured grid code are shown for the first-order HLL solver.
  The thin solid line is included to guide the eye and indicates errors $\propto h$.}
\label{figNohError}
\end{figure}

\subsubsection{Noh problem}
\label{secNoh}

Here we discuss a demanding problem originally solved by \cite{noh87}, of which we consider the two-dimensional version. The initial conditions are uniform density $\rho=1$, and a velocity of magnitude 1 everywhere in the direction of the origin. If we take the initial pressure to be zero, the problem has an analytic solution consisting of a shock of formally infinite Mach number moving radially outward. Defining $r=\sqrt{x^2+y^2}$, and taking the ratio of specific heats $\gamma=5/3$, the solution reads:
\begin{equation}
\rho(r, t)=\left\{\begin{array}{ll}
16 & r < t/3\\
1+t/r & r \geq t/3
\end{array}\right.,
\end{equation}
\begin{equation}
{\bf v}(r, t)=\left\{\begin{array}{ll}
{\bf 0} & r < t/3\\
-(x, y)^T/r & r \geq t/3
\end{array}\right.,
\end{equation}
\begin{equation}
p(r, t)=\left\{\begin{array}{ll}
16/3 & r < t/3\\
0 & r \geq t/3
\end{array}\right. .
\end{equation}
We take the computational domain to be $-1 \leq x \leq 1$ and $-1 \leq y \leq 1$, and as boundary conditions we impose the analytic solution. Since both {\sc astrix} and the Roe solver are based on a characteristic decomposition, they can not handle zero pressure, and therefore, following \cite{liska03}, we set it to $10^{-6}$ initially. The solution is evolved up to $t=2$, at which point the shock is located at $r=2/3$.

Results are shown in Fig. \ref{figNohcont}. The top panels show results obtained with the first order Roe scheme, which clearly shows the carbuncle instability \citep{peery88, quirk94}. This well-known instability that occurs for strong shocks that are aligned with the grid is usually associated with multidimensional flows, but a one-dimensional version does exist as well \citep[e.g.][]{dumbser04}. In this case, the instability may be associated with the nonlinearity of the Hugoniot locus \citep{zaide12}. In multidimensional flows, it is thought to be associated with using one-dimensional fluxes \citep{stone08}, and the instability can be corrected efficiently by adding extra multidimensional dissipation \citep{quirk94, stone08}, a feature not implemented in the Roe scheme used here. Going to second order makes the results worse rather than better and therefore these are not shown.  The HLLC solver suffers from similar problems.

Results obtained with {\sc astrix} do not show the carbuncle instability. This is largely due to the use of an unstructured grid: there is no place where the shock can be said to be aligned with the grid. However, the multidimensional upwind nature of {\sc astrix} appears to play a role as well. Results for structured grids do show some artifacts along the coordinate directions, but they remain confined to the postshock region as in the top left panel of Fig. \ref{figNohcont}, with no real carbuncles developing even at high resolution.

A numerical artifact that can be seen especially in the lower left panel is ``wall heating" \citep{noh87}, resulting in an underdense region in the centre while the pressure is constant. This rise in temperature is seen in most Riemann solver codes \citep{liska03, stone08}, and even though a complete understanding of this phenomenon is still lacking, it may be due in part by an ambiguity in the position of a shock within a cell \citep{zaide12}. We get a minimum density in the central region of $14$, which makes it comparable to the schemes tested in \cite{liska03} (better than PPM, slightly worse than the WENO scheme). While we only show results for the first order $N$ scheme, since the solution is dominated by a strong shock a blended second order scheme give exactly the same result as the $N$ scheme in the case where we take the blending coefficient as the maximum over all equations. Less diffusive blended schemes either fail on this problem, or need backup fluxes produced by the $N$ scheme so that again the results look exactly the same as in Fig. \ref{figNohcont}.

In Fig. \ref{figNohError} we show the $L_1$ density error for those algorithms that did not show any carbuncle instabilities. As a comparison to the {\sc astrix} results we also show results obtained with the original Harten-Lax-van Leer solver \citep[HLL,][]{hll}, which is known to be carbuncle-free \citep{pandolfi01}. Even the HLL method does show some artefacts around the coordinate axes; however, these have very limited impact on the total error. The errors are very comparable between the three schemes and show linear convergence at low resolution. At high resolution, a departure from linearity can be seen, which is probably due to the wall heating phenomenon.

\begin{figure}
\centering
\resizebox{\hsize}{!}{\includegraphics[]{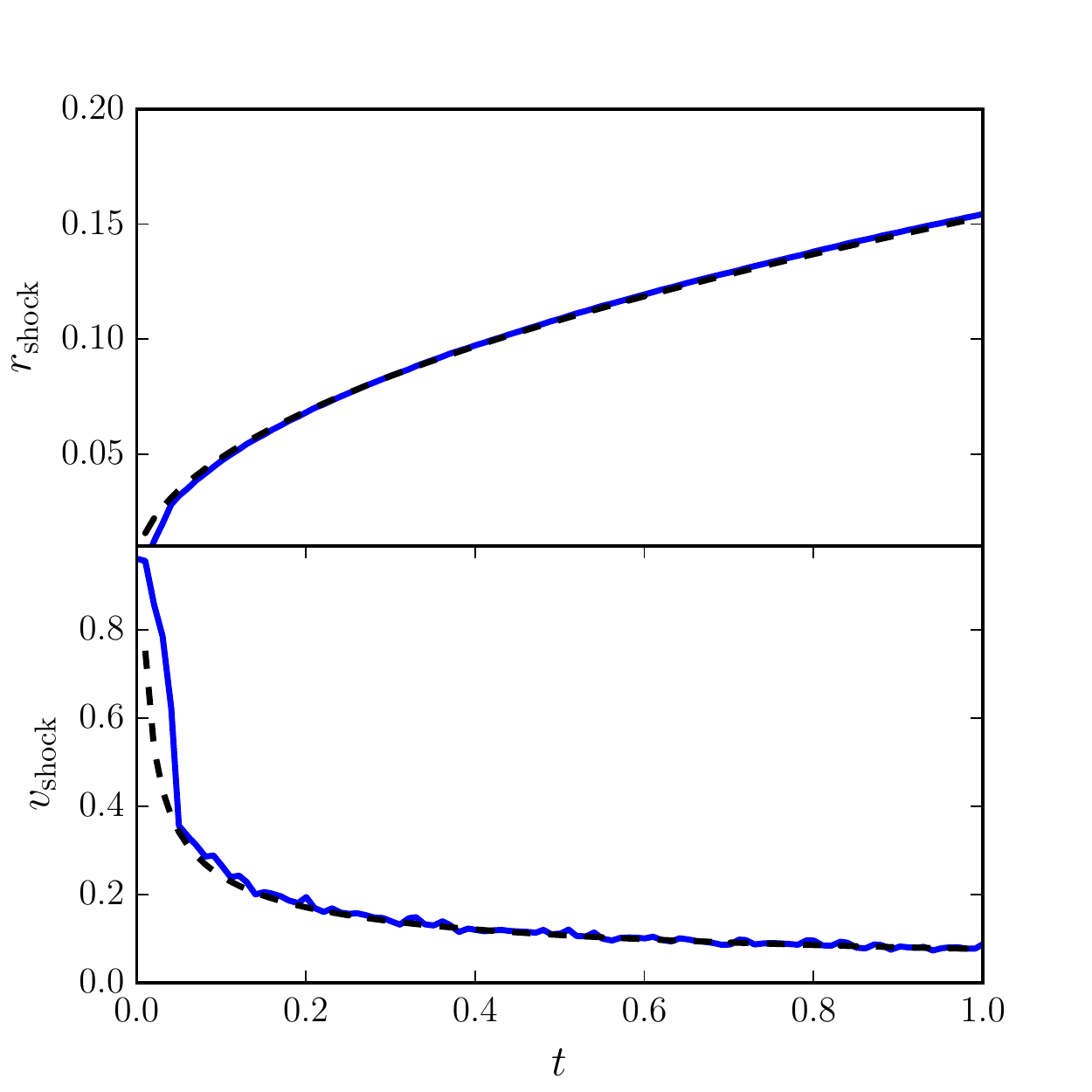}}
\caption{Shock position (top panel) and shock velocity (bottom panel) for the cylindrical blast wave problem. Shown is the result obtained with the $N$-scheme with an equivalent resolution of $100\times 100$ grid cells within $|x|, |y| < 0.2$ (blue solid curve), together with the analytical expectations (black dashed curve).}
\label{figSedovShock2D}
\end{figure}

\begin{figure}
\centering
\resizebox{\hsize}{!}{\includegraphics[]{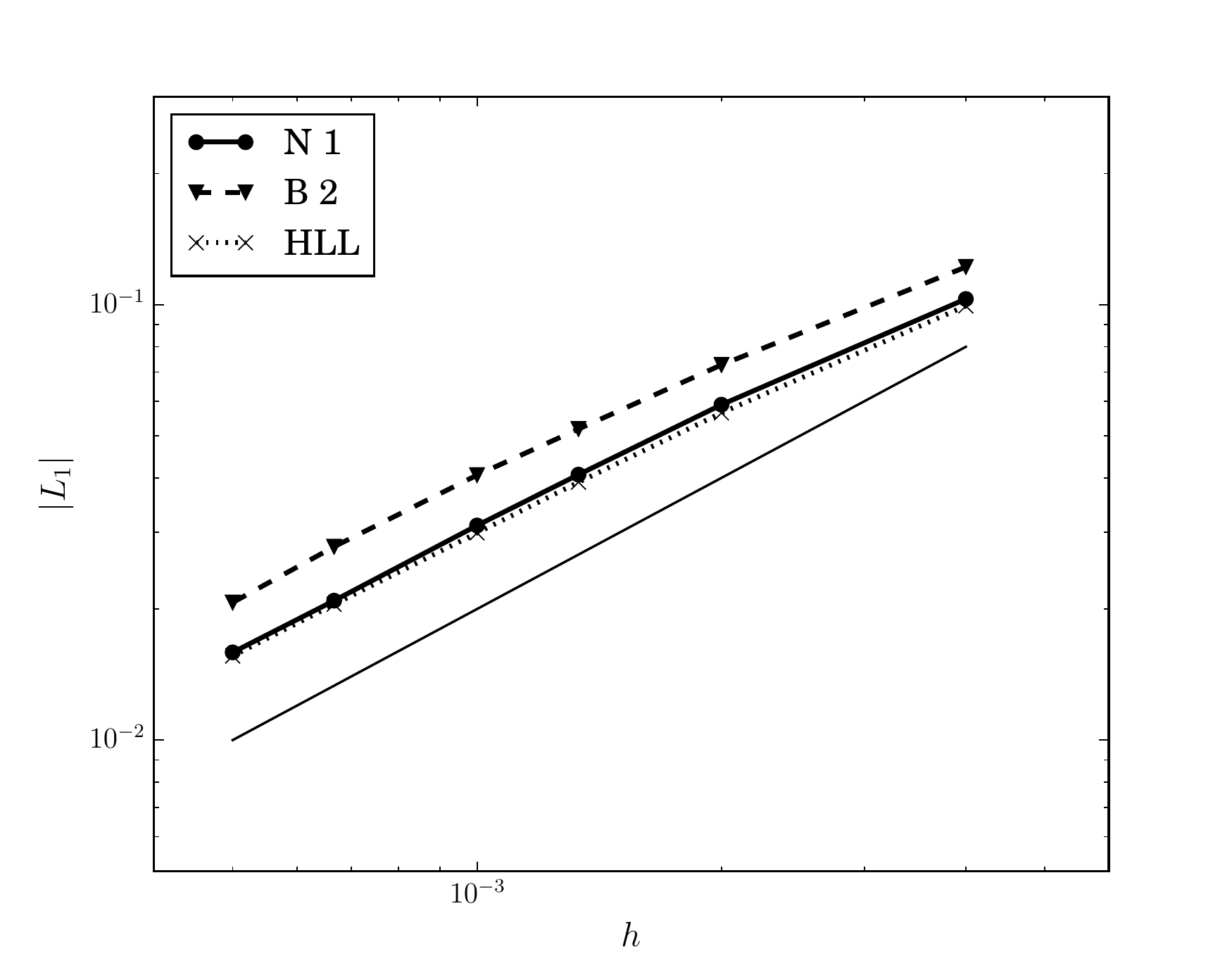}}
\caption{$L_1$ error norm for the cylindrical blast wave problem as a function of cell size $h$ for  the same algorithms as in Fig. \ref{figNohError}.  The thin solid line is included to guide the eye and indicates errors $\propto h$.}
\label{figSedovError2D}
\end{figure}

\subsubsection{Cylindrical blast wave}

We now consider a blast wave in cylindrical geometry. The initial conditions consist of uniform density $\rho_0=1$, zero velocity and negligible pressure $p_0=10^{-6}$. To this we add a Gaussian pressure perturbation of the form
\begin{equation}
p_1 = (\gamma - 1)\left(\frac{0.004}{h}\right)^2\exp\left(-\frac{x^2+y^2}{6.25 h^2}\right)
\end{equation}
with ratio of specific heats $\gamma=5/3$ and $h$ is a linear measure of the cell size. In the case of an unstructured grid this is taken to be the average over the whole domain of the square root of the cell volume. The total energy injected is equal to an order unity perturbation in a disc with radius $1/100$. A reference solution is calculated as outlined in appendix \ref{secSedovSolution}. The computational domain is taken to be $|x|,|y| < 0.2$ and the solution is evolved until $t=1$.

As in the planar case (see section \ref{secBlastPlanar}), a strong shock emerges from the location of the energy perturbation. As in the case of the Noh problem (see section \ref{secNoh}), both the Roe solver and the HLLC solver suffer from a carbuncle instability, while the {\sc astrix} results are carbuncle-free because of the use of an unstructured grid.

In Fig. \ref{figSedovShock2D} we show the shock position (top panel) and shock velocity (bottom panel) as a function of time as obtained with the $N$ scheme at an equivalent resolution of $100\times 100$, together with the analytical expectation. As in the planar case, good agreement is found. Since the shock position is now obtained by an angular average, the result is less noisy and no smoothing is necessary to obtain the shock velocity, which again shows good agreement with analytical expectations.

In Fig. \ref{figSedovError2D} we show the resulting $L_1$ error in the density for three algorithms that did not suffer from numerical instabilities. The $N$ scheme produces results that are almost indistinguishable from those obtained with the HLL scheme. Unlike for the Noh problem, the HLL solution looks perfectly smooth and with the largest part of the error originating from the strong shock it is hard to do any better on this problem. The $B$ scheme shows larger errors but linear convergence as do the $N$ and HLL schemes. Note that in this problem wall heating is not an issue, and the errors show a linear decrease for all resolutions considered.

\begin{figure*}
\centering
\resizebox{\hsize}{!}{\includegraphics[]{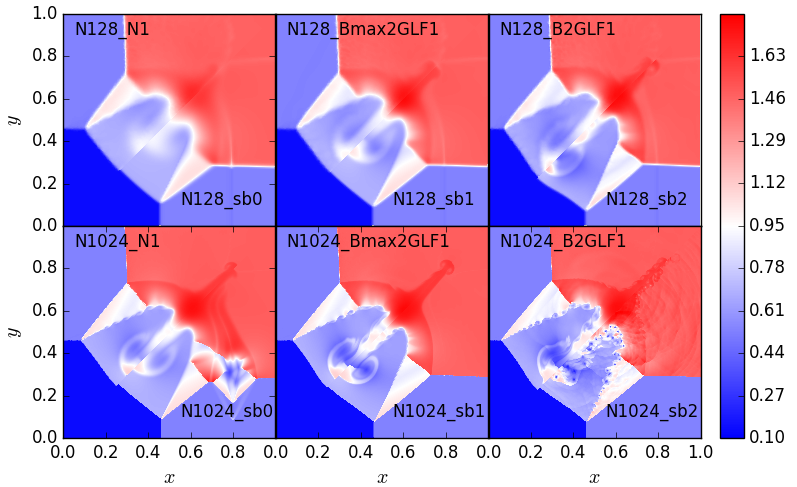}}
\caption{Results for the 2D Riemann problem at $t=0.8$. Each panel contains a result obtained with {\sc astrix} for $y > x$ and a result obtained with the Roe solver for $y < x$. The labels in the panels denote the method and resolution used: {\tt N128\_sb*} denotes the Roe method at resolution $128\times 128$ at first order (*=0) or minmod flux limiter (*=1) or superbee flux limiter (*=2). For results obtained using {\sc astrix}, the equivalent resolution is shown, and we can have the first-order $N$ scheme ({\tt N1}), the classical blended scheme ({\tt B2}), and the blended scheme where we pick the maximum blend coefficient ({\tt Bmax2}). The last two for completeness also mention they use global lumping and the first mass matrix ({\tt GLF1}).}
\label{figRiemann}
\end{figure*}

\subsubsection{Riemann problem}

Here we study the two-dimensional Riemann problem originally introduced by \cite{schulz93}. The problem is defined on a square $0\leq x< \leq 1$ and $0\leq y\leq 1$, with initial conditions
\begin{equation}
{\bf U} = \left\{\begin{array}{ll}
{\bf U}_1 & x \leq 0.8, y>0.8,\\
{\bf U}_2 & x > 0.8, y>0.8,\\
{\bf U}_3 & x \leq 0.8, y\leq 0.8,\\
{\bf U}_4 & x > 0.8, y\leq 0.8,
\end{array}\right.
\end{equation}
where ${\bf U}=(\rho, v_x, v_y, p)^T$ is the vector of primitive variables and
\begin{eqnarray}
{\bf U}_1&=&(0.5322581, 1.2060454, 0, 0.3)^T\\
{\bf U}_2&=&(1.5, 0, 0, 1.5)^T\\
{\bf U}_3&=&(0.1379928, 1.2060454, 1.2060454, 0.0290323)^T\\
{\bf U}_4&=&(0.5322581, 0, 1.2060454, 0.3)^T.
\end{eqnarray}
The ratio of specific heats is taken to be $\gamma=1.4$. The solution consists of four shocks travelling along the wall with a complex interaction region with shear flow, susceptible to Kelvin-Helmholtz instabilities \citep[e.g.][]{san14}. The solution along the walls are moving single shocks, for which the speeds can be computed easily from the Rankine-Hugoniot jump conditions. We use a one-dimensional shock solution with the computed shock speeds as boundary conditions, and compute the solution until $t=0.8$,

Results are shown in Fig. \ref{figRiemann}. Each panel contains a result obtained with {\sc astrix} for $y > x$ and a result obtained with the Roe solver for $y < x$. All schemes agree on the position of the shocks, as is to be expected for conservative schemes. The jet along the diagonal gets longer with increasing resolution and decreasing numerical dissipation. The left panels show results obtained with first-order schemes at resolutions $128\times128$ and $1024\times 1024$. It is clear from the position of the jet that the $N$ scheme has lower dissipation than the first order Roe scheme. Moreover, the Roe scheme shows an instability at $x=0.8$ at the position of the shock in the lower left panel. The seeds of this instability can be seen in most panels as light horizontal streaks near $y=0.8$ and light vertical streaks near $x=0.8$. These streaks are present in almost all codes, regardless of Riemann solver \citep{liska03}, and are present even if the solution between for example ${\bf U}_2$ and ${\bf U}_4$ is computed in one dimension. Their source is purely numerical and is a consequence of a ``startup error" \citep{jin96, zaide12}: the initial conditions are not a solution to the modified equation (including numerical dissipation) the code is solving. The adjustment to the numerical solution leads to small artifacts that in this case stay put as the post-shock velocity is close to zero. Remarkably, this leads to a numerical instability for the first-order Roe scheme. The second-order Roe schemes, possibly because it increases the stencil, are able to correct this instability very efficiently. The results obtained with the minmod limiter (middle panels in Fig. \ref{figRiemann}) still show the streaks at a resolution of $128\times 128$, albeit at lower amplitude compared to the first-order Roe scheme, but at a resolution of $1024\times 1024$ they have disappeared.  Similar features show up with the HLLC solver, with some subtle differences. No instability is observed in the first order HLLC scheme, but on the other hand the streaks remain far more pronounced in the second order case.

A second interesting feature is the appearance of Kelvin-Helmholtz vortices in the shear flow past the shocks. Except for the first order Roe scheme, which becomes dominated by the numerical instability due to the start-up error, these vortices show up in all schemes at sufficient resolution. It has been argued that these are physical rather than numerical \citep[e.g.][]{san14}, and the fact that they show up in all schemes lends support to this conclusion. However, this suggests that for higher and higher resolutions, the schemes will not converge to a well-defined solution, making the problem less suited as a numerical test (a similar problem haunts the Kelvin-Helmholtz instability, see section \ref{secKHI}). This is especially apparent from the lower right panel of Fig. \ref{figRiemann}, where the solutions look very noisy, especially in the case of the Roe solver with superbee flux limiter.  The same holds for results obtained with the HLLC solver.

\begin{figure}
\centering
\resizebox{\hsize}{!}{\includegraphics[]{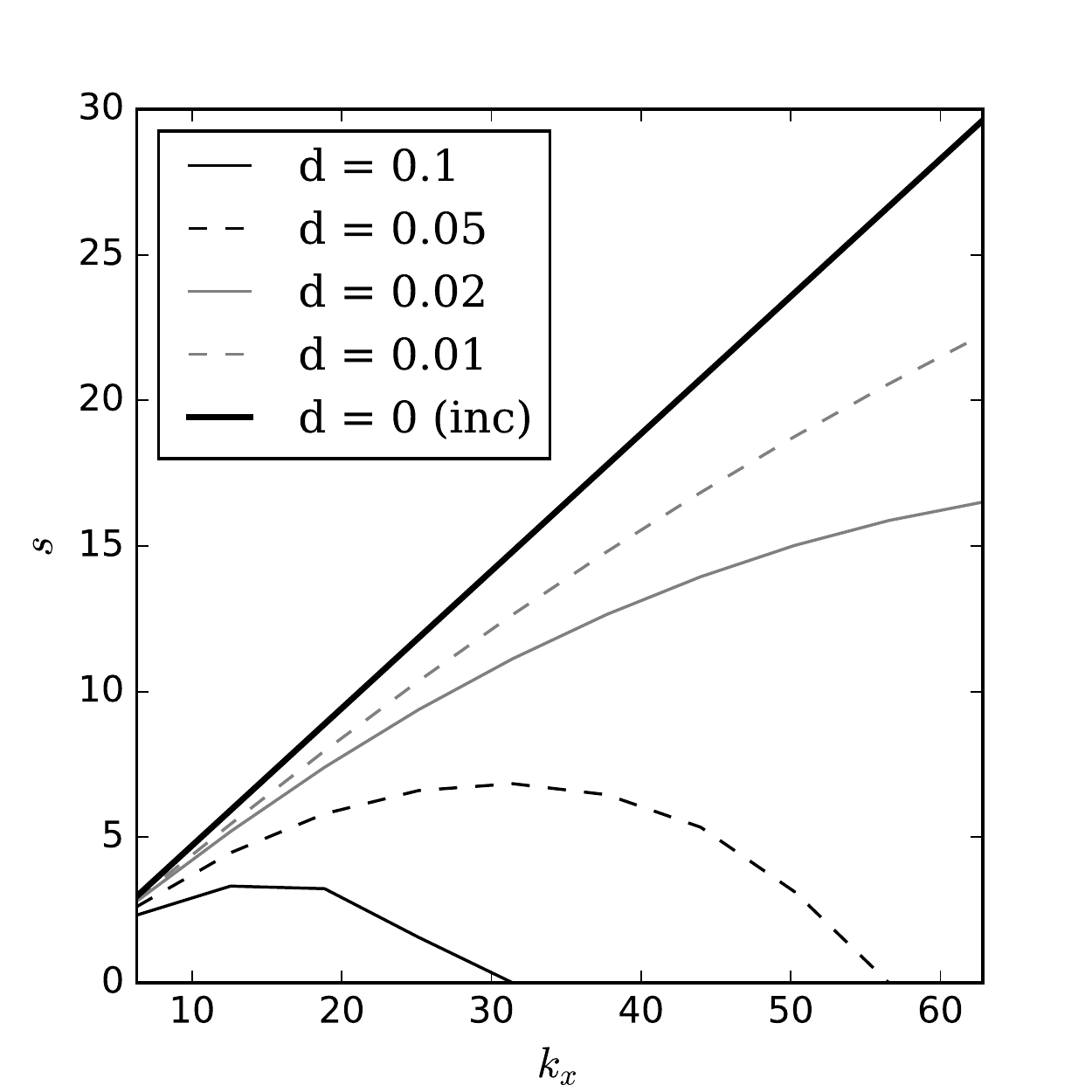}}
\caption{Growth rates as a function of horizontal wave number for the smooth KHI for various smoothing parameters $d$. Also shown is the incompressible result without smoothing.}
\label{figKHlinear}
\end{figure}

\begin{figure}
\centering
\resizebox{\hsize}{!}{\includegraphics[]{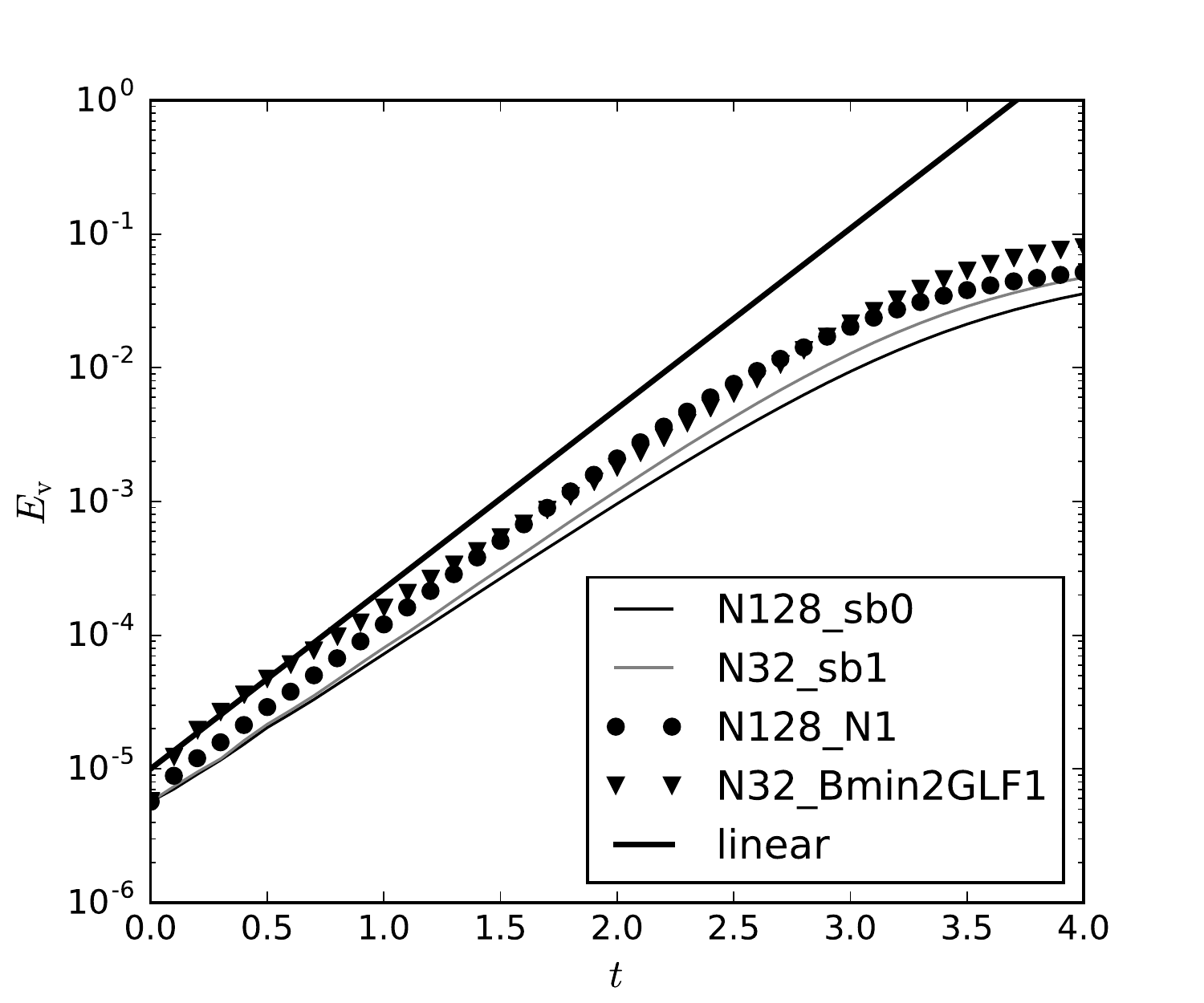}}
\caption{Total vertical kinetic energy $\rho v_y^2/2$ for the Kelvin-Helmholtz problem with $d=0.25$. The labels denote the method and resolution used: {\tt N128\_sb*} denotes the Roe method at resolution $128\times 128$ at first order (*=0) or minmod flux limiter (*=1). For results obtained using {\sc astrix}, the equivalent resolution is shown, and we can have the first-order $N$ scheme ({\tt N1}), the blended scheme where we pick the minimum blend coefficient ({\tt Bmin2}), which for completeness also mentions it uses global lumping and the first mass matrix ({\tt GLF1}). The thick solid line shows the growth rate calculated through a linear analysis.}
\label{figKHgrow}
\end{figure}

\begin{figure*}
\centering
\resizebox{\hsize}{!}{\includegraphics[]{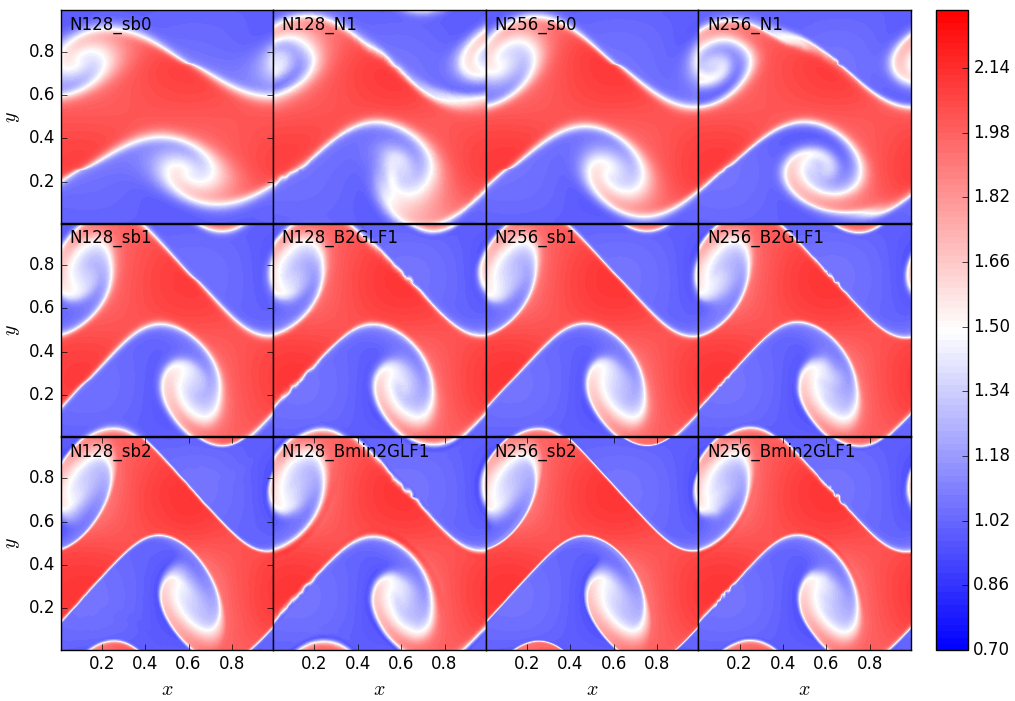}}
\caption{Contour plots of the density at $t=3.5$ for the Kelvin-Helmholtz problem with $d=0.25$. The label of each panel denotes the method and resolution used: {\tt N128\_sb*} denotes the Roe method at resolution $128\times 128$ at first order (*=0), minmod flux limiter (*=1) or superbee flux limiter (*=2). For results obtained using {\sc astrix}, the equivalent resolution is shown, and we can have the first-order $N$ scheme ({\tt N1}), the blended scheme $B$ ({\tt B2}) and the blended scheme where we pick the minimum blend coefficient ({\tt Bmin2}). For completeness the {\sc astrix} panels also specify that all results were obtained using global lumping and the first mass matrix ({\tt GLF1}).}
\label{figKHcont35}
\end{figure*}

\begin{figure*}
\centering
\resizebox{\hsize}{!}{\includegraphics[]{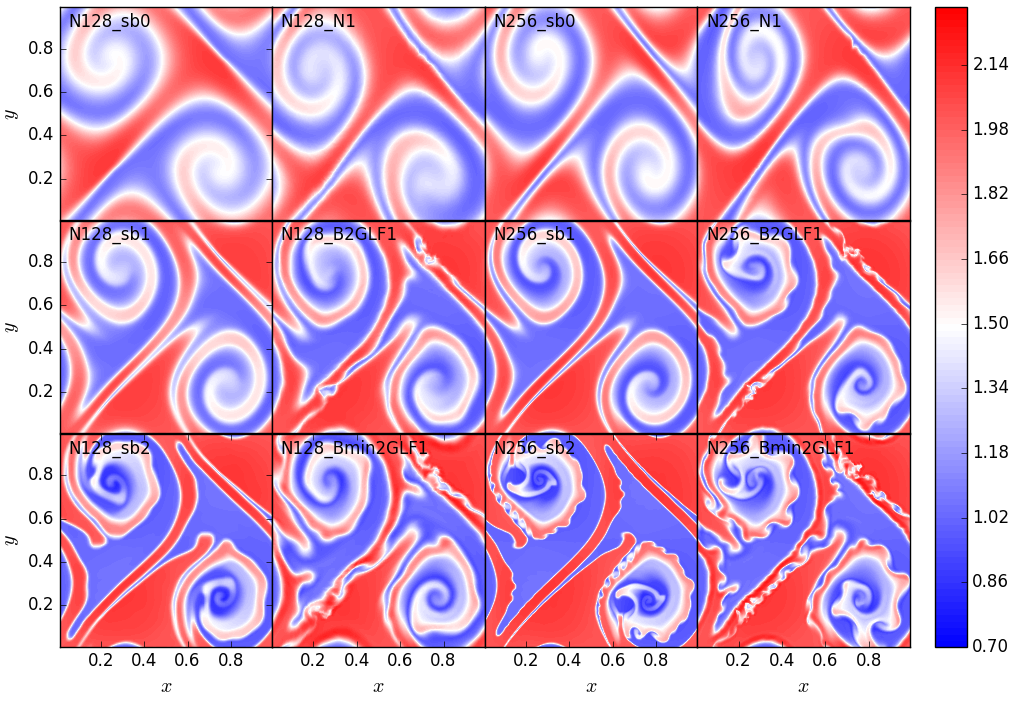}}
\caption{Same as Fig. \ref{figKHcont35}, but at a later time $t=5$.}
\label{figKHcont50}
\end{figure*}

\begin{figure*}
\centering
\resizebox{\hsize}{!}{\includegraphics[]{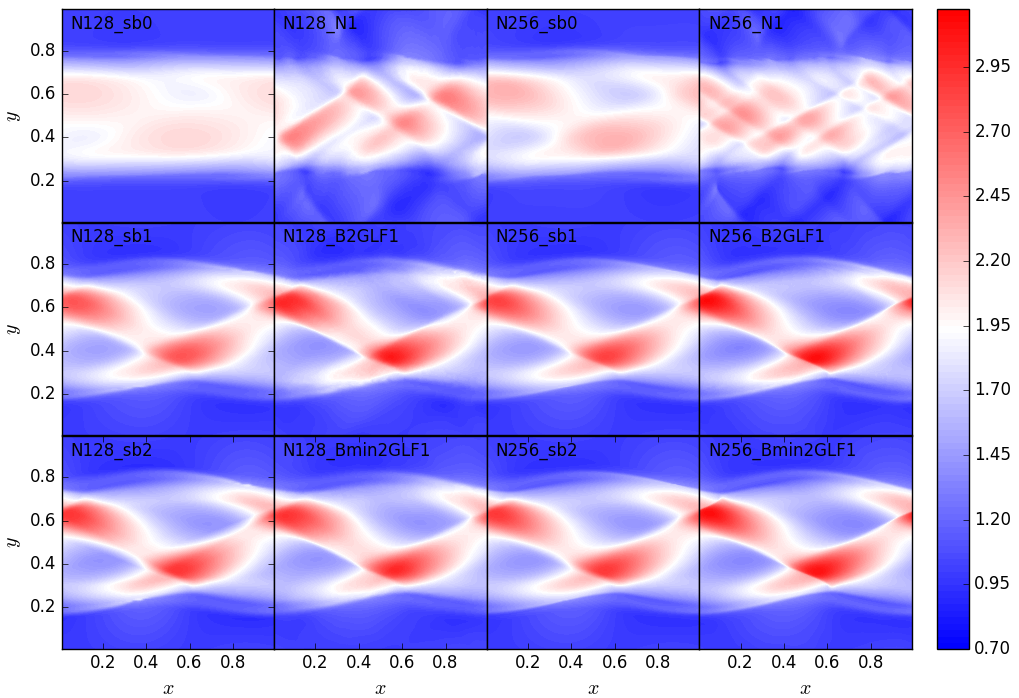}}
\caption{Same as Fig. \ref{figKHcont35}, but for supersonic initial conditions at time $t=8.5$.}
\label{figKHcontsup85}
\end{figure*}

\subsubsection{Kelvin-Helmholtz instability}
\label{secKHI}

The Kelvin-Helmholtz instability (KHI) is a classical hydrodynamical instability \citep[e.g.][]{chandra61} that operates in systems where a velocity shear is present. It has attracted significant attention as a test problem ever since \cite{agertz07} showed that SPH has difficulties resolving the KHI. In addition, \cite{springel10} claimed that grid-based methods violated Galilean invariance using the KHI with an added bulk velocity. This makes it an interesting test problem to consider.

The simplest two-dimensional setup demonstrating the KHI consists of two fluids shearing past each other on a square $0 < (x,y) \leq 1$, periodic in both $x$ and $y$, so that the unperturbed velocity ${\bf v}=(u, v)^T=(u_0(y), 0)^T$, with
\begin{equation}
u_0(y)=\left\{\begin{array}{ll}
U_0 & \frac{1}{4} < y \leq \frac{3}{4}\\
-U_0 & \mathrm{otherwise,}\end{array}\right.
\label{eqKHU}
\end{equation}
where $U_0$ is a constant. While strictly not necessary for the KHI, we take a density profile of similar profile in order to highlight the mixing properties of the schemes:
\begin{equation}
\rho_0(y)=\left\{\begin{array}{ll}
\rho_\mathrm{H} & \frac{1}{4} < y \leq \frac{3}{4}\\
\rho_\mathrm{L} & \mathrm{otherwise,}\end{array}\right.
\end{equation}
where $\rho_\mathrm{H}$ and $\rho_\mathrm{L}$ are a constant high and low density, respectively.
Taking a constant pressure $p_0$, we have an equilibrium solution that is however unstable to the KHI.

Analysis of the incompressible problem \citep[e.g.][]{chandra61} reveals that perturbations grow $\propto \exp(st)$ with growth rate
\begin{equation}
s=2U_0 \frac{\sqrt{\rho_\mathrm{H}\rho_\mathrm{L}}}{\rho_\mathrm{H}+\rho_\mathrm{L}} k_x,
\end{equation}
where $k_x$ is the horizontal wave number. Interestingly, \emph{all} wavenumbers are unstable, and the highest wave numbers have the largest growth rates. On an unstructured grid, initial perturbations are present on the grid scale, and because they grow fastest they come to dominate the solution. Fortunately, in the case of the KHI there is a simple way to regularize the problem by considering smooth profiles of velocity and density in stead of step functions. As shown in \cite{chandra61}, if we take the transition between $U_0$ and $-U_0$ to be linear in $y$ over a distance $d$, unstable wave numbers must satisfy
\begin{equation}
k_x < \frac{\kappa_0}{2d},
\end{equation}
where $\kappa_0\approx 1.27846$ is the solution of $\kappa=1+\exp(-\kappa)$. Roughly speaking, wavelengths smaller than $d$ are stable. Smooth initial conditions therefore lead to a better-posed problem, at least at early times \citep{mcnally12}, but in order to make the analysis more quantitative we need linear growth rates to compare against, which are computed using the method described in appendix \ref{secKHgrow}.

In order to obtain a smooth profile, we first define a function
\begin{equation}
f(t)=\left\{\begin{array}{ll}
\exp(-1/t) & t > 0 \\
0 & \mathrm{otherwise,}\end{array}\right.
\end{equation}
so that
\begin{equation}
g(t)=\frac{f(t)}{f(t)+f(1-t)}
\end{equation}
describes a smooth transition from $0$ to $1$ over an interval $0 \leq t \leq 1$. A smooth version of the discontinuous velocity profile (\ref{eqKHU}) is then given by
\begin{equation}
u_0(y)=2U_0g\left(\frac{1}{2}+\frac{y-1/4}{d}\right)g\left(\frac{1}{2}-\frac{y-3/4}{d}\right) - U_0,
\end{equation}
where the discontinuity has been spread over a distance $d$. Similarly for the initial density profile:
\begin{equation}
\rho_0(y)=(\rho_\mathrm{H}-\rho_\mathrm{L})g\left(\frac{1}{2}+\frac{y-1/4}{d}\right)g\left(\frac{1}{2}-\frac{y-3/4}{d}\right) +\rho_\mathrm{L}.
\end{equation}
Growth rates were computed for $U_0=1/2$, $\rho_\mathrm{L}=1$, $\rho_\mathrm{H}=2$, $p_0=2.5$ and $\gamma=1.4$ for various values of $d$ and $k_x$. The results are shown in Fig. \ref{figKHlinear}. It is clear that, as expected from the incompressible analysis, for finite $d$ there is a maximum wave number for which the flow is unstable. The effect of compressibility is to reduce the growth rate, but with $p_0=2.5$ the velocities are all subsonic and for $d \rightarrow 0$ the growth rates come close to the incompressible result.

We now choose a relatively large value of $d=0.25$, which makes for only one unstable mode with $k_x=2\pi$. The linear growth rate was computed to be $s=1.551$. Following \cite{mcnally12}, we monitor the kinetic energy using the vertical velocity only, but integrated over the entire computational domain:
\begin{equation}
E_\mathrm{v}=\int \frac{\rho v^2}{2} dxdy.
\end{equation}
This quantity is expected to grow at a rate $2s$. As initial conditions, on top of the background flow as specified above, we put in the eigenvector belonging to the growth rate $s$ with a velocity amplitude of $\sim 10^{-3}$.

Results are shown in Fig. \ref{figKHgrow} for first order schemes at a resolution of $128\times 128$ and second order schemes at a resolution of $32 \times 32$. These resolutions were chosen as to highlight differences between the various schemes; at higher resolutions they all converge to the linear result, which is shown by the thick solid line. The linear phase lasts until roughly $t=3.5$, after which the instability saturates. As an average growth rate between $t=0$ and $t=2$, we measure $s=0.6$ for the first order Roe scheme and $s=0.9$ for the first order $N$ scheme. This trend of the $N$ scheme showing higher growth rates continues towards higher resolution until both first order schemes approach the theoretical growth rates. In order for the first order Roe scheme to produce similar growth rates, we need roughly a factor of 2 higher resolution compared to the $N$ scheme. A similar story, but less dramatic, holds for the second order schemes: at a resolution of $32\times 32$, we measure $s=1.4$ for the $B$min scheme and $s=1.3$ for the second order Roe scheme. For resolutions $64 \times 64$ and higher, the results of the two second order schemes become indistinguishable, at least in the linear phase.

The KHI saturates by perturbations rolling up into large vortices. The onset of this phase at $t=3.5$ is depicted in Fig. \ref{figKHcont35}. While at this early non-linear stage, all schemes still give fairly similar results, two things are worth pointing out. First of all, it is clear from the top row, which shows results for the first order schemes, that the $N$ scheme outperforms the first order Roe scheme at a resolution of $128\times 128$. This was clear from the growth rates in Fig. \ref{figKHgrow}, but the top left panels really drive this point home. The first order Roe scheme at $256\times 256$ gives comparable results to the $N$ scheme at resolution $128\times 128$. All second order schemes show very similar amplitudes at this stage at the resolutions considered. The second point is the small artifact around $(x,y)=(0.6, 0.7)$ seen in the first-order Roe scheme as well as the {\sc astrix} results. This kink occurs at the density interface where the $y$-velocity is close to zero, which means the eigenvalue corresponding to a contact discontinuity in the $y$-direction is close to zero, which leads to low numerical dissipation. In this particular case, the numerical dissipation is in fact too low for this kink to disappear. Numerical experiments show that this artifact disappears when adding a constant $y$-velocity to the whole domain.

It is a well-known problem of linearised Riemann solvers like the Roe solver that eigenvalues close to zero can lead to numerical artifacts. Most notably, the Roe solver can not deal properly with transsonic rarefaction waves, where the eigenvalue corresponding to the slow sonic wave passes through zero, which requires some form of entropy fix \citep{harten83}. While entropy fixes have been designed for multidimensional upwind methods \citep{sermeus05}, these have not been implemented in {\sc astrix} at present and in any case they would not remove the kinks seen in Fig. \ref{figKHcont35}.  In fact, results obtained with the HLLC solver, which in general does not require an entropy fix to deal with transsonic rarefactions, show exactly the same feature.  Fortunately, its effect is much less severe than errors in transsonic rarefactions, as it appears to be limited to a few wiggles in the density profile and does not grow with time.

The KHI for the same setup but at at later time $t=5$ is shown in Fig. \ref{figKHcont50}. From the first-order schemes in the top row, it is clear that indeed the kinks visible in Fig. \ref{figKHcont35} do not grow. The most obvious differences between the various panels in Fig. \ref{figKHcont50} are the secondary instabilities seen in the second-order {\sc astrix} results as well as in the results obtained with the Roe solver using the superbee flux limiter. {\color {red} The same holds for results obtained with the HLLC solver.} There has been a lot of discussion in the literature on these instabilities \citep{mcnally12, lecoanet16}. While it is clear that there is a relation between the occurrence of these secondary instabilities and numerical diffusion (for the Roe  and HLLC  solver, they only show up when using the least-diffusive flux limiter), if there is no convergence with resolution to a well-defined solution this relation is meaningless. And this appears to be the case with the KHI without any physical dissipation (i.e. viscosity), even with smooth initial conditions as applied here. While the Roe solver with the minmod flux limiter does not show small-scale structure at the resolutions shown in Fig. \ref{figKHcont50}, they do show up at later times at higher resolution. Therefore, while smoothing the initial conditions leads to well-posed problem in the linear phase, where growth rates can be compared to results from linear calculations, at later times physical dissipation is needed to regularise the solution \citep{lecoanet16}.

Interestingly, there is a variant of the KHI that appears to be well-posed even at later times. The KHI as discussed until now has velocities that are subsonic: $U_0^2 < \gamma p_0/\rho_\mathrm{L}$. It is known that for supersonic velocities, the KHI changes character drastically \citep{karimi16}, with no roll-up into large vortices. In order to study the supersonic KHI, we take the same initial conditions but lower the initial pressure to $p_0=0.1$. While again linear growth rates can be computed, it is more difficult to focus on a single growing mode, since even at $d=0.25$ growth rates are positive up to at least $k_x=20\pi$, although the growth rates decrease rapidly with $k_x$.

The non-linear phase for the supersonic KHI at $t=8.5$ is shown in Fig. \ref{figKHcontsup85}. The top row shows results for the first-order schemes, and while the $N$-scheme shows stronger growth for $k_x=2\pi$, it is also clear that other modes are present, unlike in the first order Roe scheme. This is again due to the unstructured grid: all modes are present initially, and they all grow. Only if the scheme gets the relative growth rates correct will the $k_x=2\pi$ mode stand out, which happens at much higher resolution than depicted in Fig. \ref{figKHcontsup85} in both the first Roe scheme and the $N$ scheme.

The results improve drastically for the second-order schemes, for which all results look remarkably similar. The few small-scale features that can be seen are agreed upon by all schemes, suggesting that the solution to the supersonic KHI is well-behaved even in the absence of physical dissipation. The maximum density increases slightly when going from $128\times 128$ to $256\times 256$ for all schemes, but the results appear to be almost converged at this resolution. Note that the {\sc astrix} results show a higher maximum density compared to the Roe results.  Results obtained with the HLLC solver are almost indistinguishable from the Roe results.  It is interesting to note that the $LDA$ scheme performs equally well on this problem despite sharp gradients present in the solution. The same can not be said for the Roe  and HLLC schemes  if we force a second-order update everywhere (not shown in Fig. \ref{figKHcontsup85}), in which case spurious oscillations appear in the solution.

\section{Discussion}
\label{secDisc}

In this paper, we have presented an implementation of a residual distribution method in an astrophysical fluid dynamics package {\sc astrix}. A key difference between {\sc astrix} and other grid based methods is its inherently multidimensional nature. While multidimensional integration schemes, usually in the form suggested by \cite{colella90}, have become a standard part of codes working on structured meshes \citep[e.g.][]{fromang06,mignone07,stone08}, for methods employing unstructured grids these are rarely mentioned. This is probably partly due to the inherent complexity of unstructured meshes, but also due to the fact that methods based on one-dimensional flux estimates perform very well even for a naive implementation of multidimensional integration. For example, the Roe solver method as tested against {\sc astrix} in this paper performs equally well when using formally second order Strang splitting \citep{strang68} as when just alternating the direction of integration. However, especially in the case of the isentropic vortex problem, there are clear advantages for a truly multidimensional update as employed in {\sc astrix}. In this section, we discuss the limitations of the current version and pathways to improvement.

From the test problems presented in section \ref{secTest} it is clear that {\sc astrix} performs at least as well as the dimensionally split Roe solver in all cases, and performs significantly better at some, notably the isentropic vortex and the Noh problem. This performance comes at a price of increased complexity of the method and increased computational time.  In order to assess this quantatively we ran a simple speed test comparing {\sc astrix} to various publicly available codes. Besides {\sc rodeo}, which was used to compare to {\sc astrix} in terms of accuracy, we took the structured grid code {\sc pluto} \citep{mignone07}\footnote{See \url{http://plutocode.ph.unito.it/}}, the meshless code {\sc gizmo} \citep{hopkins14,hopkins15}\footnote{See \url{http://www.tapir.caltech.edu/~phopkins/Site/GIZMO.html}} and the moving mesh code {\sc rich} \citep{yalinewich15}\footnote{see \url{https://github.com/bolverk/huji-rich}}. We chose a particularly simple test problem for the speed comparison: constant density, constant pressure and zero velocity on the unit square. The main reason for this choice it that it makes the Lagrangian and Eulerian approaches equal, allowing for a fairer comparison of the different codes. For example, the public version of {\sc gizmo} only allows for Lagrangian meshless integration, either Meshless Finite Mass (MFM) or Meshless Finite Volume (MFV), see \cite{hopkins15} for details. In this case, a hydrostatic calculation should eliminate overhead for example from finding new neighbours, or, in the case of {\sc rich}, determining the new Voronoi tesselation. A second reason for this simple problem is that all codes should be able to do this problem without any special effort. For example, {\sc rich} can only do the Noh problem when the "cold flow" option is activated, which makes the code three times slower.

\begin{table}
        \centering
        \caption{Speed comparison of various codes on a hydrostatic test problem (constant density, constant pressure, zero velocity) on the unit square at a resolution equivalent to $128\times128$. First column lists the name of the code, the second column lists relevant parameters, and the third column gives the computing time spent per cell per time step. Only a single CPU was used in all cases. Tests were performed on an Intel $2.2$ GHz Core i7 with 8 GB DDR3 memory.}
        \begin{tabular}{lcc}
                \hline
                Code & Parameters & Time/cell/step\\
                & & ($\mathrm{\mu s}$)\\
                \hline
                {\sc astrix} & $N$1 & 0.888\\
                {\sc astrix} & $B$2 & 2.09\\
                {\sc rodeo} & sb1 & 0.938\\
                {\sc pluto} & RK2, Roe, linear & 0.864\\
                {\sc gizmo} & MFM & 9.42\\
                {\sc gizmo} & MFV & 10.2\\
                {\sc rich} & HLLC, Eulerian & 19.6\\
                \hline
        \end{tabular}
        \label{tabSpeed}
\end{table}

The results of the speed test are shown in Table \ref{tabSpeed}. All codes except {\sc astrix} with the $N$ scheme were run in order to achieve second-order accuracy. The third column lists the CPU time per cell per time step, as measured on an Intel Core i7 2.2 GHz CPU with 8 GB DDR3 memory. These timings represent the speed of the algorithms themselves, and ignore for example the fact that {\sc astrix} can run with larger time steps than a dimensionally split scheme such as {\sc rodeo}. They also ignore  the initial construction of the mesh. While this takes virtually no time in the case of the Roe solver, for {\sc astrix} it is a more significant effort. As a rule of thumb, the creation of the mesh takes about as much time as taking ten time steps with second order $LDA$. In all of the test problems performed here, this means mesh creation takes roughly $1\%$ of the total time spent. While this is clearly not an issue for a static mesh, if the mesh is to be updated frequently this can quickly become a bottleneck.

The results in Table \ref{tabSpeed} show that the CPU version of {\sc astrix} is roughly $2$-$2.5$ times slower per cell per time step than structured grid codes such as {\sc rodeo} and {\sc pluto}. This is due to the increased complexity of a multidimensional upwind update. On the other hand, {\sc astrix} compares favourably to methods that do not employ a structured grid: it is roughly $5$ times faster per cell per time step than {\sc gizmo}, and roughly $10$ times faster per cell per time step than {\sc rich}.

We have focused on static uniform meshes, to simplify the error analysis and to make a fair comparison to the Roe solver. One of the beautiful features of unstructured mesh solvers is that no extra effort is required to run on a mesh with varying resolution, unlike for example in traditional AMR, where interpolation at resolution jumps is necessary. As explained in section \ref{secRemoveLowQuality}, it is straightforward to generate a mesh with spatially varying resolution, which therefore makes running on non-uniform meshes a trivial matter. However, this is only useful if it is known in advance the location where high resolution is going to be necessary, and if that location is fixed in time. While such problems do exist (these can be tackled in traditional methods using static mesh refinement or nested grids), in general it is necessary to be able to dynamically update the mesh. This can be done by adding and removing vertices, but since these operations change the triangulation and therefore the connectivity of the mesh these tasks are not trivial and will be the subject of future work. Here we also note that a moving mesh approach is also possible \citep{michler03}.

The current implementation works in two spatial dimensions only. This is the simplest setup in which to demonstrate the power of multidimensional upwind methods, but the approach of section \ref{secResDist} can be generalised in a straightforward way to three dimensions, at the expense of adding a row and a column to all matrices, and the resulting increase in complexity of the matrix element computations. Most of the work in going to three dimensions will go into the generation of the mesh. Some of the nice properties of Delaunay triangulations do not generalise to three dimensions, and in particular some low-quality tetrahedra can survive Delaunay refinement \citep[see e.g.][]{shewchuk02}. Fortunately there are ways to remove these \citep{cheng00}.

We have considered Cartesian coordinates only. For many astrophysical problems, spherical or cylindrical coordinates are better choices to represent the flow. Astrophysical discs, for example, in traditional Eulerian methods are best described in cylindrical coordinates, so that the flow is mostly aligned with the mesh. A Cartesian frame leads to excessive diffusion as angular momentum is not conserved, which is particularly a problem in the field of disc-planet interactions \citep{devalborro06}. Even when using cylindrical coordinates, care must be taken to conserve angular momentum to machine precision \citep{kley98}. The test problem of the isentropic vortex clearly shows that angular momentum is much better conserved using a multidimensional upwind method compared to the dimensionally split Roe solver on a Cartesian grid. However, when exact angular momentum conservation is required, this is possible in two steps. First, there is no reason why a Delaunay triangulation can not be used to tesselate a region in cylindrical coordinates $(r,\varphi)$: just replace $x$ with $r$ and $y$ with $\varphi$ in Fig. \ref{figMesh2D} (and remove the periodicity in $x$). Second, solve the Euler equations in cylindrical coordinates. The main problem here is that while a conservative linearisation in general coordinates does exist \citep{eulderink95}, this does not conserve angular momentum to machine precision \citep{paardekooper06}. Fortunately, it is possible to formulate the residual distribution schemes presented here even when a conservative linearisation is not available \citep{csik02}. This formulation can bring the power of multidimensional upwind methods and unstructured grids to simulations of astrophysical discs and this will be considered in a future work.

While we have considered hydrodynamics only, it is possible to add additional physics to the residual distribution methods presented here. For example, \cite{csik02} use their residual distribution formulation without a conservative linearisation to study the case of ideal magnetohydrodynamics, which included a source term to clean the divergence of the magnetic field. Constrained transport algorithms on unstructured meshes were recently presented in \cite{mocz14, mocz16}. A general diffusion solver in residual distribution form was presented in \cite{nishikawa07}, which can for example be used to include self-gravity or radiative transport in the diffusion limit. A more accurate radiative transfer method on unstructured grid can for example be found in \cite{ritzerveld07} and \cite{paardekooper10}. For examples of (general) relativistic solvers on unstructured grids see \cite{anninos05,duffell11}. Therefore, while unstructured meshes call for different implementations, they impose no limit of the amount of physics that can be included in simulations.

We have only considered schemes that are at most second order accurate in space and time. While higher-order explicit residual distribution schemes are not available as yet, it is expected that by considering higher order elements the extension of the current method to higher order would be straightforward (\cite{ricchiuto10}, see also \cite{cohen01,jund07}).

\section{Conclusions}
\label{secCon}

We have presented a GPU implementation of a multidimensional upwind, or residual distribution, method acting on an unstructured Delaunay mesh in two spatial dimensions. While more expensive than than traditional grid-based methods, it has clear advantages when dealing with multidimensional flows over methods that use one-dimensional flux estimates as building blocks for a multidimensional integration. This is exemplified in several test problems, most notably the problem of a stationary isentropic vortex. Our GPU implementation shows speedups of $\sim 100$ for mesh generation and $\sim 250$ for the hydrodynamics, with the most expensive kernels coming close to the theoretical performance limit of our Tesla K20m GPU.

\section*{Acknowledgements}
SJP is supported by a Royal Society University Research Fellowship.

\bibliography{paardekooper}

\appendix
\onecolumn
\section{Multidimensional upwind matrices}
\label{appMat}
Consider the Euler equations in two spatial dimensions in the form:
\begin{equation}
\frac{\partial \bW}{\partial t} + \mathcal{A}\frac{\partial \bW}{\partial x}+\mathcal{B}\frac{\partial \bW}{\partial y}=0,
\end{equation}
then the matrix $\mathcal{K}=\mathcal{A} n_x + \mathcal{B} n_y$ is given by
\begin{equation}
\mathcal{K}=\left(\begin{array}{cccc}
0 & n_x & n_y & 0\\
\alpha n_x-uw & w-\gamma_2un_x & un_y-\gamma_1 vn_x & \gamma_1 n_x\\
\alpha n_y-vw & vn_x-\gamma_1 un_y & w-\gamma_2vn_y &  \gamma_1 n_y\\
(\alpha-h)w & hn_x-\gamma_1 uw & hn_y-\gamma_1 vw & \gamma w
\end{array}\right),
\end{equation}
with $w=un_x+vn_y$, $\alpha=\gamma_1(u^2+v^2)/2$ and $\gamma_1=\gamma-1$, and has eigenvalues $\lambda_1=w+c$, $\lambda_2=w-c$, $\lambda_3=\lambda_4=w$, and right eigenvectors
\begin{eqnarray}
{\bf e}_1&=&\left(1,u+cn_x,v+cn_y,h+cw\right)^T\\
{\bf e}_2&=&\left(1,u-cn_x,v-cn_y,h-cw\right)^T\\
{\bf e}_3&=&\left(0,-n_y,n_x,vn_x-un_y\right)^T\\
{\bf e}_4&=&\left(1,u,v,\alpha/\gamma_1\right)^T.
\end{eqnarray}
Putting the eigenvectors as columns in a matrix R:
\begin{equation}
\mathcal{R}=\left(\begin{array}{cccc}
1 & 1 & 0 & 1\\
u+cn_x & u - cn_x & -n_y & u\\
v+cn_y & v-cn_y & n_x & v \\
h+cw & h-cw & vn_x-un_y & \alpha/\gamma_1
\end{array}\right),
\end{equation}
then
\begin{equation}
\mathcal{R}^{-1}=\frac{1}{2c}\left(\begin{array}{cccc}
\alpha_c-\frac{w}{n_x^2+n_y^2} & \frac{n_x}{n_x^2+n_y^2} -\gamma_1u_c & \frac{n_y}{n_x^2+n_y^2}-\gamma_1v_c & -\frac{\gamma_1}{c}\\
\alpha_c+\frac{w}{n_x^2+n_y^2} & -\frac{n_x}{n_x^2+n_y^2}-\gamma_1u_c & -\frac{n_y}{n_x^2+n_y^2}-\gamma_1v_c & -\frac{\gamma_1}{c}\\
2c\frac{un_y-vn_x}{n_x^2+n_y^2} & -\frac{2cn_y}{n_x^2+n_y^2} & \frac{2cn_x}{n_x^2+n_y^2} & 0\\
2c-2\alpha_c & 2\gamma_1u_c & 2\gamma_1v_c & \frac{2\gamma_1}{c}
\end{array}\right),
\end{equation}
and $\mathcal{K}$ can be diagonalised with $\mathcal{K}=\mathcal{R}^{-1}\Lambda \mathcal{R}$,  where $\Lambda$ is a diagonal matrix with the eigenvalues on the diagonal. If we put generic values $l_1$, $l_2$, $l_3=l_4$ in $\Lambda$, we get
\begin{equation}
\Lambda\mathcal{R}=c\left(\begin{array}{cccc}
l_1/c & l_1/c & 0 & l_1/c\\
(u_c+n_x)l_2 & (u_c-n_x)l_2 & -n_yl_2/c & u_cl_2\\
(v_c+n_y)l_3 & (v_c-n_y)l_3 & n_xl_3/c & v_cl_3 \\
(h_c+w)l_3 & (h_c-w)l_3 & (v_cn_x-u_cn_y)l_3 & \alpha l_3/(\gamma_1c)
\end{array}\right),
\end{equation}
and matrix entries
\begin{eqnarray}
\mathcal{K}_{1,1}&=&\frac{\alpha_c}{c}l_{123}-\frac{w}{c}l_{12}+l_3,\\
\mathcal{K}_{1,2}&=&-\frac{\gamma_1 u_c}{c}l_{123}+\frac{n_x}{c}l_{12},\\
\mathcal{K}_{1,3}&=&-\frac{\gamma_1 v_c}{c}l_{123}+\frac{n_y}{c}l_{12},\\
\mathcal{K}_{1,4}&=&\frac{\gamma_1}{c^2}l_{123},
\end{eqnarray}
\begin{eqnarray}
\mathcal{K}_{2,1}&=&(\alpha_c u_c-wn_x)l_{123}+(\alpha_c n_x-u_cw)l_{12},\\
\mathcal{K}_{2,2}&=&\left(n_x^2-\gamma_1 u_c^2\right)l_{123}-\gamma_2u_cn_xl_{12}+l_3,\\
\mathcal{K}_{2,3}&=&(n_xn_y-\gamma_1 u_cv_c)l_{123}+(u_cn_y-\gamma_1 v_cn_x)l_{12},\\
\mathcal{K}_{2,4}&=&\frac{\gamma_1 u_c}{c}l_{123}+\frac{\gamma_1 n_x}{c}l_{12},
\end{eqnarray}
\begin{eqnarray}
\mathcal{K}_{3,1}&=&(\alpha_c v_c-wn_y)l_{123}+(\alpha_c n_y-v_cw)l_{12},\\
\mathcal{K}_{3,2}&=&(n_xn_y-\gamma_1 u_cv_c)l_{123}+(v_cn_x-\gamma_1 u_cn_y)l_{12},\\
\mathcal{K}_{3,3}&=&\left(n_y^2-\gamma_1 v_c^2\right)l_{123}-\gamma_2v_cn_yl_{12}+l_3,\\
\mathcal{K}_{3,4}&=&\frac{\gamma_1 v_c}{c}l_{123}+\frac{\gamma_1 n_y}{c}l_{12},
\end{eqnarray}
\begin{eqnarray}
\mathcal{K}_{4,1}&=&(\alpha_c h_c-w^2)l_{123}+w\left(\alpha_c-h_c\right)l_{12},\\
\mathcal{K}_{4,2}&=&\left(wn_x-u-\alpha_c u_c\right)l_{123}+(h_cn_x-\gamma_1 u_cw)l_{12},\\
\mathcal{K}_{4,3}&=&\left(wn_y-v-\alpha_c v_c\right)l_{123}+(h_cn_y-\gamma_1 v_cw)l_{12},\\
\mathcal{K}_{4,4}&=&\frac{\gamma_1 h_c}{c}l_{123}+\frac{\gamma_1 w}{c}l_{12}+l_3,
\end{eqnarray}
with $\alpha_c=\alpha/c$, $u_c=u/c$, $v_c=v/c$, $\gamma_2=\gamma-2$, $l_{123}=(l_1 + l_2-2l_3)/2$ and $l_{12}=(l_1-l_2)/2$. If we set $l_k = \lambda_k$ for $k=1,2,3,4$, we recover $\mathcal{K}$, while if we set $l_k=\lambda_k^\pm$ we get $\mathcal{K}^\pm$.

\section{Blast wave solutions}
\label{secSedovSolution}

Analytical blast wave solutions emerged shortly after the Second World War \citep{bethe47, sedov46, taylor50} and make useful test problems for numerical gas dynamics codes. Most well known in the astrophysical community is the spherical blast wave, which is a useful model for a supernova explosion. However, in this work we are concerned with at most two spatial dimensions, and therefore we need lower-dimensional analogs of the spherical blast wave. Fortunately, these exist \citep[e.g.][]{sedov59,kamm07}, and below we briefly describe the procedure of obtaining reference blast wave solutions to compare to hydrodynamical calculations.

Consider a point explosion in spherical geometry, a line explosion in cylindrical geometry and a plane explosion in Cartesian geometry. Assume a strong shock, so that the preshock pressure $P_1$ plays no role. The only parameters in the problem are the preshock density $\rho_1$ and a measure of the total input energy $E$, which in the cylindrical case is an energy per unit length, and in the Cartesian case an energy per unit area. This leads to the definition of a similarity variable
\begin{equation}
\eta \equiv r \left(\frac{Et^2}{\rho_1}\right)^\frac{-1}{3+a},
\end{equation}
where $r$ denotes the distance to the initial energy release, and $a=0$ for Cartesian, $a=1$ for cylindrical, and $a=2$ for spherical coordinates.

Jump conditions for a strong shock give
\begin{eqnarray}
\rho_2&=&\frac{\gamma+1}{\gamma-1}\rho_1,\\
P_2&=&\frac{2\rho_1 v_s^2}{\gamma+1},\\
v_2 &=&\frac{2v_s}{\gamma+1},
\end{eqnarray}
where $v_s$ is the velocity of the shock and $\gamma$ is the ratio of specific heats. In the region behind the shock, use the ansatz
\begin{eqnarray}
\rho(r,t) &=& \rho_2 \hat\rho(\eta),\\
P(r,t)&=&\frac{8}{(3+a)^2(\gamma+1)} \rho_1\frac{r^2}{t^2}\hat P(\eta),\\
v(r,t)&=&\frac{4}{(3+a)(\gamma+1)}\frac{r}{t}\hat v(\eta),
\end{eqnarray}
Note that $\hat\rho(\eta_s)=\hat P(\eta_s)=\hat v(\eta_s)=1$, where $\eta_s$ denotes the position of the shock.

The continuity equation reads
\begin{equation}
\partial_t \rho + v\partial_r\rho + \rho\partial_r v + a\rho v/r=0,
\end{equation}
where again $a=0$ for Cartesian, $a=1$ for cylindrical, and $a=2$ for spherical coordinates. Transforming to $(\eta,t)$ gives:
\begin{equation}
-\eta d_\eta \hat\rho  + \frac{2}{\gamma+1}d_\eta \left(\eta \hat\rho \hat v\right) + \frac{2a}{\gamma+1} \hat\rho \hat v=0,
\end{equation}
The momentum equation reads
\begin{equation}
\partial_t v +  v\partial_r v+\partial_r P/\rho=0.
\end{equation}
Transforming to $(\eta,t)$ gives:
\begin{equation}
-(3+a)\hat v-
2\eta d_\eta \hat v +
\frac{4}{\gamma+1}\hat v d_\eta \left(\eta \hat v\right)+
\frac{\gamma-1}{\gamma+1}\frac{2}{\hat\rho }\left[2 \hat P  + \eta d_\eta \hat P \right]=0.
\end{equation}
The energy equation reads
\begin{equation}
\partial_t\left(\frac{\rho v^2}{2}+\frac{P}{\gamma-1}\right)+r^{-a}\partial_r\left(r^a v\left(\frac{\rho v^2}{2}+\frac{\gamma P}{\gamma-1}\right)\right)=0
\end{equation}
Transforming to $(\eta,t)$ gives:
\begin{eqnarray}
-(3+a)\left(\hat\rho  \hat v^2+\hat P \right)-
\eta d_\eta\left(\hat\rho  \hat v^2+\hat P \right)+
\frac{2}{\gamma+1}\left[(a+3)\hat v\left(\hat\rho  \hat v^2+\gamma \hat P \right) + \eta d_\eta\left(\hat v\left(\hat\rho  \hat v^2+\gamma \hat P \right)\right)\right]=0
\end{eqnarray}
The three equations for $\hat\rho$, $\hat P$ and $\hat v$ need to be solved subject to the constraint that the total energy of the solution equals the initial energy input:
\begin{equation}
b \int_0^{R(t)} \left(\frac{\rho v^2}{2}+\frac{P}{\gamma-1}\right)r^a dr = E,
\end{equation}
where $b=4\pi$ for spherical coordinates, $2\pi$ in cylindrical coordinates, and $1/2$ in Cartesian coordinates. In dimensionless form:
\begin{equation}
\frac{8b}{(3+a)^2(\gamma^2-1)}  \int_0^{\eta_s}\left(\hat\rho  \hat v^2+ \hat P \right)\eta^{2+a} d\eta = 1.\label{eqSedovEnCons}
\end{equation}

\begin{figure*}
\centering
\resizebox{\hsize}{!}{\includegraphics[]{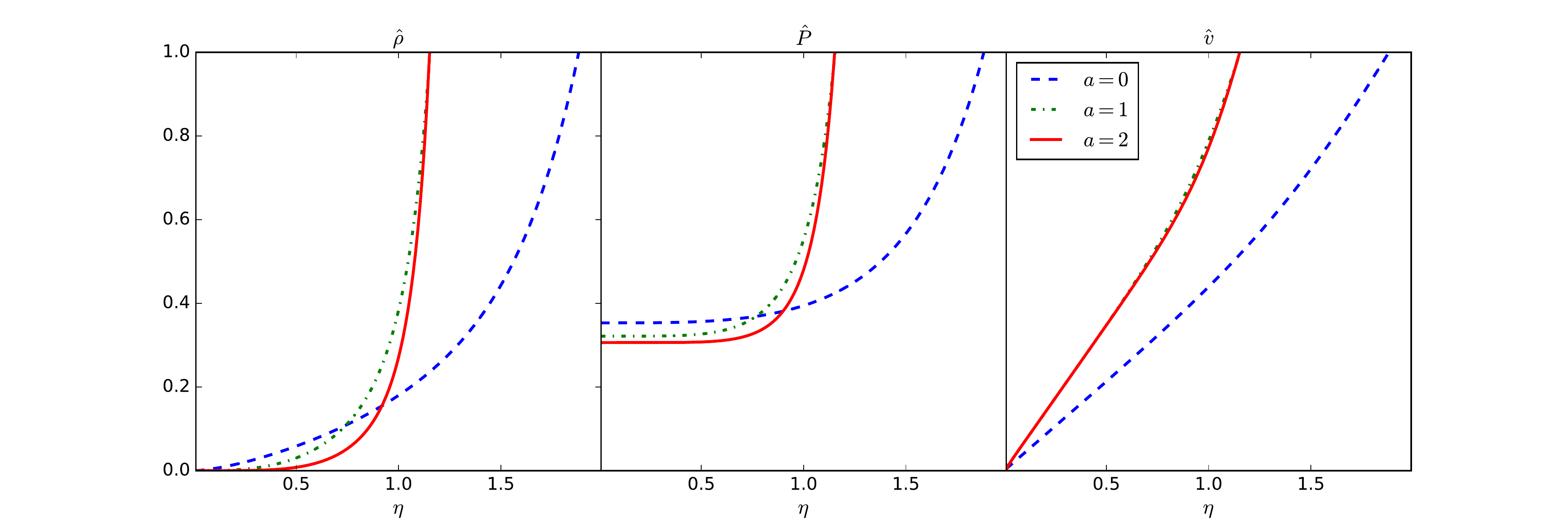}}
\caption{Dimensionless density (left panel), pressure (middle panel) and velocity (right panel) for the post-shock blast wave solution in three different geometries.}
\label{figSedovAnalytic}
\end{figure*}

In order to facilitate numerical integration, we rewrite the equations of mass and momentum conservation as:
\begin{eqnarray}
\left[\frac{2\hat v}{\gamma+1}-1\right]\eta d_\eta \hat\rho  &=&- \frac{2\hat\rho }{\gamma+1} \eta d_\eta \hat v  - \frac{2a+2}{\gamma+1} \hat\rho \hat v, \label{eqSedovMass}\\
\eta d_\eta \hat P &=&\frac{\gamma+1}{\gamma-1}\frac{(3+a)}{2}\hat\rho \hat v  -
\frac{2\hat\rho \hat v^2}{\gamma-1} - \left[\frac{2\hat\rho \hat v}{\gamma-1} -\hat\rho \frac{\gamma+1}{\gamma-1}\right]\eta d_\eta \hat v -2\hat P.\label{eqSedovMom}
\end{eqnarray}
Use these two equations in the energy equation to obtain
\begin{eqnarray}
\left[\frac{4\hat\rho \hat v^2+2\gamma \hat P }{\gamma+1}-2\hat\rho \hat v - \left[\frac{2\gamma \hat v}{\gamma+1}-1\right]\left[\frac{2\hat\rho \hat v}{\gamma-1} -\hat\rho \frac{\gamma+1}{\gamma-1}\right]\right]\eta d_\eta \hat v =\nonumber\\
(3+a)\left(\hat\rho  \hat v^2+\hat P \right)-\frac{4}{\gamma+1}\hat\rho  \hat v^3-\frac{2(a+3)}{\gamma+1}\gamma \hat P \hat v - 
\left[\frac{2\gamma \hat v}{\gamma+1}-1\right] \left\{\frac{\gamma+1}{\gamma-1}\frac{(3+a)}{2}\hat\rho \hat v  -
\frac{2\hat\rho \hat v^2}{\gamma-1} -2\hat P \right\}.\label{eqSedovEnergy}
\end{eqnarray}
The resulting expression for $d_\eta \hat v$ can be used in equations (\ref{eqSedovMass}) and (\ref{eqSedovMom}) so that equations for $d_\eta\hat\rho$, $d_\eta\hat P$ and $d_\eta \hat v$ in terms of $\hat\rho$, $\hat P$ and $\hat v$ result. These can be solved numerically using standard techniques. First, one needs to guess a value of $\eta_s$, integrate the equations from $\eta=\eta_s$ to $\eta=0$, and check energy conservation (\ref{eqSedovEnCons}). This process is repeated for different $\eta_s$ until the energy constraint is met. The resulting solutions for $a=0,1,2$ are shown in Fig. \ref{figSedovAnalytic}.

\section{Compressible Kelvin-Helmholtz growth rates}
\label{secKHgrow}

Consider a 2D domain, periodic in $x$ and $y$ with periods $L_x$ and $L_y$ and no gravity:
\begin{eqnarray}
\frac{\partial \rho}{\partial t} + \nabla\cdot\rho {\bf v}&=&0\\
\frac{\partial{\bf v}}{\partial t}+({\bf v}\cdot \nabla){\bf v} + \frac{\nabla p}{\rho}&=&0
\end{eqnarray}
Give the fluid a smooth profile in $x$-velocity and density that depends only on $y$: $\rho_0=\rho_0(y)$, $u_0=u_0(y)$, $v_0=0$, and constant pressure $p_0$. The linear perturbation equations read:
\begin{eqnarray}
\frac{\partial\rho_1}{\partial t}+ u_0\frac{\partial\rho_1}{\partial x} + \rho_0\frac{\partial u_1}{\partial x}+ \rho_0\frac{\partial v_1}{\partial y}+v_1\frac{d\rho_0}{dy}&=&0\\
\frac{\partial u_1}{\partial t}+ u_0\frac{\partial u_1}{\partial x} + v_1\frac{\partial u_0}{\partial y}+\frac{1}{\rho_0}\frac{\partial p_1}{\partial x}&=&0\\
\frac{\partial v_1}{\partial t}+ u_0\frac{\partial v_1}{\partial x} +\frac{1}{\rho_0}\frac{\partial p_1}{\partial y}&=&0,
\end{eqnarray}
together with adiabatic pressure perturbations $p_1=c_0^2\rho_1$ with $c_0^2=\gamma p_0/\rho_0$ the square of the unperturbed sound speed. Take perturbations $\propto \exp(ik_x x - i\omega t)$:
\begin{eqnarray}
(\omega - k_xu_0)\rho_1 -k_x\rho_0u_1+ i\frac{d}{d y}(\rho_0v_1)&=&0\\
(\omega -k_x u_0)u_1+ iv_1\frac{d u_0}{d y}-\frac{k_x p_1}{\rho_0}&=&0\\
i(\omega -k_x u_0)v_1&=&\frac{1}{\rho_0}\frac{d p_1}{d y},
\end{eqnarray}
Write in terms of momenta $a_1 = \rho_0 u_1$ and $b_1 = \rho_0 v_1$:
\begin{eqnarray}
 k_xu_0\rho_1 +k_xa_1-  i\frac{db_1}{dy}&=&\omega\rho_1\\
k_x u_0a_1- i\frac{d u_0}{d y}b_1+k_x p_1&=&\omega a_1\\
ik_x u_0b_1+\frac{d p_1}{d y}&=&i\omega b_1.
\end{eqnarray}
Write all quantities, both perturbed and unperturbed, as a Fourier series, e.g.:
\begin{equation}
u_0=\sum_{n=0}^{N-1}u_{0n}\exp(2\pi i ny/L_y).
\end{equation}
The equations for component $n$ read:
\begin{eqnarray}
 k_x(u_0\rho_1)_n +k_xa_{1n}+\frac{2\pi n}{L_y} b_{1n}&=&\omega\rho_{1n}\\
k_x (u_0a_1)_n- i\left(\frac{d u_0}{d y}b_1\right)_n+k_x \left(c_0^2\rho_1\right)_n&=&\omega a_{1n}\\
k_x (u_0b_1)_n+\frac{2\pi n}{L_y}\left(c_0^2\rho_1\right)_n&=&\omega b_{1n},
\end{eqnarray}
Fourier components of products are of course convolutions, so that for example
\begin{equation}
(u_0\rho_1)_n=\sum_{m=0}^n u_{0(n-m)}\rho_{1m}={\bf u}_n\cdot {\bf d}_1,
\end{equation}
where ${\bf d}_1$ is a vector of length $N$ with entries $\rho_{1m}$ and ${\bf u}_n$ is a vector with entries $u_0(n-m)$, with $0 \leq m < N$. We can therefore write all $N$ equations for the components in the form of matrices:
\begin{eqnarray}
 k_x {\bf Ud}_1+k_x{\bf Ia}_1+ \frac{2\pi}{L_y} {\bf Nb}_1&=&\omega {\bf d}_1\\
k_x {\bf Ua}_1+\frac{2\pi}{L_y}{\bf U'b}_1+k_x {\bf Cd}_1&=&\omega {\bf a}_1\\
k_x {\bf Ub}_1+\frac{2\pi}{L_y}{\bf NCd}_1&=&\omega {\bf b}_1,
\end{eqnarray}
where matrix ${\bf U}$ has entries
\begin{equation}
u_{ij}=\left\{\begin{array}{ll} u_{0(i-j)} & \mathrm{i \geq j}\\ 0 & \mathrm{otherwise}\end{array}\right.
\end{equation}
and matrix ${\bf U'}$ has entries:
\begin{equation}
u'_{ij}=\left\{\begin{array}{ll} (i-j)u_{0(i-j)} & \mathrm{i \geq j}\\ 0 & \mathrm{otherwise}\end{array}\right.
\end{equation}
and matrix ${\bf N}$ has entries:
\begin{equation}
n_{ij}=i\delta_{ij}
\end{equation}
and matrix ${\bf C}$ has entries:
\begin{equation}
c_{ij}=\left\{\begin{array}{ll} c_{0(i-j)}^2 & \mathrm{i \geq j}\\ 0 & \mathrm{otherwise}\end{array}\right.
\end{equation}
Combining ${\bf d}_1$, ${\bf a}_1$ and ${\bf b}_1$ into a single vector ${\bf e}$ of length $3N$:
\begin{equation}
\left(\begin{array}{ccc}
k_x{\bf U} & k_x{\bf I} & \frac{2\pi}{L_y}{\bf N}\\
k_x{\bf C} & k_x{\bf U} & \frac{2\pi}{L_y} {\bf U'}\\
\frac{2\pi}{L_y}{\bf NC} & {\bf 0} & k_x {\bf U}\end{array}\right){\bf e}=\omega {\bf e}
\end{equation}
It therefore remains to find the eigenvalues and eigenvectors of the $3N$ by $3N$ matrix
\begin{equation}
 {\bf A}=\left(\begin{array}{ccc}
k_x{\bf U} & k_x{\bf I} & \frac{2\pi}{L_y}{\bf N}\\
k_x{\bf C} & k_x{\bf U} & \frac{2\pi}{L_y} {\bf U'}\\
\frac{2\pi}{L_y}{\bf NC} & {\bf 0} & k_x {\bf U}\end{array}\right)
\end{equation}
If the maximum of the imaginary parts of the eigenvalues is larger than zero, the wavenumber $k_x = 2\pi m/L_x$ for some integer $m$ is unstable, with growth rate equal to this imaginary part.
\label{lastpage}

\end{document}